\documentclass[a4paper]{article}
\usepackage[utf8]{inputenc}
\usepackage{amsmath}
\usepackage{amsfonts}
\usepackage{amssymb}
\usepackage{amsthm}
\usepackage{bbm}
\usepackage{xcolor}
\usepackage{graphicx}
\usepackage{hyperref}
\numberwithin{equation}{section}
\usepackage{geometry}
\usepackage{caption}
\usepackage{subcaption}

\newtheorem{theorem}{Theorem}[section]

\newtheorem*{notation*}{Notation}

\theoremstyle{definition}
\newtheorem{remark}[theorem]{Remark}

\title{From constant to rough: A survey of continuous volatility modeling}

\author{
    G. Di Nunno$^{1,2}$\\ \href{mailto:giulian@math.uio.no}{giulian@math.uio.no}
    \and
    K. Kubilius$^3$ \\ \href{mailto:kestutis.kubilius@mif.vu.lt}{kestutis.kubilius@mif.vu.lt}
    \and
    Yu. Mishura$^{4,5}$ \\ \href{mailto:yuliyamishura@knu.ua}{yuliyamishura@knu.ua} 
    \and 
    A. Yurchenko-Tytarenko$^1$ \\ \href{mailto:antony@math.uio.no}{antony@math.uio.no}
}

\date{%
    $^1$Department of Mathematics, University of Oslo\\
    $^2$Department of Business and Management Science, NHH Norwegian School of Economics, Bergen\\
    $^3$Faculty of Mathematics and Informatics, Vilnius University\\
    $^4$Department of Probability, Statistics and Actuarial Mathematics, Taras Shevchenko National University of Kyiv\\
    $^5$Division of Mathematics and Physics, M\"alardalen University\\
    [2ex]%
    August 20, 2025
}

\begin{document}

\maketitle

\begin{abstract}
    In this paper, we present a comprehensive survey of continuous stochastic volatility models, discussing their historical development and the key stylized facts that have driven the field. Special attention is dedicated to fractional and rough methods: without advocating for either roughness or long memory, we outline the motivation behind them and characterize some landmark models. In addition, we briefly touch the problem of VIX modeling and recent advances in the SPX-VIX joint calibration puzzle. 
\end{abstract}

\textbf{Acknowledgements.} The present research is carried out within the frame and support of the ToppForsk project nr. 274410 of the Research Council of Norway with the title STORM: Stochastics for Time-Space Risk Models. The third author is supported by The Swedish Foundation for Strategic Research, grant Nr. UKR22-0017, and by Japan Science and Technology Agency CREST, project reference number JPMJCR2115. We are grateful to Åsmund Hausken Sande for his advice on a piece of data analysis as well as to Guido Gazzani and Lorenzo Torricelli for their feedback and excellent reference suggestions.

\section{Introduction}

Ultimately, finance evolves around the interplay between the expected return and risk. The established benchmark for the latter is \textit{volatility}: loosely defined, this term refers to the degree of variability of an asset $S$ in time which is traditionally quantified as
\begin{equation}\label{eq: realized volatility}
    \sqrt{\frac{1}{T} \sum_{t=1}^T \left(\log \frac{S(t)}{S(t-1)} \right)^2}.
\end{equation}
The rationale behind using the \textit{log}-returns in \eqref{eq: realized volatility} is simple and can be traced back to P. Samuelson: the prices are strictly positive and hence it is reasonable to represent them as an exponential of some random variable. For example, if one models the returns as
\[
    \frac{S(t)}{S(t-1)} = e^{\xi_t}, \quad t = 1,..., T,
\]
where $\xi_t \sim \mathcal N(0, \sigma^2)$, $t = 1,..., T$, are centered i.i.d. random variables, the value \eqref{eq: realized volatility} becomes a strongly consistent estimator of the \textit{standard deviation} parameter $\sigma$.

However, it turns out that the metric \eqref{eq: realized volatility} is not as innocuous as it may seem at first sight. For example, if one computes \eqref{eq: realized volatility} using samples from distinct time periods, the results can be \textit{drastically} different. In some sense, it is not a big surprise: there are periods on the market when prices exhibit minimal variation and other times when they fluctuate with extreme intensity (especially during crises). Moreover, as noted in e.g. \cite{Cont_2005, Cutler_Poterba_Summers_1989}, the variability in asset prices cannot be fully explained only by changes in ``fundamental'' economic factors. The important conclusion from this observation is as follows: \textbf{volatility changes over time in an unpredictable manner}.

In this survey, we offer an overview of one approach aimed at addressing this phenomenon: \textit{stochastic volatility modeling}. This methodology, which can be traced back to the early discrete-time model of Clark (1973) \cite{Clark_1973}, has evolved into an enormous, deep, and highly engaging area of research. Contrary to the older surveys of Skiadopoulous (2001) \cite{Skiadopoulos_2001}, Shepard \& Andersen (2009) \cite{Shephard_Andersen_2009} and Duong \& Swanson (2011) \cite{Duong_Swanson_2011}, we put a significant emphasis on fairly recent topics of fractional and rough volatility. We tried to make the presentation as friendly as possible, with the hope that readers familiar with stochastic analysis will find most parts of the material accessible.

Before proceeding to the main part of the paper, let us make two important disclaimers.
\begin{itemize}
    \item First of all, apart from more straightforward applications in, e.g., mean-variance portfolio theory originating from the paper \cite{Markowitz_1952} by H. Markowitz, the concept of volatility has a crucial role in another significant area within finance: \textit{non-arbitrage pricing theory} conceptualized by F. Black, M. Scholes, and R. Merton in \cite{BS_1973, Merton_1973}. This theory has a unique and intricate perspective on volatility and has been the major driving force of volatility modeling for the past four decades. Therefore, in this survey, we consider the volatility predominantly from the option pricing viewpoint.

    \item Our focus primarily centers on \textit{continuous stochastic volatility models in continuous time}. This means that we will not discuss various discrete-time models such as ARCH, GARCH or EARCH of Engle \cite{Engle_1982}, Bollerslev \cite{Bollerslev_1986} and Nelson \cite{Nelson_1991} as well as multiple jump-diffusion models including the approaches of Duffie et. al. \cite{Duffie_Pan_Singleton_2000} or Barndorff-Nielsen \& Shephard \cite{Barndorff-Nielsen_Shephard_2001}. The omission of jumps seems to be the most notable gap in light of the realism this approach provides. However, given our desire to highlight fractional and rough volatility in more detail, the inclusion of discrete-time and jump-diffusion models with the level of detail they deserve would substantially inflate the size of this survey and result in a notable change of emphasis. Therefore, we leave these two topics for a separate work. Readers with a specific interest in discrete-time models are referred to \cite{Bollerslev_Chou_Kroner_1992, Francq_2019}; for jump models, see the specialized books by Barndorff-Nielsen \& Shepard \cite{Barndorff-Nielsen_Shephard_2011}, Gatheral \cite{Gatheral_2006}, Rachev et. al. \cite{Rachev_Kim_Bianchi_Fabozzi_2011} or Tankov \& Cont \cite{Tankov_Cont_2003}. For other modeling viewpoints that are different from stochastic volatility, see the overviews given by Shiryaev \cite{Shiryaev_1999} and Mariani \& Florescu \cite{Mariani_2020}.
\end{itemize}

The paper is structured as follows. In Section \ref{sec: stylized facts}, we present a historical context for stochastic volatility modeling as well as list several relevant stylized facts that are aimed to be reproduced. We devote separate attention to the notion of \textit{implied volatility} since the latter serves as the focal point of the field. Section \ref{sec: models} gives a detailed survey of continuous stochastic volatility models from the constant elasticity of variance (CEV) model proposed by J. C. Cox in 1975 to the modern fractional and rough volatility approaches. For each class of models, we provide the context of their applicability and characterize their advantages and shortcomings. In Section \ref{sec: VIX}, we touch on selected aspects of a special topic in volatility modeling: VIX. After a brief detour in its computation, we describe two problems within VIX modeling: reproducing positive VIX skew and the SPX-VIX joint calibration problem. In addition, we analyze advances in continuous stochastic volatility modeling for solving these problems. Section \ref{sec: conclusion} concludes our presentation.

\section{Market volatility: empirical challenges and stylized facts}\label{sec: stylized facts}

\subsection{From random walk to geometric Brownian motion}

Probability theory and stochastic analysis have evolved into irreplaceable tools for economics and finance. The first attempts to model financial markets using probabilistic methods can be traced back to the French economist Jules Regnault\footnote{For more details about this lesser known and somewhat underestimated figure and his contributions to finance, the reader is referred to the paper \cite{Jovanovic_Le_Gall_2001}.}: as early as in 1863 \cite{Regnault_1863}, he proposed modeling stock prices with, what we call nowadays, symmetric random walks. However, the foundation for employing probability in finance is commonly attributed to another French mathematician, Louis Bachelier, and his 1900 dissertation titled "\textit{Théorie de la spéculation}" \cite{Bachelier_1900}, where he considered market modeling and even derivative pricing in a very comprehensive manner. Naturally, the tools of stochastic calculus had not yet been formulated at that time, which means that the dissertation might lack some modern aspects of mathematical rigor. Nonetheless, if one translates Bachelier's reasoning into today's mathematical language\footnote{An excellent job in contextualizing Bachelier's work from the perspective of modern mathematics was done by M. Davis and A. Etheridge in \cite{Bachelier_1900_eng}.}, it turns out that Bachelier essentially modeled stock price dynamics with a prototype of a standard Brownian motion, 5 years before A. Einstein and his famous paper \cite{Einstein_1905}.  

Unfortunately, Bachelier's ideas did not gain immediate recognition and went relatively unnoticed. Only more than 50 years after the publication of the original dissertation, statistician Jimmy Savage inexplicably stumbled upon this work and brought it to the attention of a number of researchers in economics. One of those researchers turned out to be Paul Samuelson who described Savage's finding in the foreword to the English translation of Bachelier's dissertation \cite{Bachelier_1900_eng} as follows:
\begin{quote}
    ``\textit{...Discovery or rediscovery of Louis Bachelier’s 1900 Sorbonne thesis} [...] \textit{initially involved a dozen or so postcards sent out from Yale by the late Jimmie Savage, a pioneer in bringing back into fashion statistical use of Bayesian probabilities. In paraphrase, the postcard’s message said, approximately, ‘Do any of you economist guys know about a 1914 French book on the theory of speculation by some French professor named Bachelier?’}
    
    \textit{Apparently I was the only fish to respond to Savage’s cast. The good MIT mathematical library did not possess Savage’s 1914 reference. But it did have something better, namely Bachelier’s original thesis itself.}
    
    \textit{I rapidly spread the news of the Bachelier gem among early finance theorists. And when our MIT PhD Paul Cootner edited his collection of worthy finance papers, on my suggestion he included an English version of Bachelier’s 1900 French text...}''
\end{quote}
Samuelson noticed that Bachelier's quasi-Brownian model could potentially take negative values, which is an unrealistic property for real-life prices. Therefore he proposed\footnote{Samuelson himself acknowledged (see e.g. his comments in a foreword to \cite{Bachelier_1900_eng}) that the same idea was independently expressed by an astronomer M. Osborne in \cite{Osborne_1956}.} a simple but very useful modification of Bachelier's original approach: Brownian dynamics should be used to model \textit{price logarithms} rather than the \textit{prices themselves}. After a small adjustment with a linear trend, Samuelson's model took the form of a \textit{geometric} or (\textit{economic relative}, the term used by Samuelson himself \cite{Samuelson_1973}) \textit{Brownian motion}
\begin{equation}\label{intro: GBM explicit}
    S(t) = S(0) \exp\left\{ \left(\mu - \frac{\sigma^2}{2}\right)t + \sigma W(t) \right\}, \quad \mu\in\mathbb R, \quad S(0),~\sigma > 0,
\end{equation}
or, as a stochastic differential equation (SDE),
\begin{equation}\label{intro: GBM sde}
    dS(t) = \mu S(t) dt + \sigma S(t) dW(t).
\end{equation}
Note that the term \textit{volatility} obtains a very specific meaning within the model \eqref{intro: GBM explicit}--\eqref{intro: GBM sde}; namely, it refers to the parameter $\sigma$. Such terminology is very intuitive and, moreover, agrees well with the metric \eqref{eq: realized volatility} mentioned above: if $0=t_0<t_1<...<t_n=T$ is a partition of the interval $[0,T]$, then
\begin{align*}
    \frac{1}{T} \sum_{k=1}^n \left(\log\frac{S(t_{k+1})}{S(t_k)}\right)^2 \to \sigma^2
\end{align*}
in $L^2$ as the diameter of the partition $\max_{k}|t_{k+1}-t_k| \to 0$ (see e.g. \cite{Revuz_Yor_1999}). Due to its simplicity and tractability, the log-normal process \eqref{intro: GBM explicit}--\eqref{intro: GBM sde}, together with the corresponding notion of volatility, subsequently became a mainstream choice for stock price models for the next couple of decades. Even now, well-informed of multiple arguments against the geometric Brownian motion, practitioners still use it as a benchmark or a reliable ``first approximation'' model. 

It is important to note that, in addition to the market model, Bachelier also considered the problem of option pricing and eventually derived an expression that can be called a precursor of the now famous \textit{Black-Scholes formula}. Of course, his rationale was not based on the no-arbitrage principle and had a number of shortcomings, as it is often the case with pioneering works. The correction of those shortcomings became the subject of a number of studies in the 1960s, among which one can mention \cite{Boness_1964, Sprenkle_1961, Thorp_Kassouf_1967}. Samuelson himself also heavily contributed to that topic, see e.g. \cite{Samuelson_1965} or his paper \cite{Samuelson_Merton_1972} (in co-authorship with Robert Merton) where it was suggested to consider a warrant/option payoff as a function of the price of the underlying asset. One could argue that these works were just a few steps away from the breakthrough made by Black, Scholes, and Merton just a couple of years later.

Here it is worth paying attention to the fact that the options market remained relatively illiquid until the end of the 60s. The reason for that was the lack of a consistent pricing methodology, and serious investors regarded options as akin to gambling rather than worthy trading instruments. It is somewhat ironic that even Robert Merton himself, right in his pivotal article \cite{Merton_1973}, wrote the following:
\begin{quote}
    ``\textit{Because options are specialized and relatively unimportant financial securities, the amount of time and space devoted to the development of a pricing theory might be questioned...}''
\end{quote}
However, in 1968, a demand for that type of contract suddenly arose from the Chicago Board of Trade. The organization observed a significant decline in commodity futures trading on its exchange and therefore opted to create additional instruments for investors. They settled on options, and, in 1973, the Chicago Board of Options Exchange commenced its activity. Precisely in that year, two revolutionary papers appeared: ``\textit{The pricing of options and corporate liabilities}'' \cite{BS_1973} by Fischer Black and Myron Scholes and ``\textit{Theory of rational option pricing}'' \cite{Merton_1973} by Robert Merton\footnote{As a side note, the publication of the Black and Scholes paper was far from a smooth process: in 1987 \cite{BS_1987}, Black recalled that the manuscript was rejected first by the \textit{Journal of Political Economy} and then by the \textit{Review of Economics and Statistics}. The paper was published only after Eugene Fama and Merton Miller personally recommended the \textit{Journal of Political Economy} to reconsider its decision (in the meanwhile, Robert Merton showed a great deal of academic integrity by delaying the publication of his own article so that Black and Scholes would be the first).}. The ideas of Black, Scholes and Merton revolutionized mathematical finance and enjoyed empirical success: Stephen Ross, for instance, claimed in 1987 \cite{Ross_1987} that
\begin{quote}
    ``\textit{When judged by its ability to explain the empirical data, option pricing theory is the most successful theory not only in finance, but in all of economics.}''
\end{quote}

\subsection{Implied volatility and the smile phenomenon}\label{subsec: IV and smile}

The main result of Black, Scholes, and Merton can be formulated as follows: if a stock follows the model \eqref{intro: GBM explicit}--\eqref{intro: GBM sde}, then, under some assumptions, the discounted \textit{no-arbitrage price} of a standard European call option $C^{\text{B-S}} = C^{\text{B-S}}(t,S)$ evolves as a function of the current time $t$ and current price $S$ and must satisfy a partial differential equation, known now as the \textit{Black-Scholes formula}, of the form
\begin{equation}\label{intro: BS equation}
    \frac{\partial C^{\text{B-S}}}{\partial t} + \frac{1}{2} \sigma^2 S^2 \frac{\partial^2 C^{\text{B-S}}}{\partial S^2} + rS \frac{\partial C^{\text{B-S}}}{\partial S} - rC^{\text{B-S}} = 0
\end{equation}
with a boundary condition\footnote{In what follows, we will utilize the standard notation $(x)_+ := \max\{x,0\}$.}
\begin{equation}\label{intro: BS equation boundary}
    C^{\text{B-S}}(T,S) = (S-K)_+,
\end{equation}
where $r$ denotes the instantaneous interest rate that is assumed to be constant, $T$ is the maturity date of the option and $K$ is its exercise price. As mentioned above, the Black--Scholes--Merton rationale relies on a number of rather abstract assumptions that do not align with real-world market conditions. In particular, their reasoning required specific price dynamics, the absence of transaction costs, and the capacity to buy and sell any quantity of assets. However, the inability to perfectly replicate the reality does not necessarily carry significant implications. Black, Scholes, and Merton themselves were aware of the limitations of their approach: for example, \cite{Fengler_2005} quotes Fisher Black on this subject:
\begin{quote}
    ``\textit{Yet that weakness is also its greatest strength. People like the model because they can easily understand its assumptions. The model is often good as a first approximation, and if you can see the holes in the assumptions you can use the model in more sophisticated ways.}''
\end{quote}
What really mattered was the successful empirical performance of the vanilla Black--Scholes--Merton model; as it was noted by J. Wiggins, one of the pioneers of continuous-time stochastic volatility modeling, ``\textit{given the elegance and tractability of the Black-Scholes formula, profitable application of alternate models requires that economically significant valuation improvements can be obtained empirically}'' \cite{Wiggins_1987}.

However, after the Black Monday market crash in 1987, it became evident that there were glaring flaws within the log-normal paradigm prompting the need for rectification. Jackwerth \& Rubinstein \cite{Jackwerth_Rubinstein_1996} described the problem as follows:
\begin{quote}
    ``\textit{Following the standard paradigm, assume that stock market returns are lognormally distributed with an annualized volatility of 20\% (near their historical realization). On October 19, 1987, the two month S\&P 500 futures price fell 29\%. Under the lognormal hypothesis, this is a -27 standard deviation event with probability $10^{-160}$. Even if one were to have lived through the entire 20 billion year life of the universe and experienced this 20 billion times (20 billion big bangs), that such a decline could have happened even once in this period is a virtual impossibility.}''
\end{quote}
Evidently, experiencing ``\textit{virtually impossible}'' price falls which rendered investors insolvent was already a good argument to reassess financial modeling approaches. Yet, apart from that single shock that, in principle, could be attributed to a single anomaly, there was another consistent phenomenon that manifested itself in the aftermath of Black Monday: \textbf{the volatility smile}.

In order to explain the problem, let us first discuss the parameters of the Black--Scholes model in more detail. Note that the sole value in the Black-Scholes formula \eqref{intro: BS equation}--\eqref{intro: BS equation boundary} that is not directly observable is the volatility parameter $\sigma$: indeed, the maturity date $T$ and exercise price $K$ are given in the specifications of the given option contract whereas the price $S(t)$ can be read from the market. The volatility $\sigma$ is unknown and necessitates some form of estimation based on data. In 1976, Latan\'e \& Rendelman \cite{Latane_Rendleman_1976} proposed an elegant method to address this problem. First of all, note that the equation \eqref{intro: BS equation}--\eqref{intro: BS equation boundary} has an explicit solution of the form
\begin{equation}\label{intro: BS explicit}
\begin{aligned}
    C^{\text{B-S}}(t, T, S(t), K, \sigma)& = S(t) \Phi\left( \frac{\log\frac{e^{r(T-t)}S(t)}{K} + \frac{\sigma^2}{2}(T-t)}{\sigma\sqrt{T - t}} \right) 
    \\
    &\quad - Ke^{-r(T-t)} \Phi\left( \frac{\log\frac{e^{r(T-t)}S(t)}{K} - \frac{\sigma^2}{2}(T-t)}{\sigma\sqrt{T - t}} \right),
\end{aligned}
\end{equation}
where $\Phi(x) := \frac{1}{\sqrt{2\pi}} \int_{-\infty}^x e^{-\frac{y^2}{2}}dy$. Now, take the \textit{actual} market price $C_t(T,K)$ of the corresponding option with payoff $K$ and maturity date $T$ at some evaluation time $t \ge 0$ and notice that, since $ C^{\text{B-S}}(t, T, S(t), K, \sigma)$ is supposed to coincide with $C_t(T,K)$, the volatility $\sigma$ can be estimated by solving the equation
\begin{equation}\label{intro: IV equation}
     C^{\text{B-S}}(t, T, S(t), K, \sigma) - C_t(T,K) = 0.
\end{equation}
Note that equation \eqref{intro: IV equation} depends on $t$, $T$ and $K$, so its solution $\widehat \sigma = \widehat \sigma_t(T,K)$, known in the literature as the \textit{implied volatility}\index{volatility!implied}, is technically a function of $t$, $T$ and $K$ as well. If the stock price model \eqref{intro: GBM explicit}--\eqref{intro: GBM sde} indeed reflects the reality well enough, we would expect $\widehat{\sigma}(t, T, K)$ to closely match the constant model parameter $\sigma$, and thus to remain approximately constant across different strikes $K$ and maturities $T$ for options on the same underlying asset.

However, this is not the case! For instance, $ \widehat \sigma(T,K)$ turns out to change with $T$ for fixed $K$ (see Fig.~\ref{fig:VaryingTTM} below). 
\begin{figure}[h!]
    \centering
    \includegraphics[width = 0.7\textwidth]{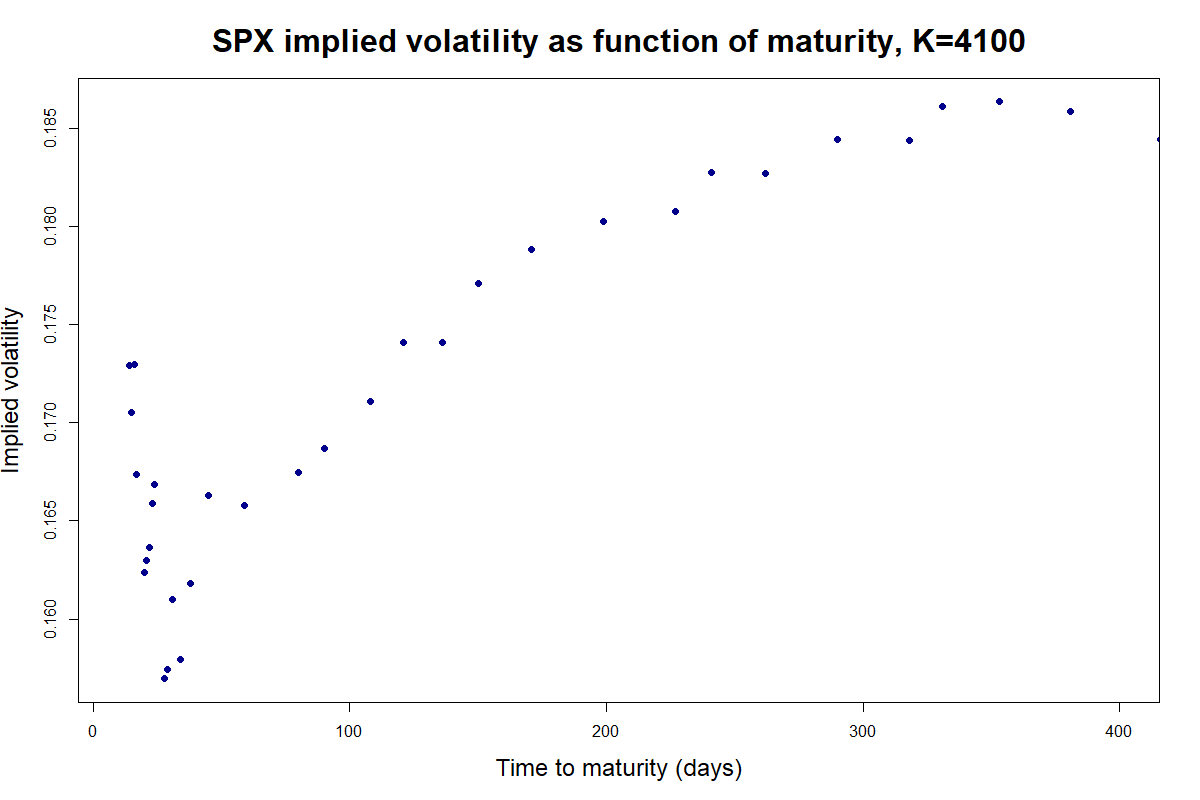}
    \caption{Variation of implied volatility with time to maturity $T$ for a fixed strike $K$. The values were calculated using S\&P500 (SPX) option prices on May 3, 2023, retrieved from \textit{Yahoo! Finance}.}
    \label{fig:VaryingTTM}
\end{figure}
A natural extension of \eqref{intro: GBM explicit}, first considered by Merton in his original paper \cite{Merton_1973}, addresses this issue by allowing the volatility $\sigma = \sigma(t)$ to be a deterministic function of time. In that case, one can obtain a version of \eqref{intro: BS explicit} of the form
\begin{equation*}
\begin{aligned}
     C^{\text{B-S}}(t, T, S(t), K, \sigma) & = S(t) \Phi\left( \frac{\log\frac{e^{r(T-t)}S(t)}{K} + \frac{1}{2}\int_t^T \sigma^2(s)ds}{\sqrt{\int_t^T \sigma^2(s)ds}} \right) 
    \\
    &\quad - Ke^{-r(T-t)} \Phi\left(\frac{\log\frac{e^{r(T-t)}S(t)}{K} - \frac{1}{2}\int_t^T \sigma^2(s)ds}{\sqrt{\int_t^T \sigma^2(s)ds}}\right)
    \\
    &:= S(t) \Phi\left( \frac{\log\frac{e^{r(T-t)}S(t)}{K} + \frac{1}{2}\overline{\sigma}^2(t,T)(T-t)}{\overline{\sigma}(t,T)\sqrt{T-t}} \right) 
    \\
    &\quad - Ke^{-r(T-t)} \Phi\left(\frac{\log\frac{e^{r(T-t)}S(t)}{K} - \frac{1}{2}\overline{\sigma}^2(t,T)(T-t)}{\overline{\sigma}(t,T)\sqrt{T-t}}\right),
\end{aligned}
\end{equation*}
where $\overline{\sigma}^2(t,T) := \frac{1}{T-t}\int_t^{T} \sigma^2(s)ds$. Then, for a fixed evaluation time $t \ge 0$, the counterpart of the equation \eqref{intro: IV equation} gets the form
\[
    C^{\text{B-S}}(t, T, S(t), K, \overline{\sigma}(t,T)) - C_t(T,K) = 0,
\]
so the solution $\widehat{\sigma}(t, T, K)$ approximates $\overline{\sigma}(t,T)$ and is therefore allowed to vary with $T$ for a fixed $K$. One could even argue that this assumption of time-varying deterministic volatility is quite natural: as noted in \cite[p. 144]{Derman_Miller_Park_2016}, ``\textit{there is nothing inconsistent about expecting high volatility this year and low volatility next year}''. 

\begin{figure}
     \centering
     \begin{subfigure}[b]{0.45\textwidth}
         \centering
         \includegraphics[width=\textwidth]{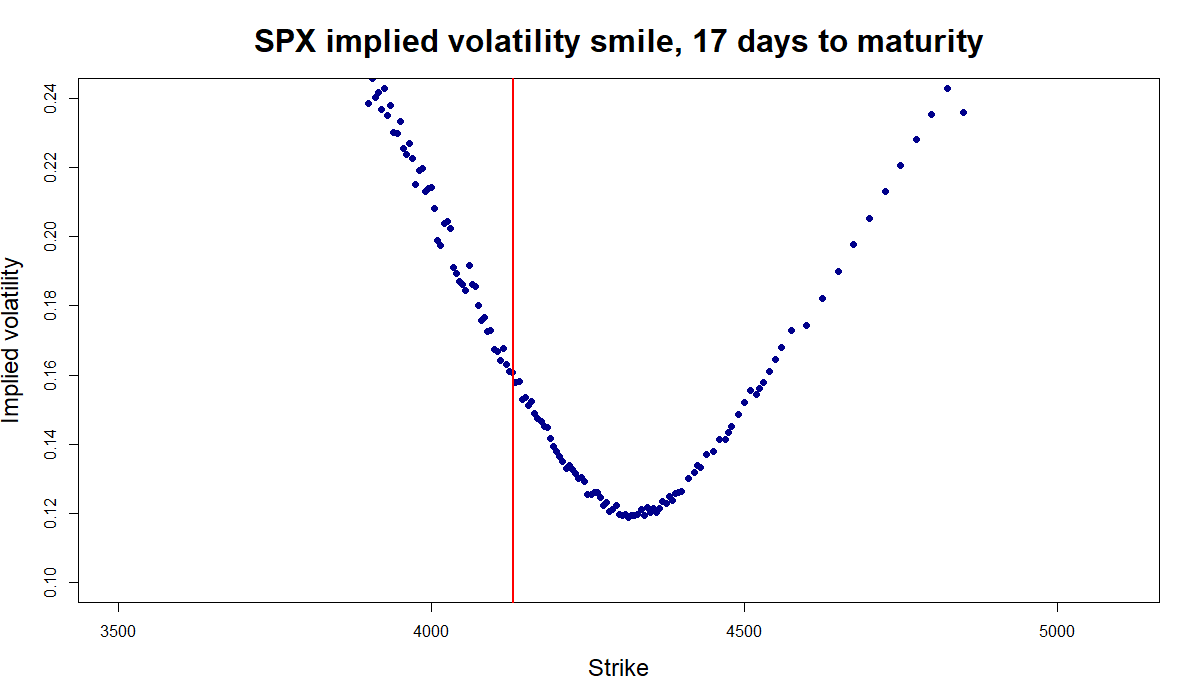}
         \caption{17 days to maturity}
     \end{subfigure}
     \hfill
     \begin{subfigure}[b]{0.45\textwidth}
         \centering
         \includegraphics[width=\textwidth]{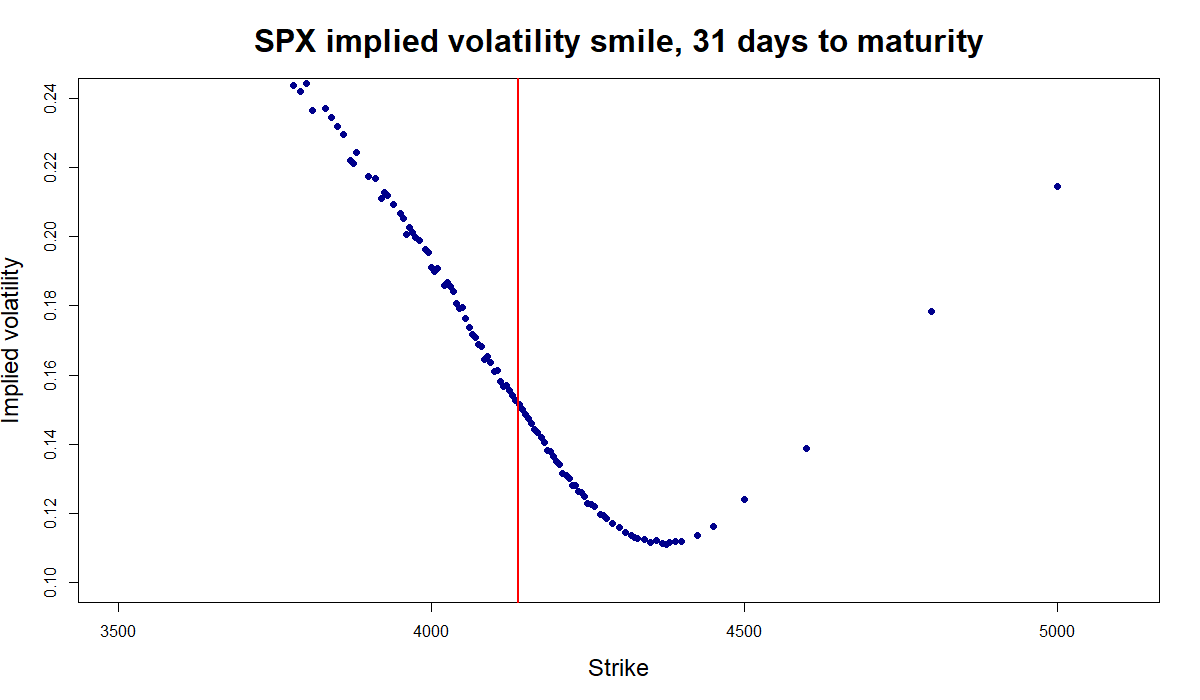}
         \caption{31 days to maturity}
     \end{subfigure}
     \hfill
     \begin{subfigure}[b]{0.45\textwidth}
         \centering
         \includegraphics[width=\textwidth]{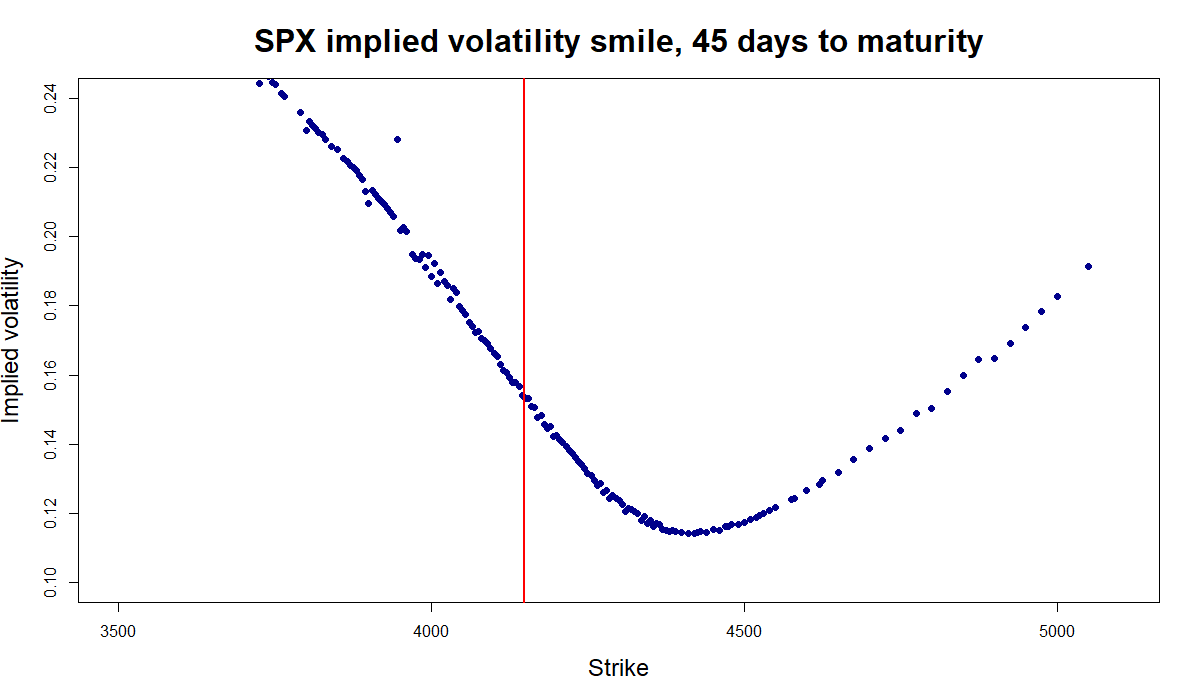}
         \caption{45 days to maturity}
     \end{subfigure}
     \hfill
     \begin{subfigure}[b]{0.45\textwidth}
         \centering
         \includegraphics[width=\textwidth]{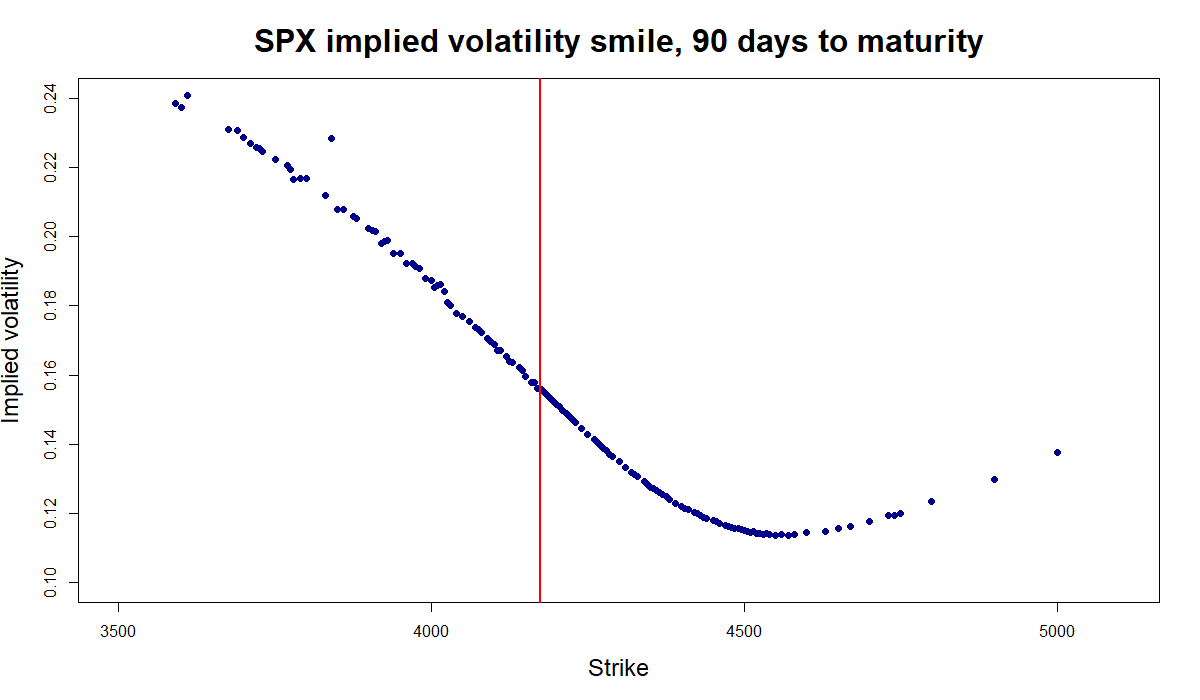}
         \caption{90 days to maturity}
     \end{subfigure}
     \hfill
     \begin{subfigure}[b]{0.45\textwidth}
         \centering
         \includegraphics[width=\textwidth]{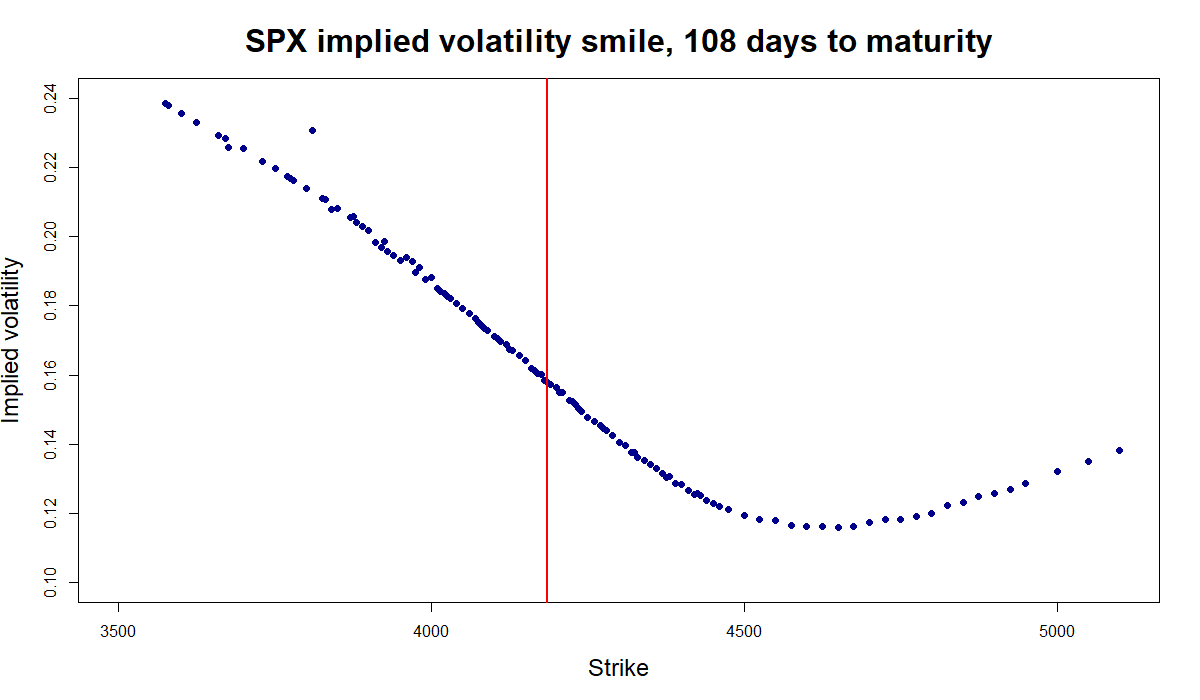}
         \caption{108 days to maturity}
     \end{subfigure}
     \hfill
     \begin{subfigure}[b]{0.45\textwidth}
         \centering
         \includegraphics[width=\textwidth]{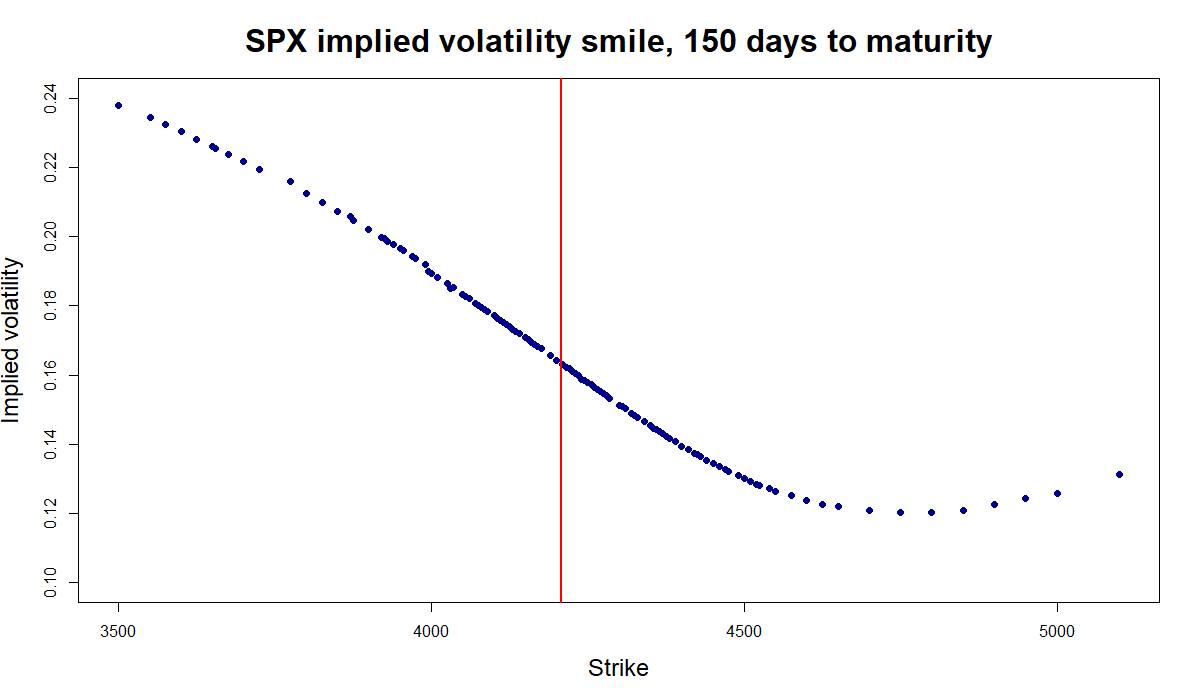}
         \caption{150 days to maturity}
     \end{subfigure}
     \hfill
     \begin{subfigure}[b]{0.45\textwidth}
         \centering
         \includegraphics[width=\textwidth]{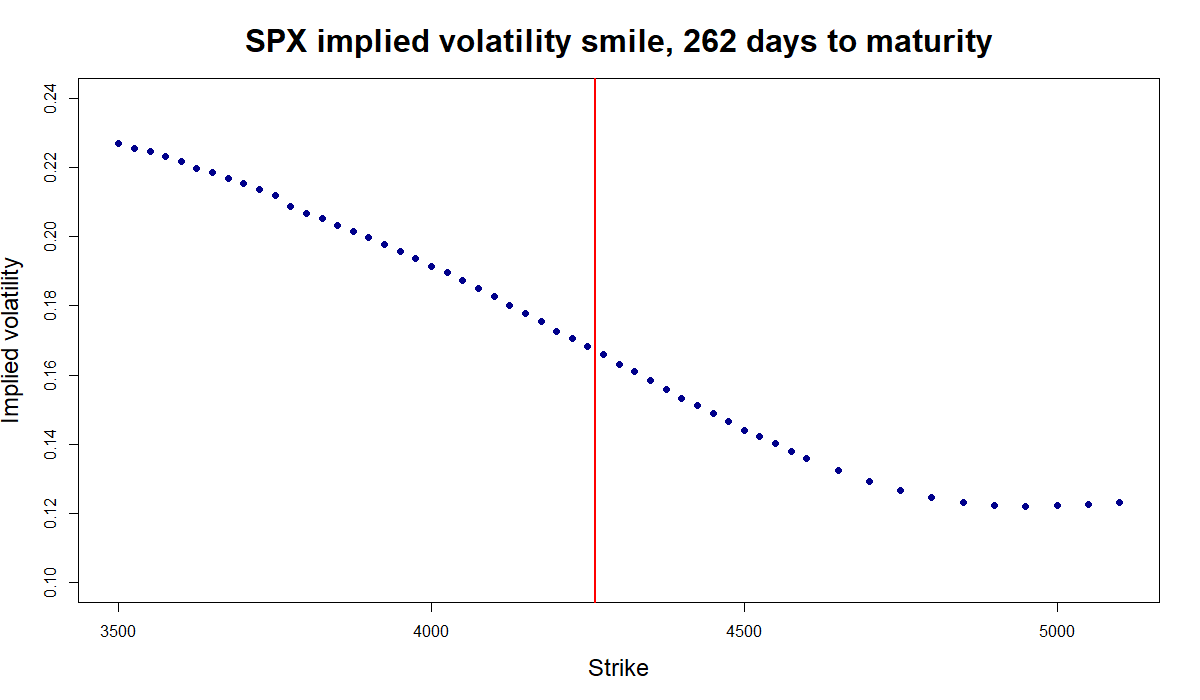}
         \caption{262 days to maturity}\label{fig:Smile262}
     \end{subfigure}
     \hfill
     \begin{subfigure}[b]{0.45\textwidth}
         \centering
         \includegraphics[width=\textwidth]{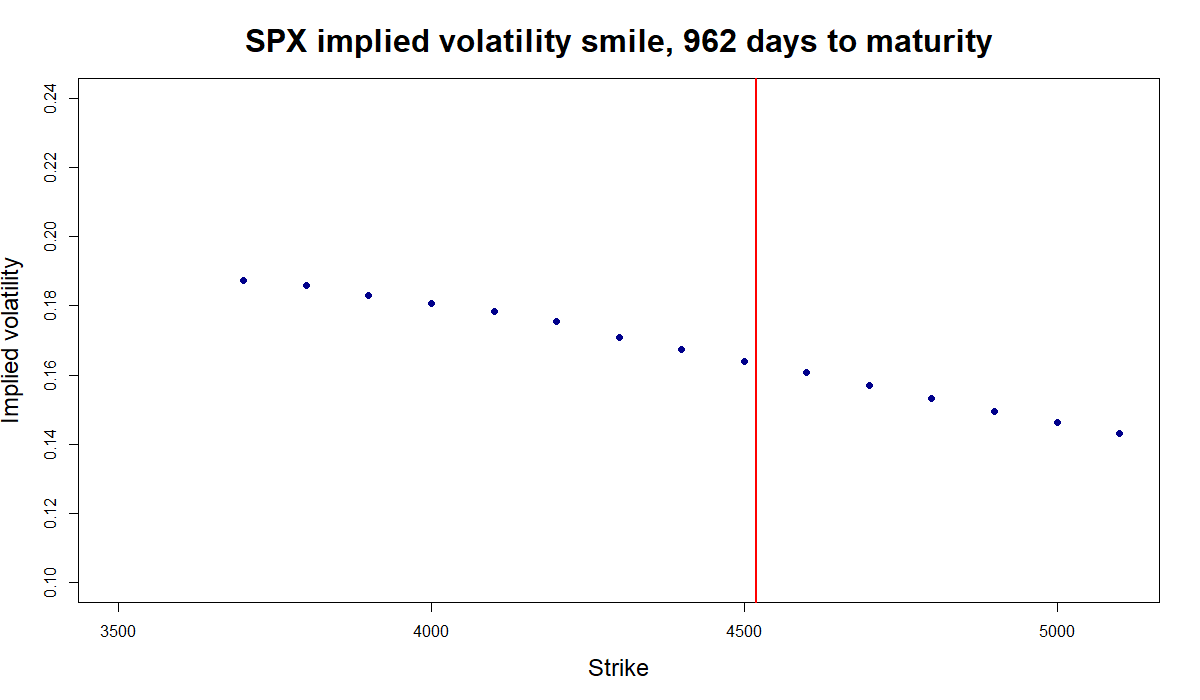}
         \caption{962 days to maturity}\label{fig:Smile962}
     \end{subfigure}
        \caption{Implied volatility smiles for different times to maturity. The values were calculated using S\&P500 (SPX) option prices on May 3, 2023, retrieved from \textit{Yahoo! Finance}. Vertical red lines correspond to the at-the-money level $K = S(0)e^{rT}$.}
        \label{fig:Smile}
\end{figure}

Regarding the variation in $K$ for fixed $T$, the implied volatility remained relatively flat\footnote{More precisely, some dependence was present, but it was subtle enough to be disregarded, see e.g. the discussion in \cite[Chapter 1]{Derman_Miller_Park_2016}.} before the above-mentioned Black Monday crash in 1987. Starting from that (terrible) date, investors observed notable variability of the implied volatility in $K$ characterized by distinct convex patterns (see Fig. \ref{fig:Smile} below as well as \cite{Das_Sundaram_1999}) which were eventually called ``\textit{volatility smiles}'' or ``\textit{volatility smirks}''. Such behavior was consistent, had a direct adverse impact on the empirical performance of the Black-Scholes formula, and could not be explained by the price dynamics \eqref{intro: GBM explicit}--\eqref{intro: GBM sde}.

Before we proceed further, let us make a remark about the notation. 

\begin{remark}
    In the literature (see e.g. \cite[p. 244]{Lee_2006}), the implied volatility $\widehat \sigma$ is often written as a function $\widehat \sigma_{\text{log-m}}$ of time to maturity $T$ and \textit{log-moneyness} $\kappa := \log\frac{K}{e^{r(T-t)}S(t)}$, i.e.
    \[
        \widehat \sigma_{\text{log-m}}(t, T, \kappa) := \widehat \sigma(t, T, S(t)e^{\kappa + r(T-t)}).
    \]
    In what follows, we will slightly abuse the notation and omit the subscript ``log-m'', using $\widehat \sigma(t,T,K)$ for the function of strike and $\widehat \sigma(t,T,\kappa)$ for the function of log-moneyness.
\end{remark}

\begin{figure}
     \centering
     \begin{subfigure}[b]{0.48\textwidth}
         \centering
         \includegraphics[width=\textwidth]{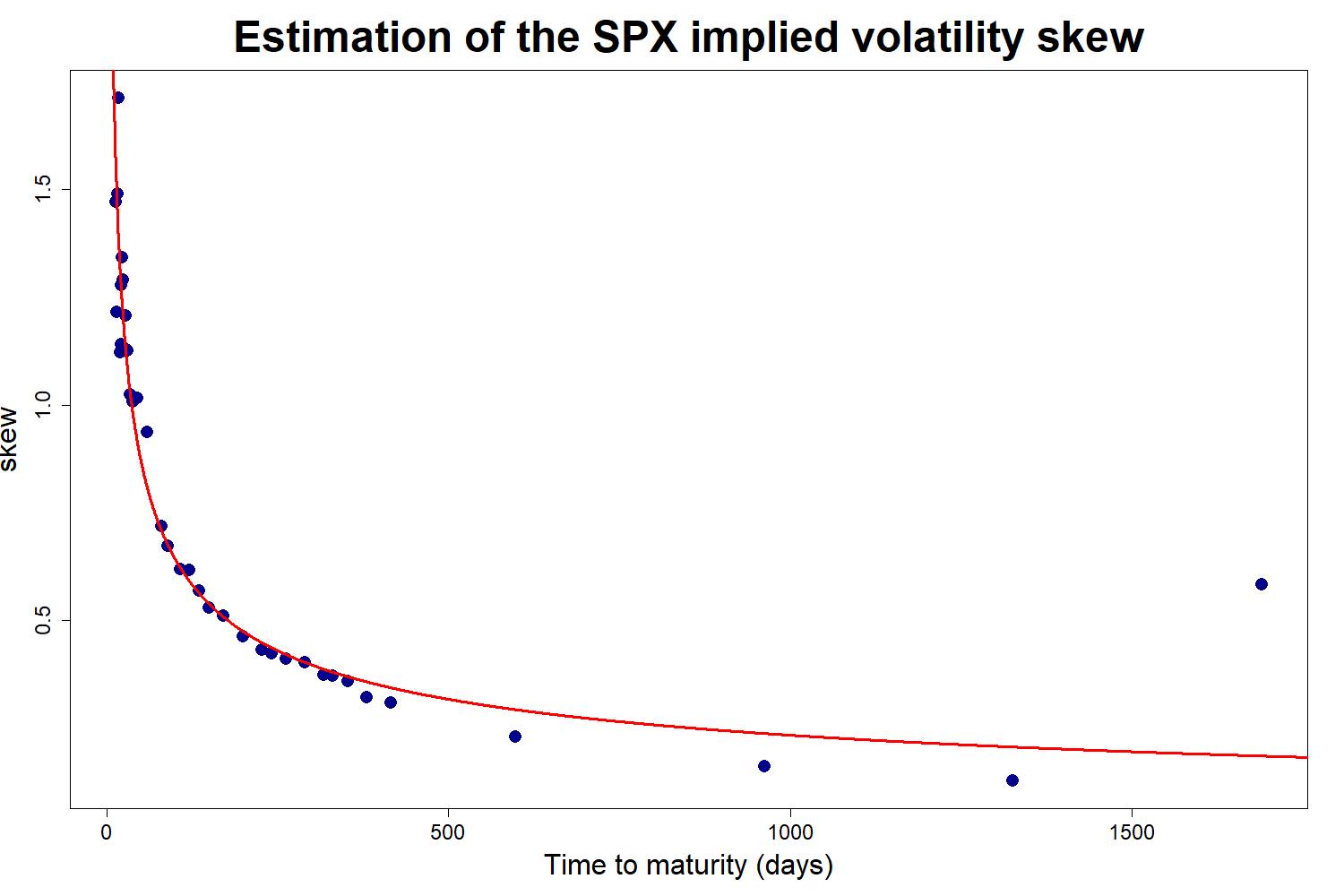}
         \caption{Regular scale}\label{fig: Skew normal}
     \end{subfigure}
     \hfill
     \begin{subfigure}[b]{0.48\textwidth}
         \centering
         \includegraphics[width=\textwidth]{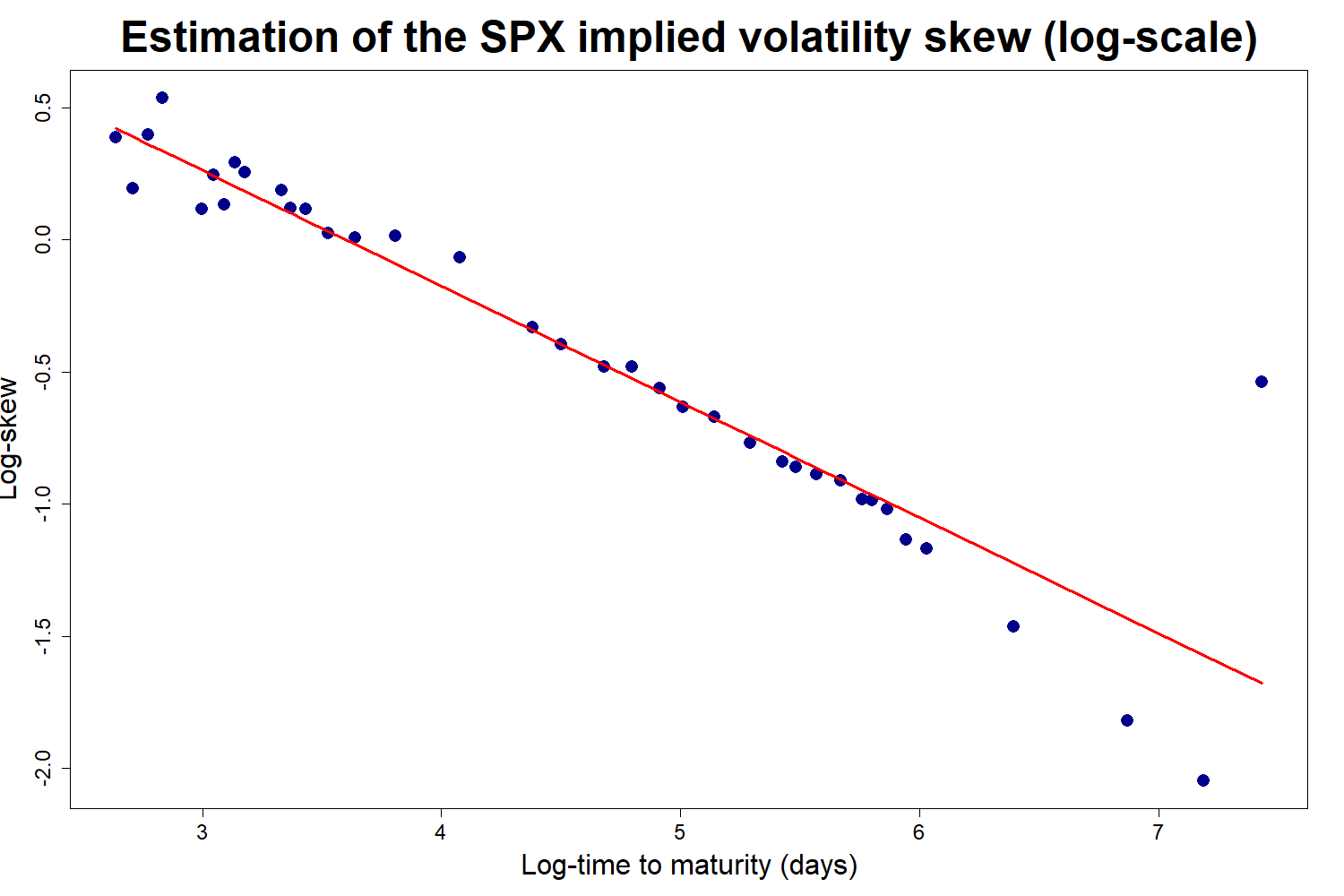}
         \caption{Logarithmic scale}\label{fig: Skew log}
     \end{subfigure}
    \caption{Absolute values of at-the-money implied volatility slopes on regular (a) and log-log (b) scales. Blue dots denote the data and red lines denote the regression fits $e^{1.5753587} T^{-0.4377343}$ (a) and $1.5753587 -0.4377343 \log T$ (b). The values were calculated using S\&P500 (SPX) option prices on May 3, 2023, retrieved from \textit{Yahoo! Finance}. }
        \label{fig: Skew}
\end{figure}

Figure \ref{fig:Smile} also reveals two important typical characteristics of implied volatility smiles related to the change of the smile shape with $T$. 
\begin{itemize}
    \item First of all, note that the smile gradually flattens out as the time to maturity $T$ increases. However, this decrease is fairly slow: for example, on Fig.~\ref{fig:Smile}, the curvature is noticeable 262 days (Fig.~\ref{fig:Smile262}) and even 962 days (Fig.~\ref{fig:Smile962}) before maturity. The latter effect turns out to require special treatment; in this regard, see Subsection \ref{subsec: fractional and rough models} below as well as \cite{Carr_Wu_2003, Comte_Renault_1998, Funahashi_Kijima_2017}.
    
    \item Second, observe the behavior of the smile \textit{at-the-money}, i.e. when $K=S(0)e^{rT}$ (or, in terms of log-moneyness, when $\kappa = 0$): as $T$ gets smaller, the smile at-the-money tends to become \textit{steeper}. Figure~\ref{fig: Skew} characterizes this phenomenon in more detail: for each $T$, we took implied volatilities of 7 options with strikes closest to $S(0)e^{rT}$, performed the least squares linear fit to them and depicted the absolute values of the resulting slopes on Fig.~\ref{fig: Skew normal}. It turns out that the variation of absolute slopes with $T$ for shorter maturities seems to be well-described by the \textit{power law} $CT^{-\frac{1}{2}+H}$ with $H\approx 0$, \textit{at least as the first approximation} (see the discussion at the end of Subsection \ref{subsec: fractional and rough models} below as well as \cite{Guyon_El_Amrani_2022} and \cite{Delemotte_Marco_Segonne_2023}). As an illustration, red lines on Fig.~\ref{fig: Skew normal} and \ref{fig: Skew log} depict power law fits for the SPX implied volatility slopes corresponding to May 3, 2023; for our dataset, $H=0.06226572$. A similar type of behavior is reported by e.g. Fouque, Papanicolaou, Sircar \& Sølna (2004) \cite{Fouque_Papanicolaou_Sircar_Solna_2004}, Gatheral, Jaisson \& Rosenbaum (2018) \cite{Gatheral_Jaisson_Rosenbaum_2014} and, more recently, by Delemotte, De Marco \& Segonne (2023) \cite{Delemotte_Marco_Segonne_2023}.
\end{itemize}

\begin{figure}
    \centering
    \includegraphics[width = 0.7\textwidth]{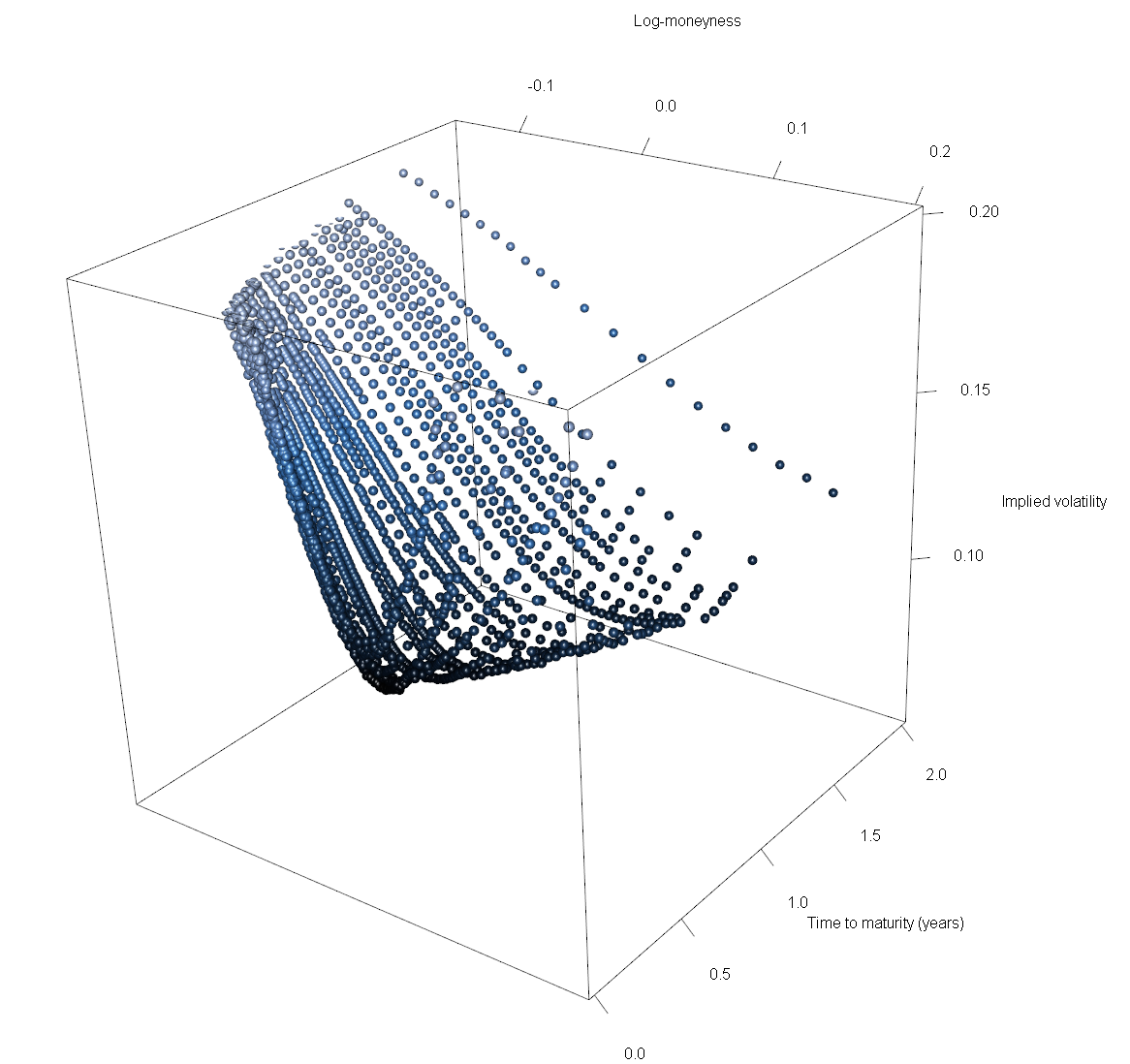}
    \caption{Implied volatility surface $(T, \kappa) \mapsto \widehat \sigma(T, \kappa)$. The values were calculated using S\&P500 (SPX) option prices on May 3, 2023, retrieved from \textit{Yahoo! Finance}.}
    \label{fig:VolatilitySurface}
\end{figure}

In principle, a ``\textit{perfect}'' model should capture\footnote{By ``\textit{capturing the volatility surface}'', we mean the established benchmark for assessing the performance of a model by its ability to reproduce the shape of $\widehat \sigma(T,K)$ or, equivalently, $\widehat \sigma(T,\kappa)$. As a rule, testing adheres to the following algorithm:
\begin{enumerate}
    \item[1)] the parameters of the model are calibrated with respect to the market data;
    \item[2)] the calibrated parameters are plugged into the model and the respective model no-arbitrage European option prices are computed;
    \item[3)] these ``synthetic'' prices are then used to construct the implied volatility surface by the procedure described here in Subsection \ref{subsec: IV and smile};
    \item[4)] finally, the obtained model-generated surface is compared to the implied volatility surface extracted from real-world option prices.
\end{enumerate}} the shape of the entire \textit{implied volatility surface} $(T, K) \mapsto \widehat \sigma(T, K)$ (or, equivalently, $(T, \kappa) \mapsto \widehat \sigma(T, \kappa)$, see Fig.~\ref{fig:VolatilitySurface}). For example, if one wants to reflect the power law described above, one may demand the model-generated \textit{at-the-money volatility skew}
\begin{equation}\label{intro: volatility skew t}
    \Psi(t, T) := \left| \frac{\partial}{\partial \kappa} \widehat \sigma(t, T, \kappa) \right|_{\kappa = 0}
\end{equation}
to have the power-law asymptotics $O((T - t)^{-\frac{1}{2}+H})$, $T \to t$. However, as we will see later, the task of constructing a model that reproduces all of the stylized facts simultaneously -- both for long and short maturities -- is not straightforward at all.

\begin{remark}
    In what follows, we often consider the case where the maturity time $T$ approaches the fixed starting point $t$. In many applications, the primary focus is on the present moment $t=0$, so we will frequently state our results for $t=0$ fixed and $T \to 0$. To simplify our notation, we will omit the 0 from the list of arguments and simply write $C^{\text{B-S}}(T, S(0), K, \sigma)$, $C(T,K)$ and $\widehat{\sigma}(T, K)$ (or $\widehat{\sigma}(T, \kappa)$) for the Black–Scholes price, the observed option price and the implied volatility respectively. The same convention applies to the skew
    \begin{equation}\label{intro: volatility skew}
        \Psi(T) := \left| \frac{\partial}{\partial \kappa} \widehat \sigma(T, \kappa) \right|_{\kappa = 0},
    \end{equation}
    for which we are interested in the asymptotic behavior as $T\to 0$.
    
    The time parameter $t$ will be written explicitly only when it is necessary to emphasize that the starting point may vary.
\end{remark}

\subsection{Fat tails, leverage, clustering and long memory}\label{subsec: stylized facts}

The empirical evidence of volatility smiles makes a spectacular point against the geometric Brownian dynamics \eqref{intro: GBM explicit}--\eqref{intro: GBM sde}. However, it certainly does not stand alone as the sole argument in this regard. In fact, objections to the log-normality of prices appeared as early as the log-normal model itself. In this section, we briefly enumerate several pivotal stylized facts concerning log-returns and the volatility of financial time series. For a more detailed discussion of this topic, we also refer our readers to \cite[Section 2.2]{Ghysels_Harvey_Renault_1996}, \cite[Section 3]{Frey_1997}, well-known survey articles \cite{Cont_2001, Cont_2005} or the book \cite{Zumbach_2012}.

\paragraph{Fat tails and non-Gaussian distribution of log-returns.} First of all, statistical analysis of price returns pointed out that their distribution has fat tails, i.e. the probabilities of extreme values tend to be significantly higher than predicted by log-normal models. In this context, we mention empirical studies of Mandelbrot \cite{Mandelbrot_1963} (1963) and Fama \cite{Fama_1965} (1965); see also \cite{Cootner_1967} for an early review. In response to this phenomenon, Mandelbrot proposed modeling price log-returns with $\alpha$-stable distributions, which are characterized by infinite variance. However, this idea seems to contradict subsequent studies (such as \cite{Cont_2001}) which suggest that the variance of returns should be finite. Overall, \cite{Cont_2001} gives the following summary regarding the properties of the ``\textit{true}'' log-returns distribution: it tends to be non-Gaussian, sharp-peaked, and heavy-tailed. Clearly, there are multiple parametric models satisfying these three properties and one can mention log-return models based on inverse Gaussian distributions \cite{Barndorff-Nielsen_1997}, generalized hyperbolic distributions \cite{Prause_1998}, truncated stable distributions \cite{Bouchaud_Potters_2000, Cont_Potters_Bouchaud_1997} and so on. For a more recent analysis of the topic, see also \cite{Eom_Kaizoji_Scalas_2019}.

\paragraph{Leverage and Zumbach effect.} Another interesting phenomenon not grasped by the geometric Brownian motion is the so-called \textit{leverage effect}: negative correlation between variance and returns of an asset. This empirical artifact, initially noticed by Black \cite{Black_1976} and then studied in more detail by Christie \cite{Christie_1982}, Cheung \& Ng \cite{Cheung_Ng_1992} and Duffee \cite{Duffee_1995}, was explained by Black himself as follows: a decrease in a stock price results in a drop of firm's equity value and hence increases its financial \textit{leverage}\footnote{In this context, the term ``leverage'' means the company's debt relative to its equity.}. This, according to Black, makes the stock riskier and hence more volatile. Interestingly, the name ``leverage effect'' stuck due to this explanation although subsequent research \cite{Ait-Sahalia_Fan_Li_2013, Figlewski_Wang_2001, Hasanhodzic_Lo_2019} pointed out that this correlation may not be connected to the leverage at all. Zumbach in \cite[Section 3.9.1]{Zumbach_2012} gives a different interpretation of this negative correlation: downward moves of stock prices are generally perceived as unfavorable, triggering many sales and thereby increasing the volatility. Conversely, upward moves do not result in such drastic changes in investors' portfolios, given that the majority of market participants already hold long positions. Zumbach \cite{Zumbach_2010} also studied another effect of the same nature now recognized as the \textit{Zumbach effect}: pronounced \textit{trends} in stock price movement, irrespective of sign, increase the subsequent volatility. This effect stems from the fact that large price moves motivate investors to modify their portfolios, unlike scenarios where prices oscillate within a narrower range.

\begin{figure}
     \centering
     \begin{subfigure}[b]{0.48\textwidth}
         \centering
         \includegraphics[width=\textwidth]{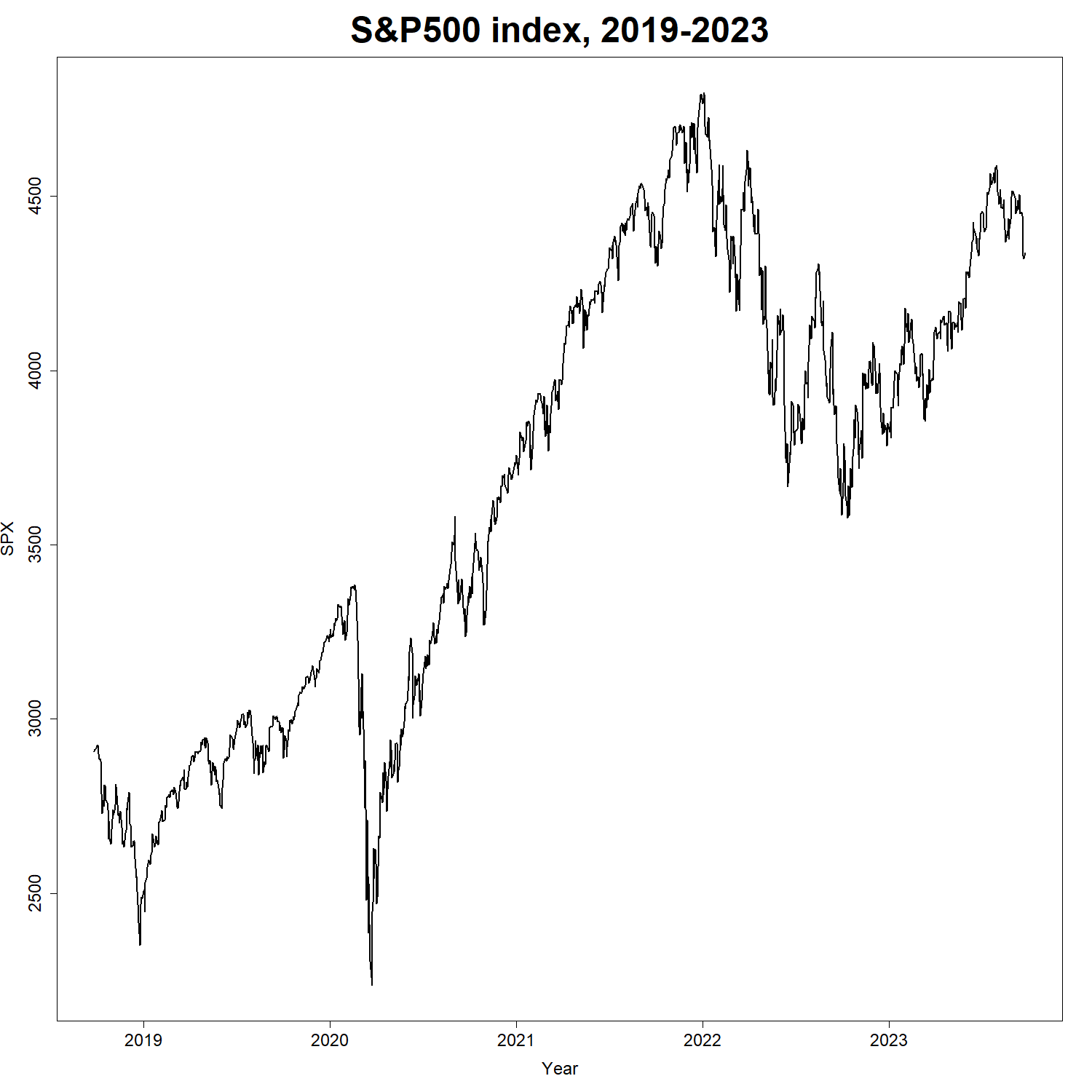}
         \caption{Daily index data}
     \end{subfigure}
     \hfill
     \begin{subfigure}[b]{0.48\textwidth}
         \centering
         \includegraphics[width=\textwidth]{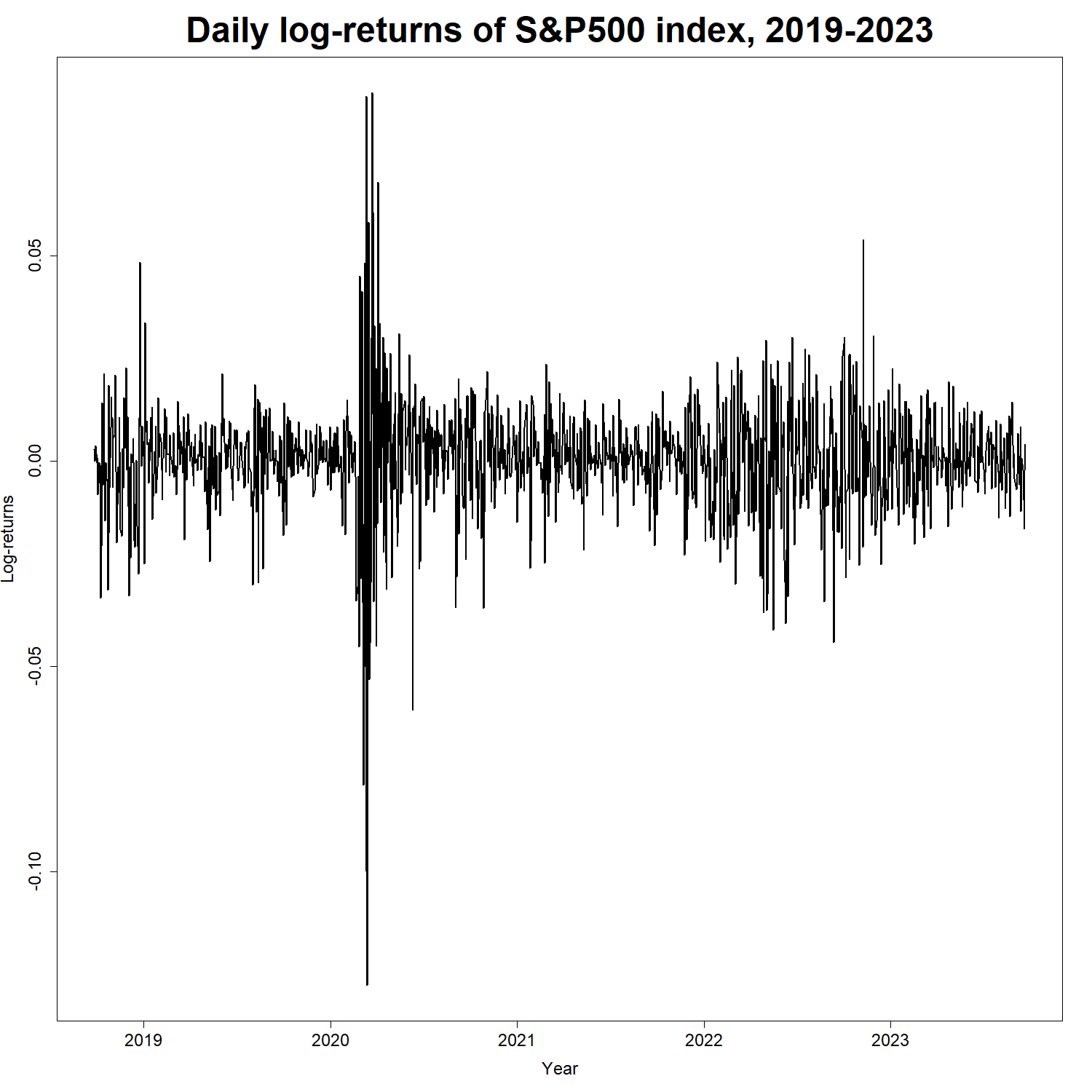}
         \caption{Logarithmic returns}\label{fig: SPXLR}
     \end{subfigure}
    \caption{Daily split-adjusted values of the S\&P500 (SPX) index (a) and the corresponding logarithmic returns (b). Note that the amplitude of fluctuations in log-returns tends to form clusters over time; the period of the highest variation in March 2020 corresponds to the shock caused by the COVID-19 pandemic. The data is retrieved from \textit{Yahoo! Finance}.  }
    \label{fig: SPX}
\end{figure}

\paragraph{Volatility clustering.} The next empirical contradiction to the Black-Scholes-Merton framework arises from a direct econometric analysis of financial time series uncovering clusters of high and low volatility episodes. As noted by Mandelbrot \cite{Mandelbrot_1963}, ``\textit{large changes tend to be followed by large changes, of either sign, and small changes tend to be followed by small changes}'' (see Fig.~\ref{fig: SPXLR}). 
There are several ways to interpret this phenomenon. Some authors (see e.g. \cite[Chapter 3]{Fouque_Papanicolaou_Sircar_2000}) identify volatility clustering with \textit{mean reversion}: in loose terms, this term refers to volatility's tendency to ``return back'' to the mean level of its long-run distribution (see e.g. \cite[Section 2.3.1 and Chapter 3]{Fouque_Papanicolaou_Sircar_2000}), with possible extended periods of staying above or below the latter. Another -- and, perhaps, more nuanced -- way to quantify clustering lies in analyzing the autocorrelation function of absolute log-returns (see e.g. an excellent review \cite{Cont_2006} on the topic), i.e.
\begin{equation}\label{intro: autocorrelation}
    \text{corr}(|R(t)|, |R(t+\tau)|),
\end{equation}
where the log-return $R(t) := \log\left(\frac{S(t+\Delta)}{S(t)}\right)$ is defined for some given time scale $\Delta$ (which may vary between a fraction of a second for tick data to several days). Multiple empirical studies \cite{Bollerslev_Chou_Kroner_1992, Bollerslev_Mikkelsen_1996, Breidt_Crato_Lima_1998, Cont_2005, Cont_2006, Cont_Potters_Bouchaud_1997, Ding_Granger_1996, Ding_Granger_Engle_1993,   Guillaume_Dacorogna_Dave_Muller_Olsen_Pictet_1997} report that the autocorrelation function \eqref{intro: autocorrelation} is consistently positive and, moreover, shows signs of slow decay of the type $O(\tau^{-\beta})$, $\tau\to\infty$, with an exponent $0 < \beta \le 0.5$. 

\paragraph{Long range dependence.} The $O(\tau^{-\beta})$ decay of \eqref{intro: autocorrelation} as $\tau\to\infty$ requires a separate discussion due to its profound implications. For $\beta\in(0,1)$, such behavior is often referred to as the \textit{long-range dependence} (see e.g. \cite{Beran, Samorodnitsky_2016}), and if its statistical significance is established, it indicates the \textit{presence of memory} on the market. Notably, confirming the long-range dependence poses a substantial challenge: by its nature, memory manifests itself when $\tau\to\infty$ which raises the concern that any statistical estimation procedure of \eqref{intro: autocorrelation} might be inconsistent due to e.g. the non-stationarity of financial time series over extended time periods (see e.g. discussion in \cite[Section 1.4]{Mikosch_Starica_2000}). Moreover, as it is shown in \cite[Section 4]{Gatheral_Jaisson_Rosenbaum_2014}, misspecification of the model in the estimation procedure can result in spurious long memory conclusion, even if the ``\textit{correct}'' model does not exhibit long memory at all. Nevertheless, several studies still report the presence of long range dependence on financial markets. For instance, Willinger et. al. \cite{Willinger_Taqqu_Teverovsky_1999} apply the so-called \textit{rescaled range ($R/S$) analysis} technique to the CRSP (Center for Research in Security Prices) daily stock time series and find some weak\footnote{As the authors write, ``\textit{...we find empirical evidence of long-range dependence in stock price returns, but because the corresponding degree of long-range dependence [...] is typically very low [...] the evidence is not absolutely conclusive}''.} evidence of memory in the data. Lobato \& Velasco \cite{Lobato_Velasco_2000} analyze volatility in connection to trading volumes and find that both of these financial characteristics exhibit the same degree of long memory. Another interesting point comes from the analysis of the implied volatility surface: Comte \& Renault \cite{Comte_Renault_1998} noticed that the decrease of the smile amplitude as time to maturity increased was much slower than many advanced market models predicted. They argued that such an effect could be mimicked by having long memory in volatility (see also a simulation study \cite{Funahashi_Kijima_2017} that directly confirms this claim).

\section{Continuous models of volatility}\label{sec: models}

As previously discussed, the real-world data does not support the standard log-normal framework of \eqref{intro: GBM explicit}--\eqref{intro: GBM sde}. Luckily, developments in the option pricing theory subsequent to the seminal Black, Scholes and Merton papers allowed for some decent flexibility in terms of the choice of price models. Namely, we refer to the gradual translation of the Black-Scholes-Merton approach into the language of martingale theory which evolved in the celebrated \textit{Fundamental Theorem of Asset Pricing} -- the result which connects non-arbitrage pricing and the existence of equivalent local martingale measures. In this regard, we mention the early research of Ross \cite{Ross_1978}, Harrison \& Kreps \cite{Harrison_Kreps_1979}, Harrison \& Pliska \cite{Harrison_Pliska_1981}, Kreps \cite{Kreps_1981} as well as subsequent seminal works of Delbaen \& Schachermayer \cite{Delbaen_Schachermayer_1994, Delbaen_Schachermayer_1998, Delbaen_Schachermayer_2006} (see also \cite{Schachermayer_2013} for a detailed historical overview on the subject). This line of research eventually evolved into a general theory allowing for quite a broad variety of price models to choose from -- hence giving researchers all the necessary tools to adjust the classical model \eqref{intro: GBM explicit} to account for volatility smiles and all other empirical inconsistencies.

As highlighted in Section \ref{sec: stylized facts}, the particular problem of \eqref{intro: GBM sde} lies in the fact that the volatility parameter $\sigma$ cannot be constant: it varies with time in an unpredictable manner, is correlated with the current price level, and has clusters of low and high values with some evidence of long-range dependence. One of the possible ways to treat this problem is by modifying \eqref{intro: GBM sde} by introducing an appropriate \textit{stochastic volatility process} $\{\sigma(t),~t \ge 0\}$ instead of the deterministic coefficient $\sigma$, i.e. by taking\footnote{In general, it is reasonable to treat the drift coefficient $\mu = \{\mu(t),~t\ge0\}$ as a stochastic process as well. However, we will not cover modeling drift in this survey.}
\begin{equation}\label{eq: general SV model}
    dS(t) = \mu(t) S(t) dt + \sigma(t) S(t) dW(t).
\end{equation}

Naturally, now the problem comes down to the selection of a particular process $\{\sigma(t),~t \ge 0\}$ that, once inserted in \eqref{eq: general SV model}, can best reproduce the behavior of real-world prices. In this section, we list prominent continuous volatility modeling approaches and briefly characterize their performance. 

\subsection{Local volatility}

\paragraph{CEV model.} The first cluster of models covered in this survey are the so-called \textit{local} or \textit{deterministic volatility models}. The core idea behind this approach is the assumption that the process $\sigma = \{\sigma(t),~t \ge 0\}$ in \eqref{eq: general SV model} is a non-random function of current price and time, i.e. $\sigma(t) = \sigma(t, S(t))$. Perhaps the first model of this kind was the \textit{constant elasticity of variance (CEV)} model suggested by Cox in his 1975 note\footnote{The original 1975 note remained unpublished since J.C Cox viewed it as a simple extension of his earlier work with S. Ross \cite{Cox_Ross_1976}. The version \cite{Cox_1997} cited here appeared in the Journal of Portfolio Management in 1997 as a tribute to F. Black and his contribution to studying the leverage effect.} \cite{Cox_1997}. CEV model assumes that $\sigma(t, S(t)) = \theta S^{\beta}(t)$, i.e. \eqref{eq: general SV model} takes the form
\begin{equation}\label{eq: CEV equation}
    dS(t) = \mu S(t) dt + \theta S^{1+\beta}(t) dW(t),
\end{equation}
where the parameter $\beta$ is called the \textit{elasticity parameter}. Initially, Cox considered $\beta\in [-1,0)$: in such case, the increase in $S(t)$ decreases the value of $\sigma(t, S(t)) = \theta S^{\beta}(t)$ and vice versa hence reproducing the leverage effect. The case $\beta >0$, more inherent to commodity prices, was discussed in \cite{Emanuel_MacBeth_1982}. For more details on the model \eqref{eq: CEV equation}, we also recommend a survey article \cite{Linetsky_Mendoza_2010}.

As a final remark regarding the CEV model, we note that the SDE \eqref{eq: CEV equation} exhibits substantially different behavior depending on the exact value of $\beta$. If $\beta \ne 0$, the diffusion coefficient in \eqref{eq: CEV equation} is not Lipschitz and hence the existence and uniqueness of the solution cannot be guaranteed by the classical result. For $\beta\in\left[-\frac{1}{2}, 0\right)$, the solution exists by the celebrated Yamada-Watanabe theorem (see e.g. \cite{YW1970}), but the cases $\beta\in\left(-1,-\frac{1}{2}\right)$ and $\beta>0$ require a separate and very careful treatment. For more details in that regard, we refer the reader to \cite[Chapter 5]{Cherny_Engelbert_2005} as well as to the discussion in \cite{Andersen_Piterbarg_2006}. 

\paragraph{Local volatility and Dupire formula.} The general local volatility was initially considered by Derman \& Kani \cite{Derman_Kani_1994} and Dupire \cite{Dupire_1994}: both assumed the risk-neutral dynamics of the form
\begin{equation}\label{eq: general local volatility}
    dS(t) = rS(t)dt + \sigma(t, S(t))S(t)dW(t),
\end{equation}
where $r$ denotes the interest rate. Note that this model is way more intricate and flexible than it seems at first glance. Unlike the CEV approach, \eqref{eq: general local volatility} does not specify any parametric form of the function $\sigma$: instead, the no-arbitrage principle and properties of diffusions allow to fully recover $\sigma$ from option prices using the celebrated Dupire formula (for its derivation, see the original paper \cite{Dupire_1994} or \cite[Subsection 2.2.1]{Bergomi_2016}): assuming no dividends,
\begin{equation}\label{eq: Dupire formula}
    \sigma(t, S) = \sqrt{2 \frac{\frac{\partial C}{\partial T}(T,K)  + r S \frac{\partial C}{\partial K}(T,K)}{S^2 \frac{\partial^2 C}{\partial K^2}(T,K)}}\Bigg|_{(T,K) = (t,S)},
\end{equation}
where $C(T,K)$, as before, denotes the price at $t=0$ of the European call-option with payoff $K$ and maturity $T$. It is possible to prove (see e.g. \cite[Subsection 2.2.2]{Bergomi_2016}) that no-arbitrage conditions imply that the expression in the right-hand side of \eqref{eq: general local volatility} is well-defined. Moreover, recalling that
\begin{equation}\label{eq: IV and option prices}
    C(T,K) = C^{\text{B-S}}(T, S(0), K, \widehat\sigma(T,K)),
\end{equation}
where $C^{\text{B-S}}$ is the Black-Scholes price, one can plug the right-hand side of \eqref{eq: IV and option prices} to \eqref{eq: Dupire formula} in place of $C(T,K)$ and obtain the theoretical relation between the implied volatility $\widehat \sigma$ and local volatility function $\sigma$ that can then be used to accurately calibrate \eqref{eq: general local volatility} to the current implied volatility smiles (see e.g. \cite[Section 2.3]{Bergomi_2016}). For more details on local volatility models, we refer the reader to \cite[Chapter 2]{Bergomi_2016}.

\paragraph{Shortcomings of local volatility.} As mentioned above, local volatility models are perhaps the simplest and the most straightforward generalizations of the original log-normal model \eqref{intro: GBM explicit}--\eqref{intro: GBM sde}. In addition to their simplicity and tractability, they inherit another convenient property of the classical geometric Brownian motion: one can show that the market produced by \eqref{eq: general local volatility} is complete, i.e. any option with $S$ as an underlying asset can be perfectly hedged by a self-financing portfolio composed exclusively of $S$ and a riskless asset. However, despite the obvious mathematical attractiveness of this feature, some sources actually view completeness as a disadvantage from the modeling viewpoint. As noted in \cite[Chapter 10]{Tankov_Cont_2003}, 
\begin{quote}
    ``\textit{While absence of arbitrage is both a reasonable property to assume in a real market and a generic property of many stochastic models, market completeness is neither a financially realistic nor a theoretically robust property. From a financial point of view, market completeness implies that options are redundant assets\footnote{Here ``redundant'' is understood in the sense that options are perfectly replicable in a complete market.} and the very existence of the options market becomes a mystery, if not a paradox in such models.}''
\end{quote}
This observation suggests that local volatility models may be too restrictive to grasp the complexity of the stock dynamics. And, importantly, such a claim is supported by multiple studies. For example, Buraschi \& Jackwerth \cite{Buraschi_Jackwerth_2001} utilize a formal statistical testing procedure to check whether options are indeed ``\textit{redundant}'', and their findings strongly reject this hypothesis questioning the viability of the model \eqref{eq: general local volatility}. Another empirical study by Dumas et. al. \cite{Dumas_Fleming_Whaley_1998} analyzes the predictive performance of \eqref{eq: general local volatility} and concludes that its out-of-sample results are no better than ones of the standard Black-Scholes approach. Gatheral et. al. \cite{Gatheral_Jaisson_Rosenbaum_2014} indicates that the dynamics \eqref{eq: general local volatility} tends to generate future volatility surfaces ``\textit{completely unlike those we observe}''. Aït-Sahalia \& Jacod \cite{AitSahalia_Jacod_2019} utilize a statistical test based on local time to check whether stock prices follow\footnote{In fact, they consider even more general framework with jumps and microstructure noise and still reject the local volatility hypothesis.} the SDE of the form
\[
    dS(t) = a(S(t))dt + b(S(t))dW(t),
\]
where $a$ and $b$ are some deterministic functions, and report a ``\textit{clear rejection}'' of such model, advocating for alternative approaches. For more details on the performance of local volatility models, see also an overview in \cite[Section 4]{Skiadopoulos_2001}.

\subsection{Stochastic volatility models}\label{subsec: classical SV models}

As indicated above, local volatility models can be subject to criticism as they might not possess sufficient flexibility to grasp the market in its full complexity. An alternative approach, known as \textit{stochastic volatility}\footnote{The term ``stochastic volatility'' may seem too vague given that the volatility $\sigma(t, S(t))$ employed in local models is also a stochastic process. Nevertheless, in the literature, the term ``stochastic'' generally refers to approaches that treat volatility separately from the price dynamics.}, involves modeling with a dedicated stochastic process that exhibits only partial dependence with $S$. For example, $\sigma = \{\sigma(t),~t\ge 0\}$ may be another diffusion driven by a separate Brownian motion $B = \{B(t),~t\ge 0\}$ correlated with $W$. Naturally, within this framework, there are countless candidates for $\sigma$, and choosing a particular one that characterizes the market well is a complex task. However, stylized facts presented in Subsection \ref{subsec: stylized facts} offer several starting points that are useful to keep in mind.
\begin{itemize}
    \item To begin with, given the nature of volatility, it seems reasonable to model it with a non-negative process. In practice, this can be achieved by modeling log volatility and then taking an exponential. Alternatively, one may utilize non-negative diffusions such as Bessel-type processes (see e.g. \cite[Chapter XI]{Revuz_Yor_1999}).

    \item Selecting an appropriate dependence structure between $B$ and $W$ from \eqref{eq: general SV model} provides all the necessary tools to account for the leverage effect: a typical assumption is $\mathbb E[W(t)B(t)] = \rho t$ with $-1< \rho < 0$.

    \item Another property shared by multiple stochastic volatility models is \textit{mean-reversion}. As mentioned above in Subsection \ref{subsec: stylized facts}, this behavior seems to be a common trait for real-life volatility. Moreover, a thoroughly chosen mean-reverting process can, to some extent, mimic the clustering effect (see e.g. \cite[Chapter 3]{Fouque_Papanicolaou_Sircar_2000}). A common approach to introduce mean reversion to the dynamics is to take a drift term of the form
    \[
        \theta_1(\theta_2 - \sigma(t))
    \]
    in the SDE for volatility. This drift ``\textit{pulls $\sigma$ back}'' to the level $\theta_2$ whenever $\sigma$ deviates from it. The parameter $\theta_1$ calibrates the speed of mean-reversion.  
\end{itemize}

Starting from the 1987 pioneering works of Hull \& White \cite{Hull_White_1987}, Wiggins \cite{Wiggins_1987} and Scott \cite{Scott_1987}, multiple generations of approaches and numerous models have emerged in the literature. While not even attempting to provide an exhaustive list, we present below a selection of notable contributions.

\begin{itemize}
    \item Hull \& White \cite{Hull_White_1987} assumed that the squared volatility $\sigma^2 = \{\sigma^2(t),~t\ge 0\}$ is itself a geometric Brownian motion, i.e. price and volatility satisfy stochastic differential equations of the form 
    \begin{equation*}
    \begin{aligned}
        dS(t) &= \mu S(t)dt + \sigma(t) S(t) dW(t),
        \\
        d\sigma^2(t) & = \theta_1 \sigma^2(t)dt + \theta_2 \sigma^2(t) dB(t)
    \end{aligned}    
    \end{equation*}
    respectively, where $B$ and $W$ are two Brownian motions that are allowed to be correlated to account for the leverage effect. Note that the volatility process is positive but not mean-reverting.
    
    \item Wiggins \cite{Wiggins_1987} suggested a slightly more general dynamics of the form
    \begin{equation*}
    \begin{aligned}
        dS(t) &= \mu S(t)dt + \sigma(t) S(t) dW(t),
        \\
        d\sigma(t) & = f(\sigma(t))dt + \theta \sigma(t) dB(t).
    \end{aligned}   
    \end{equation*}
    
    \item Scott \cite{Scott_1987} and Stein \& Stein \cite{Stein_Stein_1991} considered the volatility to be an Ornstein-Uhlenbeck process, i.e.
    \begin{equation}\label{eq: classical Stein-Stein}
    \begin{aligned}
        dS(t) &= \mu S(t)dt + \sigma(t) S(t) dW(t),
        \\
        d\sigma(t) & = \theta_1(\theta_2 - \sigma(t))dt + \theta_3 dB(t).
    \end{aligned}   
    \end{equation}
    Note that $\sigma$ is not positive: Ornstein-Uhlenbeck process is Gaussian and hence can take negative values with positive probability. In practice, this issue is treated by either taking the absolute value of $\sigma$ or introducing a reflecting barrier to the volatility dynamics \cite{SZh1999}.
    
    \item Heston \cite{Heston_1993} introduced the SDE of the form 
    \begin{equation}\label{eq: Heston model}
    \begin{aligned}
        dS(t) &= \mu S(t)dt + \sqrt{\sigma(t)} S(t) dW(t),
        \\
        d\sigma(t) & = \theta_1(\theta_2 - \sigma(t))dt + \theta_3 \sqrt{\sigma(t)}dB(t).
    \end{aligned}   
    \end{equation}
    In this model, now commonly referred to as the \textit{Heston model}, the volatility follows the so-called \textit{Cox-Ingersoll-Ross} or \textit{square root} process (see also \cite{Cox_Ingersoll_Ross_1985}) which enjoys strict positivity provided that $2\theta_1\theta_2 \ge \theta_3^2$. Benhamou et. al. \cite{Benhamou_Gobet_Miri_2010} considered a modification of \eqref{eq: Heston model} of the form
    \begin{align*}
        dS(t) &= \sqrt{\sigma(t)} S(t) dW(t),
        \\
        d\sigma(t) & = \theta_1(\theta_2(t) - \sigma(t))dt + \theta_3(t) \sqrt{\sigma(t)}dB(t),
        \\
        d\langle W, B\rangle_t &= \rho(t)dt,
    \end{align*}
    with time-dependent $\theta_2$, $\theta_3$ and correlation $\rho$ to account for structural changes on the market. Another modification of \eqref{eq: Heston model} was considered in \cite{Goutte_Ismail_Pham_2017}: there, the discounted price dynamics is assumed to follow
    \begin{align*}
        dS(t) &= \sqrt{\sigma(t)} S(t) dW(t),
        \\
        d\sigma(t) & = \theta_1(Z_t)(\theta_2(Z_t) - \sigma(t))dt + \theta_3(Z_t) \sqrt{\sigma(t)}dB(t),
        \\
        d\langle W, B\rangle_t &= \rho(Z_t)dt,
    \end{align*}
    where $Z$ is a homogeneous continuous-time Markov chain that represents market switching between different regimes.

    \item Melino \& Turnbull \cite{Melino_Turnbull_1990, Melino_Turnbull_1995} considered the model of the form
    \begin{equation}\label{eq: MelinoTurnbull}
    \begin{aligned}
        dS(t) &= (a+bS(t))dt + \sigma(t)S^{\beta}dW(t),
        \\
        d\log(\sigma(t)) &= \left(\theta_1 + \theta_2 \log(\sigma(t))\right)dt + \theta_3 dB(t),
    \end{aligned}    
    \end{equation}
    where $\beta\in\left[\frac{1}{2},1\right]$, i.e. the log volatility is a mean-reverting Ornstein-Uhlenbeck process and $S$ can be regarded as a combination of the CEV equation \eqref{eq: CEV equation} with stochastic volatility.

    \item Hagan et. al. \cite{Hagan} proposed the stochastic \textit{alpha-beta-rho (SABR) model} of the form
    \begin{equation}\label{eq: SABR}
    \begin{aligned}
        dS(t) &= \sigma(t) S^\beta(t) dW(t),
        \\
        d\sigma(t) &= \alpha \sigma(t) dB(t).
    \end{aligned}    
    \end{equation}
    where $\alpha \ge 0$, $0 \le \beta \le 1$ and $\mathbb E[W(t)B(t)] = \rho t$, $-1 < \rho < 1$. Note that the case $\beta\in\left(0,\frac{1}{2}\right)$ is quite intricate from the mathematical perspective and requires a special treatment; see \cite{Cherny_Engelbert_2005} for more details.

    \item  Lewis \cite{Lewis_2000} and, later, Carr \& Sun \cite{Carr_Sun_2007} (see also Baldeaux \& Badran \cite[Section 2]{Baldeaux_Badran_2014}) considered the so-called \textit{3/2-model}
    \begin{equation}\label{eq: 3/2 model}
    \begin{aligned}
        dS(t) & =  S(t)\sqrt{\sigma(t)} dW(t), 
        \\
        d\sigma(t) & = \kappa \sigma(t)(\theta - \sigma(t))dt + \varepsilon \sigma^{3/2}(t) dB(t).
    \end{aligned} 
    \end{equation}
    The motivation behind the model partially comes from the statistical analysis of volatility models performed by Jawaheri \cite{Javaheri_2004} and Bakshi et. al. \cite{Bakshi_Ju_Ou-Yang_2006} as well as from its ability to replicate the VIX skew (see Section \ref{sec: VIX} for further details). For more details on the existence and properties of the solution to \eqref{eq: 3/2 model}, we refer the reader to \cite{Cherny_Engelbert_2005}.
    
    \item Gatheral \cite{Gatheral2008ConsistentMO} presented the double CEV dynamics of the form
    \begin{equation}\label{eq: double CEV}
    \begin{aligned}
        dS(t) &= S(t)\sqrt{\sigma(t)}dW(t), 
        \\
        d\sigma(t) &= -\kappa(\sigma(t) - \sigma'(t))dt + \eta \sigma^\alpha(t) dB(t), 
        \\
        d\sigma'(t)  &= -\kappa' (\sigma'(t) - \theta)dt + \eta' {\sigma'^\beta(t)} dB'(t),
    \end{aligned}   
    \end{equation}
    where $\alpha, \beta \in \left[\frac{1}{2}, 1\right]$;

    \item Fouque et. al. \cite{Fouque_Papanicolaou_Sircar_Solna_2003} propose a \textit{multiscale} stochastic volatility. Their approach lies in modeling volatility with two diffusions, one fluctuating on a fast ``time scale'', and the other fluctuating on a ``slow'' time scale. Namely, their model takes the form
    \begin{align*}
        dS(t) &= \mu S(t)dt + \sigma(t)S(t)dW(t),
        \\
        \sigma(t) &= f(Y(t), Z(t)),
        \\
        dY(t) &= \frac{1}{\varepsilon} (\theta_1 - Y(t))dt + \frac{\theta_2\sqrt{2}}{\sqrt{\varepsilon}} dB_1(t)
        \\
        dZ(t) &= \delta c(Z(t)) + \sqrt{\delta} g(Z(t)) dB_2(t),
    \end{align*}
    where $f$ is a bounded positive function, $\delta,\varepsilon>0$ are assumed to be small, $Y$ is the fast scale volatility factor, i.e. a fast mean-reverting process, and $Z$ is the slow scale volatility factor. Later in \cite{Fouque_Saporito_2018}, Fouque \& Saporito employ the same multiscale paradigm to model the Heston-type stochastic volatility-of-volatility parameter:
    \begin{equation}\label{eq: Heston vol-of-vol}
    \begin{aligned}
        dS(t) &=  S(t)\sqrt{\sigma(t)} dW(t), 
        \\
        d\sigma(t) &= \kappa(\theta - \sigma(t))dt + \eta(t)\sqrt{\sigma(t)}dW'(t),
        \\
        \eta(t) &= f(Y(t), Z(t)),
        \\
        dY(t) &= \frac{\sigma(t)}{\varepsilon} \alpha(Y(t))dt + \sqrt{\frac{\sigma(t)}{\varepsilon}} \beta(Y(t))dB(t), 
        \\
        dZ(t) &= \sigma(t) \delta c(Z(t))dt + \sqrt{\delta \sigma(t)} g(Z(t))dB'(t).
    \end{aligned}     
    \end{equation}
    
\end{itemize}

In addition to reproducing the leverage effect and clustering via mean reversion, Brownian stochastic volatility models turn out to have an additional important advantage: they have an ability to capture, to some extent, ``smiley'' patterns of the implied volatility (see e.g. \cite{Renault_Touzi_1996}, \cite[Section 2.8.2]{Fouque_Papanicolaou_Sircar_2000} or \cite{Gatheral_2006}). However, by design, parametric stochastic volatility models impose structural constraints on the relationship between the dynamics of the spot and implied volatilities and hence may not be able to capture the exact shape of the implied volatility surface (see e.g. \cite[Figure 3.6]{Gatheral_2006}).

A detailed analysis on that matter can be found in \cite{Lee_2006} and \cite{Alos_Leon_Vives_2007}: in particular, as stated in \cite[Section 7.1]{Alos_Leon_Vives_2007} or \cite[Remark 11.3.21]{Lee_2006}, Brownian diffusion models of volatility tend to produce the at-the-money skew $\Psi(T) = O(1)$, $T\to 0$, where $\Psi$ is defined by \eqref{intro: volatility skew}. This directly contradicts the empirical behavior of $O(T^{-\beta})$, $\beta\approx\frac{1}{2}$, mentioned in Subsection \ref{subsec: stylized facts}. In addition, as highlighted by Comte \& Renault \cite{Comte_Renault_1998} (see also \cite{Carr_Wu_2003}), the decrease of the real-life volatility smile amplitude as $T\to\infty$ seems to be much slower than predicted by the classical Brownian diffusions. Comte \& Renault connect this phenomenon to the \textit{long-range dependence} in the volatility dynamics, which is well-aligned with several empirical studies listed above in Subsection \ref{subsec: stylized facts} that also report long memory on the market.

In other words, the inherent properties of Brownian motion limit the modeling capabilities of diffusion models and it is no surprise that a lot of effort was made to advance the stochastic volatility framework further to account for the mentioned inconsistencies.

\subsection{Fractional and rough models}\label{subsec: fractional and rough models}

One of the ways to extend stochastic volatility models described in Subsection \ref{subsec: classical SV models} involves substituting the standard Brownian driver $B$ with an alternative process possessing the attributes to capture the intended stylized facts. Perhaps the most common option utilized in the literature is \textit{fractional Brownian motion} $B^H = \{B^H(t), t\ge 0\}$ defined as a Gaussian process with $B^H(0) = 0$ a.s., $\mathbb E[B^H(t)] = 0$ for all $t\ge 0$ and
\begin{equation}\label{eq: cov of fbm}
    \mathbb E\left[B^H(t) B^H(s)\right] = \frac{1}{2}\left(t^{2H} + s^{2H} - |t-s|^{2H}\right), \quad s,t \ge 0,
\end{equation}
where $H$ can take values in $(0,1)$ and is called the \textit{Hurst index}\footnote{This parameter is named after Harold Edwin Hurst (1880--1978) who studied long-range dependence in fluctuations of the water level in the Nile River.}. Fractional Brownian motion was initially considered by Kolmogorov \cite{Kolmogorov_1940} and later reintroduced by Mandelbrot and van Ness \cite{Mandelbrot_Van_Ness_1968} who also obtained its Volterra-type representation
\begin{equation}\label{intro: FBM}
\begin{aligned}
    B^H(t) &:= \frac{1}{\Gamma\left(H+\frac{1}{2}\right)} \int_{-\infty}^0 \left( (t-s)^{H-\frac{1}{2}} - (-s)^{H-\frac{1}{2}} \right)dB(s) + \frac{1}{\Gamma\left(H+\frac{1}{2}\right)} \int_0^t (t-s)^{H-\frac{1}{2}} dB(s),
\end{aligned}
\end{equation}
where $B = \{B(t),~t\ge 0\}$ is a standard Brownian motion. In the literature, a truncated version of \eqref{intro: FBM} called the \textit{Riemann-Liouville fractional Brownian motion} is often used:
\begin{equation}\label{eq: RL-FBM}
    B^H_{RL}(t) := \frac{1}{\Gamma\left(H+\frac{1}{2}\right)} \int_0^t (t-s)^{H-\frac{1}{2}} dB(s),
\end{equation}
and the Volterra kernel 
\begin{equation}\label{eq: fractional kernel}
    \mathcal K(t,s) := \frac{1}{\Gamma\left(H+\frac{1}{2}\right)} (t-s)^{H-\frac{1}{2}}\mathbbm 1_{s<t}
\end{equation}
in \eqref{eq: RL-FBM} is called the \textit{fractional} kernel.

Fractional Brownian motion is extremely convenient from the mathematical perspective: it is the only stochastic process that simultaneously \cite{Mishura_2007}
\begin{itemize}
    \item is Gaussian,
    \item has stationary increments and
    \item is \textit{self-similar}, i.e.
    \begin{equation}\label{intro: self-similarity}
        B^H(at) \stackrel{\text{Law}}{=} a^H B^H(t), \quad \forall a \ge 0.
    \end{equation}
\end{itemize}
In addition, if $H=1/2$, $B^H$ coincides with a standard Brownian motion and hence can be viewed as a broad generalization of the latter. Interestingly, the value $H=1/2$ turns out to be the boundary between two distinct volatility modeling paradigms: one favoring $H>1/2$ and the other advocating for $H<1/2$, and the aim of this subsection is to characterize both of them.

\paragraph{Fractional models with long memory: $H>1/2$.} First of all, observe that
\begin{equation}\label{eq: autocorrelation of FBM}
\begin{aligned}
    \mathbb E\left[B^H(1)\left(B^H(n) - B^H(n-1)\right)\right] &= \frac{1}{2} \left( n^{2H} - 2(n-1)^{2H} +(n-2)^{2H} \right) 
    \\
    &\sim H(2H-1) n^{2H-2}, \quad n\to\infty.
\end{aligned}
\end{equation}
Therefore, if $H\in (1/2,1)$, the autocorrelation function behaves as $O(n^{-\beta})$ with $\beta \in (0,1)$, revealing the long memory property. In particular, if $H\in(3/4,1)$, the behavior of \eqref{eq: autocorrelation of FBM} matches the empirical estimates $0 < \beta \le \frac{1}{2}$ for absolute log-returns highlighted in Subsection \ref{subsec: stylized facts}.

Historically, the long memory of fractional Brownian motion for $H>1/2$ was the original reason to employ the latter in volatility modeling. Namely, in 1998, Comte \& Renault  \cite{Comte_Renault_1998} suggested the first continuous time fractional volatility model of the form
\begin{equation}\label{eq: RC}
    \sigma(t) = \theta_1 \exp\left\{ \theta_2 \int_{0}^{t} e^{-\theta_3(t-s)}dB^H(s)  \right\},
\end{equation}
which can mimic volatility persistence and explain slow decays in the smile amplitude of implied volatility surfaces when $T \to \infty$ (in this regard, we also recommend the simulation study \cite{Funahashi_Kijima_2017} that illustrates this phenomenon numerically). Other contributions studying stochastic volatility driven by fractional Brownian motion with $H>1/2$ include:
\begin{itemize}
    \item \cite{Rosenbaum_2008}, which considers the model of the form
    \begin{align*}
        dS(t) & = \sigma(t)dW(t),
        \\
        \sigma(t) &= F\left( \int_0^t a(t,u)dB^H(u) + f(t)\xi_0 \right)
    \end{align*}
    where $F$, $a$ and $f$ are nuisance parameters and $\xi_0$ is a random initial condition;

    \item \cite{ChronopoulouViens2012} and \cite{BMdP2018}, which discuss a fractional counterpart of the model \eqref{eq: classical Stein-Stein}:
    \begin{equation}\label{eq: fractional Stein-Stein}
    \begin{aligned}
        dS(t) & = \mu S(t)dt + \sigma(Y(t))S(t)dW(t),
        \\
        Y(t) &= -\theta_1 Y(t)dt + \theta_2 dB^H(t),
    \end{aligned}    
    \end{equation}
    where $\mu$, $\theta_1$, $\theta_2 > 0$ and $\sigma$ is a deterministic function that is additionally assumed to have sublinear growth in  \cite{BMdP2018};

    \item the series of papers \cite{MYuT2018, MYuT2019, MYT2020} that proposes
    \begin{equation}\label{eq: fractional Heston}
    \begin{aligned}
        dS(t) & = \mu S(t)dt + \sigma(Y(t))S(t)dW(t),
        \\
        dY(t) &= \frac{1}{2}\left(\frac{\theta_1}{Y(t)} - \theta_2 Y(t)\right)dt + \frac{\theta_3}{2} dB^H(t),
    \end{aligned}   
    \end{equation}
    where $\mu$, $\theta_1$, $\theta_2$, $\theta_3 > 0$ and $\sigma$ is a given function with sublinear growth. Note that \eqref{eq: fractional Heston} can be regarded as a fractional extension of the Heston model \eqref{eq: Heston model} since the process $X(t) = Y^2(t)$ satisfies the pathwise SDE
    \[
        dX(t) = (\theta_1 - \theta_2 X(t))dt + \theta_3 \sqrt{X(t)}dB^H(t).
    \]
\end{itemize}
In some models, long memory is incorporated not directly through fractional Brownian motion, but rather by combining standard Brownian diffusions with the fractional kernel \eqref{eq: fractional kernel}. Examples of this approach can be found in e.g. \cite{ChronopoulouViens2012} or \cite{Bauerle_Desmettre_2020}, where the prices are assumed to follow
\begin{align*}
    dS(t) & = \mu(t) S(t)dt + \sigma(t)S(t)dW(t),
    \\
    \sigma^2(t) &= \theta + \frac{1}{\Gamma\left(H+\frac{1}{2}\right)}\int_{-\infty}^t (t-s)^{H-\frac{1}{2}} X(s)ds,
    \\
    dX(t) &= \theta_1(\theta_2 - X(t))dt + \theta_3 \sqrt{X(t)}dB(t),
\end{align*}
where $W$ and $B$ are correlated standard Brownian motions.

As a final remark, we mention the special interplay between the long memory and self-similarity \eqref{intro: self-similarity} properties of fractional Brownian motion. As previously discussed in Subsection \ref{subsec: stylized facts}, any measures of long-range dependence are hard to estimate statistically: the analysis of time series during overly extended time periods naturally raises concerns about data non-stationarity. However, if the data is additionally assumed to display self-similarity, its long-term behavior can be inferred from high-frequency observations over shorter periods. For a detailed discussion on self-similarity in financial time series as well as its connection with long memory, we refer the reader to \cite[Subsections 2.3 and 2.4]{Cont_2005}.

\paragraph{Rough revolution: $H<1/2$.} As highlighted in the simulation study \cite{Funahashi_Kijima_2017}, fractional Brownian motion with $H>1/2$ indeed allows to capture the behavior of implied volatility surfaces for longer maturities. However, a high Hurst index does not seem to have any positive impact on the short-term fit. For example, Alòs, León \& Vives in Section 7.2.1 of their 2007 paper \cite{Alos_Leon_Vives_2007} analyze the behavior of the short-term implied volatility skew \eqref{intro: volatility skew t} generated by a variation of the model \eqref{eq: fractional Stein-Stein} with $H>1/2$. They analytically prove that this skew evaluated at a fixed time $t\ge 0$ does not behave as $O((T-t)^{-\beta})$, $\beta\approx \frac{1}{2}$, for $T\downarrow t$. Interestingly, in Section 7.2.2 of the same paper, they notice that the stochastic volatility model driven by a Riemann-Liouville fractional Brownian motion with $H\in (0,1/2)$ produces the required skew asymptotics $O((T-t)^{-\frac{1}{2}+H})$ as $T\downarrow t$.

The arguments in \cite{Alos_Leon_Vives_2007} were based on \textit{Malliavin calculus} (see e.g. \cite{Alos_Garcia_Lorite_2021, Nualart_2006}): under certain regularity assumptions, the short-term explosion of the implied volatility skew translates to the explosion of the Malliavin derivative of volatility, the property which holds for e.g. fractional Brownian motion with $H<1/2$. In 2014\footnote{Although paper \cite{Gatheral_Jaisson_Rosenbaum_2014} appeared in \textit{Quantitative finance} in 2018, the preprint had been available on SSRN and ArXiv since 2014.}, Gatheral, Jaisson \& Rosenbaum \cite{Gatheral_Jaisson_Rosenbaum_2014}  advocated for $H<1/2$ from a different perspective: using estimation techniques based on power variations, they came to a conclusion that volatility must have H\"older regularity of order $\approx 0.1$. Using \eqref{eq: cov of fbm}, it is easy to check that
\[
    \mathbb E[|B^H(t) - B^H(s)|^2] = |t-s|^{2H}, \quad s,t\ge 0,
\]
so, by the Kolmogorov-Chentsov theorem (see e.g. \cite[p. 192]{Gihman_Skorokhod_2004}), the paths of $B^H$ are H\"older continuous up to the order $H$. In other words, \cite{Gatheral_Jaisson_Rosenbaum_2014} concluded that fractional Brownian motions with very small Hurst index $H$ are preferable choices for stochastic volatility modeling. Later in 2021, Fukasawa \cite{Fukasawa_2021} re-visited the interplay between roughness and power-law of at-the-money volatility skew. The main theoretical result presented in \cite{Fukasawa_2021} is as follows: \textit{if} the price $S$ is a positive \textit{continuous} semimartingale and \textit{if} the at-the-money implied volatility skew \eqref{intro: volatility skew t} evaluated at $t>0$ exhibits the power-law behavior of the form $(T-t)^{-\frac{1}{2} + H}$ as $t \uparrow T$ with $H\approx 0$, then $H_0$-H\"older continuity of the quadratic variation derivative
\begin{equation}\label{eq: quadratic variation}
    \frac{d}{dt} \langle \log S \rangle_t
\end{equation}
leads to an arbitrage opportunity if $H_0 > H$. Note that \eqref{eq: quadratic variation} exactly coincides with $\sigma^2 = \{\sigma^2(t),~t \ge 0\}$ in the general stochastic volatility model \eqref{eq: general SV model} and hence \cite{Fukasawa_2021} implies that, in a continuous setting with power law of the implied volatility skew, the volatility process ``\textit{has to}'' be rough to avoid arbitrage.

These collective observations rapidly developed into a vast research field known as ``\textit{rough volatility}'', with hundreds of papers published over the years. For an extensive literature list on the topic that is regularly updated by specialists in the field, we refer the reader to \cite{Rough_volatility_literature}. Some notable models include (in all cases, $H\in\left(0, \frac{1}{2}\right)$):
\begin{itemize}
    \item the rough fractional stochastic volatility (RFSV) model \cite{Gatheral_Jaisson_Rosenbaum_2014}
    \[
        \sigma(t) = \exp\left\{\theta_1 +\theta_2\int_{-\infty}^t e^{-\theta_3 (t-s)}dB^H(s) \right\},
    \]
    which can be regarded as a rough counterpart of the model \eqref{eq: RC};
    
    \item the rough Bergomi model \cite{Bayer_Friz_Gatheral_2016, Jacquier_Martini_Muguruza_2018}
    \begin{equation}\label{eq: RB}
    \begin{aligned}
        dS(t) &= \sqrt{\sigma(t)}S(t)dW(t),
        \\
        \sigma(t) & = \theta_{0}(t) \exp\left\{ 2\theta_1 \int_0^t (t-s)^{H-\frac{1}{2}}dB(s) - \theta_1^2 t^{2H}  \right\},
    \end{aligned}
    \end{equation}
    where $\theta_1>0$ and $\theta_0$ is a deterministic function;

    \item the mixed rough Bergomi model \cite{Guyon_RBM, Lacombe_Muguruza_Stone_2021}
    \begin{equation}\label{eq: MRB}
    \begin{aligned}
        dS(t) &= S(t)\sqrt{\sigma(t)}dB(t), 
        \\
        \sigma(t) = \theta_0(t) &\bigg( \delta \exp\left\{ 2\theta_1 \int_0^t (t-s)^{H-\frac 1 2} dW(s) - \theta_1^2 t^{2H}\right\}
        \\
        &\quad + (1 - \delta) \exp\left\{ 2\theta_2 \int_0^t (t-s)^{H-\frac 1 2} dW(s) - \theta_2^2 t^{2H}\right\}  \bigg),
    \end{aligned}    
    \end{equation}
    where $\delta\in[0,1]$, $\theta_1$, $\theta_2 >0$ and $\theta_0$ is a deterministic function;
    
    \item the rough SABR model \cite{Fukasawa_Gatheral_2022} where the price is given by
    \begin{equation*}
    \begin{aligned}
        dS(t) &= \sqrt{\sigma(t)} \beta(S(t)) dW(t),
        \\
        \sigma(t) & = \theta_0(t)\exp\left\{ \theta\sqrt{2H} \int_0^t (t-u)^{H-\frac 1 2}dB(s) - \frac{1}{2} \theta^2 t^{2H} \right\} dB(t),
    \end{aligned}
    \end{equation*}
    with $0 \le t \le s$, $\theta_0$ being a positive random process and $\beta$ being a positive continuous function;
    
    \item the rough Stein-Stein model \cite{Harms_Stefanovits_2019}
    \begin{equation*}
    \begin{aligned}
        dS(t) &= \theta \sigma(t) S(t) dW(t),
        \\
        \sigma(t) & = \frac{1}{\Gamma\left(H+\frac{1}{2}\right)} \int_0^t (t-s)^{H-\frac{1}{2}}dB(s);
    \end{aligned}
    \end{equation*}

    \item the fast-varying rough volatility \cite{Garnier_Sølna_2020}
    \begin{equation}
    \begin{aligned}
        dS(t) &= \mu(t)dt + \sigma^\varepsilon(t)dW(t),
        \\
        \sigma^\varepsilon(t) &= F(Z^\varepsilon(t)),
        \\
        Z^\varepsilon(t) &= \frac{\theta}{\sqrt{\varepsilon}} \int_{-\infty}^t \mathcal K_H\left(\frac{t-s}{\varepsilon}\right)dB(s), 
    \end{aligned}
    \end{equation}
    where $\mathcal K_H(t) := C_H\left( t^{H-\frac{1}{2}} - \int_0^t (t-s)^{H-\frac{1}{2}} e^{-s}ds \right)$, $C_H>0$, $F$ is positive and bounded one-to-one function, $\theta > 0$ and $\varepsilon>0$ is assumed to be small; 
    
    \item the rough Heston model \cite{El_Euch_Rosenbaum_2019}
    \begin{equation}\label{eq: rough Heston}
    \begin{aligned}
        dS(t) &= \sqrt{\sigma(t)} S(t) dW(t),
        \\
        \sigma(t) & = \sigma(0) +  \int_0^t \frac{(t-s)^{H-\frac{1}{2}}}{\Gamma\left(H+\frac{1}{2}\right)} \left(\theta_1(\theta_2 - \sigma(s))ds + \theta_3 \sqrt{\sigma(s)}dB(s)\right)
    \end{aligned}
    \end{equation}
    and its modification, the {quadratic rough Heston model} \cite{Gatheral_Jusselin_Rosenbaum_2020, Rosenbaum_Zhang_2021},
    \begin{equation}\label{eq: quadratic rough Heston}
    \begin{aligned}
        dS(t) &= S(t)\sqrt{\sigma(t)}dW(t),
        \\
        \sigma(t) &= a(Z(t) - b)^2 + c,
        \\
        Z(t) = \int_0^t \theta_1 \frac{(t-s)^{H-\frac{1}{2}}}{\Gamma\left(H+\frac{1}{2}\right)}& (\theta_0(s) - Z(s))ds + \int_0^t \theta_2 \frac{(t-s)^{H-\frac{1}{2}}}{\Gamma\left(H+\frac{1}{2}\right)} \sqrt{a(Z(s)-b)^2+c}dW(s),
    \end{aligned}
    \end{equation}
    where $B$, $W$ are Brownian motions, $a$, $b$, $c$, $\theta_1$, $\theta_2$, $\theta_3 >0$ and $\theta_0$ is a deterministic function.
\end{itemize}
The rough Heston approach \eqref{eq: rough Heston} is especially interesting from the modeling perspective: in addition to the power law behavior of \eqref{intro: volatility skew}, it can reproduce a special form of Zumbach effect \cite{Dandapani_Jusselin_Rosenbaum_2021, El_Euch_Gatheral_Radoicic_Rosenbaum_2020} and can be interpreted as the limit of a reasonable tick-by-tick price model based on two-dimensional Hawkes processes \cite{El_Euch_Fukasawa_Rosenbaum_2018}, i.e. deduces roughness of the volatility directly from the market microstructure.

\paragraph{Puzzles of fractionality and roughness.} In principle, one can divide the majority of arguments in favor of rough volatility into two distinct categories:
\begin{itemize}
    \item econometric studies like \cite{Gatheral_Jaisson_Rosenbaum_2014} which involve regularity estimations of volatility from historical high-frequency samples of an asset (e.g.\ S\&P500 index);

    \item analyses from the options pricing perspective such as \cite{Alos_Leon_Vives_2007, Fukasawa_2021} that justify roughness by its ability to reproduce e.g. the power law behavior of implied volatility skews \eqref{intro: volatility skew}.   
\end{itemize}
However, as is often the case when modeling extremely complex systems like the financial market, neither of these arguments is convincing enough to be unequivocally declared undeniable and immune to valid criticism. For example, the methodology used in \cite{Gatheral_Jaisson_Rosenbaum_2014} was examined by Rogers \cite{Rogers_2019}, Cont \& Das \cite{Cont_Das_2022} as well as Fukasawa, Takabatake \& Westphal \cite{Fukasawa_Takabatake_Westphal_2019, Fukasawa_Takabatake_Westphal_2022}: they apply the same regularity estimation procedure as in \cite{Gatheral_Jaisson_Rosenbaum_2014} to \textit{synthetic} datasets and report that the estimator tends to produce low values of $H$ \textit{regardless of the true parameter} used in the simulation. This issue is consensually explained as follows in the abovementioned literature: since the volatility $\sigma$ is not observable directly, one must first extract some volatility proxy from the stock data and then use \textit{it} as the sample for regularity estimation. However, this procedure results in additional approximation errors that bias the estimator of $H$ towards zero. The authors of \cite{Gatheral_Jaisson_Rosenbaum_2014} acknowledge this issue themselves writing that ``\textit{estimation errors when estimating volatility can be quite significant for some models, leading to downward biases in the measurement of the smoothness}'' \cite[p.~946]{Gatheral_Jaisson_Rosenbaum_2014}.

Interestingly, there were several subsequent attempts to account for the problem described above. For example, Fukasawa, Takabatake \& Westphal \cite{Fukasawa_Takabatake_Westphal_2019, Fukasawa_Takabatake_Westphal_2022} develop a more robust estimation technique and still report that $H < 0.1$ is indeed the best fit for the volatility model
\begin{equation}\label{eq: log-fOU}
    d\log\sigma^2(t) = \theta_1(t)dt + \theta_2dB^H(t)
\end{equation}
with $\theta_1$ being an unknown adapted càdlàg process. Bolko et. al. \cite{Bolko_Christensen_Pakkanen_Veliyev_2023} perform the generalized method of moment (GMM) estimation and also confirm the roughness of volatility for thirty-one leading stock indexes under the assumption that $\log\sigma^2$ is a fractional Ornstein-Uhlenbeck process. 

Next, the claim that the implied volatility skew follows the power law can also be challenged. For example, Guyon \& El Amrani \cite{Guyon_El_Amrani_2022} investigate the volatility skew behavior in more detail and conclude that 
\begin{quote}
    ``\textit{...power law fits the term-structure of ATM skew of equity indexes quite well over a large range of maturities, from 1 month to a few years\footnote{In the dataset we used to produce Fig.~\ref{fig: Skew normal} above, the shortest maturity was 10 days.}}. \textit{But this should not lead one to conclude that the ATM skew blows up like this power law at zero maturity. Zooming in on the short maturities shows that the power-law extrapolation fits the data poorly.}''
\end{quote}
In summary, Guyon \& El Amrani \cite[p.~14]{Guyon_El_Amrani_2022} argue that ``\textit{far from being infinite, the zero-maturity extrapolation of the skew\footnote{I.e. the value $\Psi(0)$ in \eqref{intro: volatility skew}.} is distributed around a typical value of 1.5}''. On the contrary, Delemotte, De Marco \& S\'{e}gonne \cite{Delemotte_Marco_Segonne_2023} investigate the behavior of average volatility skews between 2007 and 2015 (a period that is different from 2020-2022 of Guyon \& El Amrani \cite{Guyon_El_Amrani_2022}) and find that a power-law-type explosive behavior of the implied volatility skew when $T\to 0$ is entirely appropriate! In addition, they argue \cite[Section 3]{Delemotte_Marco_Segonne_2023} that the average skew behaves differently for shorter and longer maturities: their analysis suggests that
\begin{equation}\label{eq: 2PL}
    \mathbb E [\Psi(T)] \propto \begin{cases}
        T^{-\frac{1}{2}+H_1}, &\quad T<T_1,
        \\
        T^{-\frac{1}{2} + H_2}, &\quad T>T_1,
    \end{cases}
\end{equation}
where $H_1 > H_2$ and $T_1$ is roughly 2 months.

What is the reason for such a discrepancy in conclusions regarding very similar (and sometimes \textit{literally the same}) datasets? Features like the behavior of the volatility skew are very intricate and their assessment based on discrete data leaves much space for interpretation and extrapolation. For example, Delemotte, De Marco \& S\'{e}gonne \cite[pp. 2--3]{Delemotte_Marco_Segonne_2023} compare their result to \cite{Guyon_El_Amrani_2022} and summarize the difference as follows:
\begin{quote}
    ``\textit{As a result, both models could very reasonably be used to extrapolate the ATM skew for very short maturities, and in any case below the first maturity quoted on the market, while leading to different skew asymptotics: finite limiting skew in the 2-factors Bergomi model\footnote{The one of \cite{Guyon_El_Amrani_2022}.}, and exploding skew in the 2PL model\footnote{The one of \cite{Delemotte_Marco_Segonne_2023}.}—indicating that \textbf{the question of the explosive or non-explosive nature of the short-end of the skew curve might be (at least in the case of the SP500 index) hard to disambiguate}.}''
\end{quote}
In some sense, this ambiguity is very natural: the volatility itself is a purely theoretical (albeit meaningful) concept, and hence there is no single ``\textit{ultimately true volatility model}''. In this context, the question ``\textit{is volatility \textbf{truly} rough}?'' lacks a definitive ``\textit{correct}'' answer. Rough volatility is a valid modeling framework with its advantages and limitations which indeed mimics some important features of the data: for example, the empirically observed steepness of smiles at-the-money does increase close to maturity (as clearly visible on Fig.~\ref{fig:Smile}), and rough models seem to capture this effect better than classical Brownian diffusions, irrespective of whether this increase ``truly'' follows the power law. In turn, econometric analyses like \cite{Bolko_Christensen_Pakkanen_Veliyev_2023, Fukasawa_Takabatake_Westphal_2019, Fukasawa_Takabatake_Westphal_2022} can be viewed as reasonable reality checks which can reveal possible points of contradiction with data and test specific models against each other: for example, the results of \cite{Bolko_Christensen_Pakkanen_Veliyev_2023, Fukasawa_Takabatake_Westphal_2019, Fukasawa_Takabatake_Westphal_2022} could mean that \eqref{eq: log-fOU} with small $H$ seems to be more consistent with some features of the data than e.g. \eqref{eq: MelinoTurnbull} with $\beta = 1$ or \eqref{eq: RC} with $H>\frac 1 2$. On the other hand, other models may also be compatible with observations: after all, according to a famous aphorism commonly attributed to the statistician George Box, \textit{all models are wrong, but some are useful}.

With this disclaimer in mind, let us finish this Subsection with a discussion of another intricate point coming from the interplay between roughness and long memory \textit{in the specific context of fractional Brownian motion $B^H$}. More precisely, \textit{long memory} requires the Hurst index $H>1/2$ whereas \textit{roughness} demands $H<1/2$. Despite some studies that demonstrate spurious long memory appearing due to model misspecifications (see e.g. \cite[Section 4]{Gatheral_Jaisson_Rosenbaum_2014}), roughness alone cannot explain the behavior of the entire volatility surface. For example, Funahashi \& Kijima \cite{Funahashi_Kijima_2017-1} demonstrate that volatility models based on fractional Brownian motion with $H<1/2$ do not give the required rate of decrease in the smile amplitude as $T\to\infty$ whereas $H>1/2$ gives this effect (see also \cite{Funahashi_Kijima_2017}). 

It should be noted that this contradiction, referred to as the ``\textit{fractional modeling puzzle}'' in \cite{Funahashi_Kijima_2017-1}, is somewhat synthetic: in general, long memory and roughness do not depend on each other and can co-exist within a single stochastic process. In other words, the reason for this ``\textit{fractional puzzle}'' comes from the properties of fractional Brownian motion itself and a different modeling framework may utilize both of these features without any structural contradictions (see e.g. the model based on Brownian semistationary processes by Bennedsen, Lunde \& Pakkanen \cite{Bennedsen_Lunde_Pakkanen_2022}). Nevertheless, in a continuous setting, $B^H$ seems to be an extremely convenient and valuable asset from the mathematical perspective: 
\begin{itemize}
    \item it is Gaussian, enabling the utilization of numerous methods from Gaussian process theory including efficient numerical methods, Malliavin calculus etc. (see e.g. \cite[Chapter 5]{Nualart_2006});
    \item it has stationary increments and is ergodic, which facilitates various statistical estimation techniques (see e.g. \cite{Kubilius_Mishura_Ralchenko_2017});
    \item despite being non-semimartingale, fractional Brownian motion has a developed stochastic integration theory \cite{Mishura_2007}.
\end{itemize}

In the literature, there are several possible approaches to incorporate long memory and roughness within the fractional framework. The most straightforward methodology is to use two fractional Brownian motions with different Hurst indices. Such model was utilized in e.g. \cite{Funahashi_Kijima_2017-1} (see also \cite[Section 7.7]{Alos_Garcia_Lorite_2021}) in the form
\begin{equation}\label{intro: long and short memory together}
\begin{aligned}
    dS(t) &= \mu S(t)dt + \sigma(X^1(t), X^2(t)) dW(t),
    \\
    dX^i(t) & = (\theta^i_1 - \theta^i_2 X^i(t))dt + \theta^i_3 dB^{H_i}(t), \quad i = 1,2,
\end{aligned}
\end{equation}
with $B^{H_1}$, $B^{H_2}$ being two fractional Brownian motions with $H_1 > 1/2$ and $H_2 < 1/2$. The authors report that this model indeed manages to grasp the implied volatility surface with the power law for short maturities and slower decay of the smile amplitude for long maturities. Such a suggestion somehow resonates with the observations made in \cite{Delemotte_Marco_Segonne_2023}: the authors also discuss the possibility of introducing several factors with different regularity into the model in order to reproduce \eqref{eq: 2PL} (although they consider $0 < H_2 < H_2 < \frac{1}{2}$). 

Another interesting possibility -- \textit{a multifractional Brownian motion} -- is advocated in Corlay et. al. \cite{CLV_2014} (see also \cite{Ayache_Peng_2012}). There, the authors estimate the local roughness of the volatility and conclude that it is heavily variable and has periods of low ($\approx 0.1$) and high ($\approx 0.8$) regularity (see \cite[Figure 2]{CLV_2014}). It is also important to note that \cite[Section 2.6]{Gatheral_Jaisson_Rosenbaum_2014} also reports some dependence of the volatility roughness on time.

As a generalization of both approaches mentioned above, one can also consider the usage of \textit{Gaussian Volterra processes} $Z(t):= \int_0^t \mathcal K(t,s) dB(s)$. Such volatility drivers were considered in e.g. \cite{Catalini_Pacchiarotti_2022, Merino_Pospisil_Sobotka_Sottinen_Vives_2021} or the series of papers \cite{DiNunno_Mishura_Yurchenko-Tytarenko_2022, DiNunno_Yurchenko-Tytarenko_2022}.

\subsection{Usability challenges of stochastic volatility models}

We finish this section with several remarks that, in our opinion, are worth the attention of the reader.

\paragraph{Positivity of volatility.} As mentioned in Subsection \ref{subsec: classical SV models}, one of the natural expectations from a ``\textit{reasonable}'' volatility model is the positivity of its paths. This requirement actually goes beyond the simple consideration that $\sigma = \{\sigma(t),~t\ge 0\}$ should resemble its original proxy \eqref{eq: realized volatility} and is connected to the procedure of transition between the physical and the pricing measures. In the Black-Scholes-type models \eqref{eq: general SV model}, martingale densities usually involve terms of the form $\int_0^T \frac{1}{\sigma(s)}dW(s)$ and $\int_0^T \frac{1}{\sigma^2(s)}ds$ (see e.g. \cite[Proposition 1.11]{BGP2000}) which can be poorly defined, if the volatility hits zero with positive probability. Of course, one may model the market under the pricing measure in the first place, but, in this situation, one sacrifices the ambition to justify their approach with econometric analysis based on historical time series a la \cite{Gatheral_Jaisson_Rosenbaum_2014} which, of course, should be performed under the physical measure.

\paragraph{Possibility of moment explosions} A common issue for the stochastic volatility framework is the possibility of moment explosions in price \cite{Andersen_Piterbarg_2006}. That means that the moment of the price $\mathbb E[S^r(t)]$ may be infinite for all time points $t$ after some $t_*$. Moment explosions can be a notable drawback from at least two perspectives. First, numerical schemes involving $L^r$-convergence become inaccessible. Another perspective is asset pricing: as it is noted in \cite[Section 8]{Andersen_Piterbarg_2006}, ``\textit{several actively traded fixed-income derivatives require at least $L^2$ solutions to avoid infinite model prices}''. In principle, both positivity and absence of moment explosions can be achieved by assuming bounds on the volatility; see e.g. \cite{Alos_Rolloos_Shiraya_2022, BMdP2018, DiNunno_Mishura_Yurchenko-Tytarenko_2022, DiNunno_Yurchenko-Tytarenko_2022, Fouque_Papanicolaou_Sircar_Solna_2003, Garnier_Sølna_2020, Ocone_Karatzas_1991} and the footnote on p. 2 in \cite{Rosenbaum_Zhang_2021}. It seems that this assumption is not too aggravating: in industry, there seems to be a consensus \cite{Edwards_Lazzara_2011} that real-world proxies of volatility are typically range-bound.
    
\paragraph{Importance of numerical methods.} As noted in \cite[p. 24]{Gatheral_2006}, the standard Heston model is still widely used despite all the empirical inconsistencies outlined in Subsection \ref{subsec: classical SV models}. The reason for that is mainly the availability of algorithms for practically all possible applications. Stochastic volatility models normally do not have closed-form expressions for option pricing, portfolio optimization, hedging etc., and therefore the development of efficient numerics for them is of acute importance. It is especially relevant for fractional/rough models which are predominantly non-Markovian and hence cannot utilize various stochastic optimization techniques developed for Markov processes. In recent years, some work in this direction was done in e.g. \cite{AbiJaber_ElEuch_2019, Jaber_Illand_Shaun_Li_2022b, DiNunno_Mishura_Yurchenko-Tytarenko_2022, DiNunno_Mishura_Yurchenko-Tytarenko_2023b, DiNunno_Yurchenko-Tytarenko_2022, Rosenbaum_Zhang_2021}.

\section{VIX and the joint calibration puzzle}\label{sec: VIX}

\begin{figure}
    \centering
    \includegraphics[width = \textwidth]{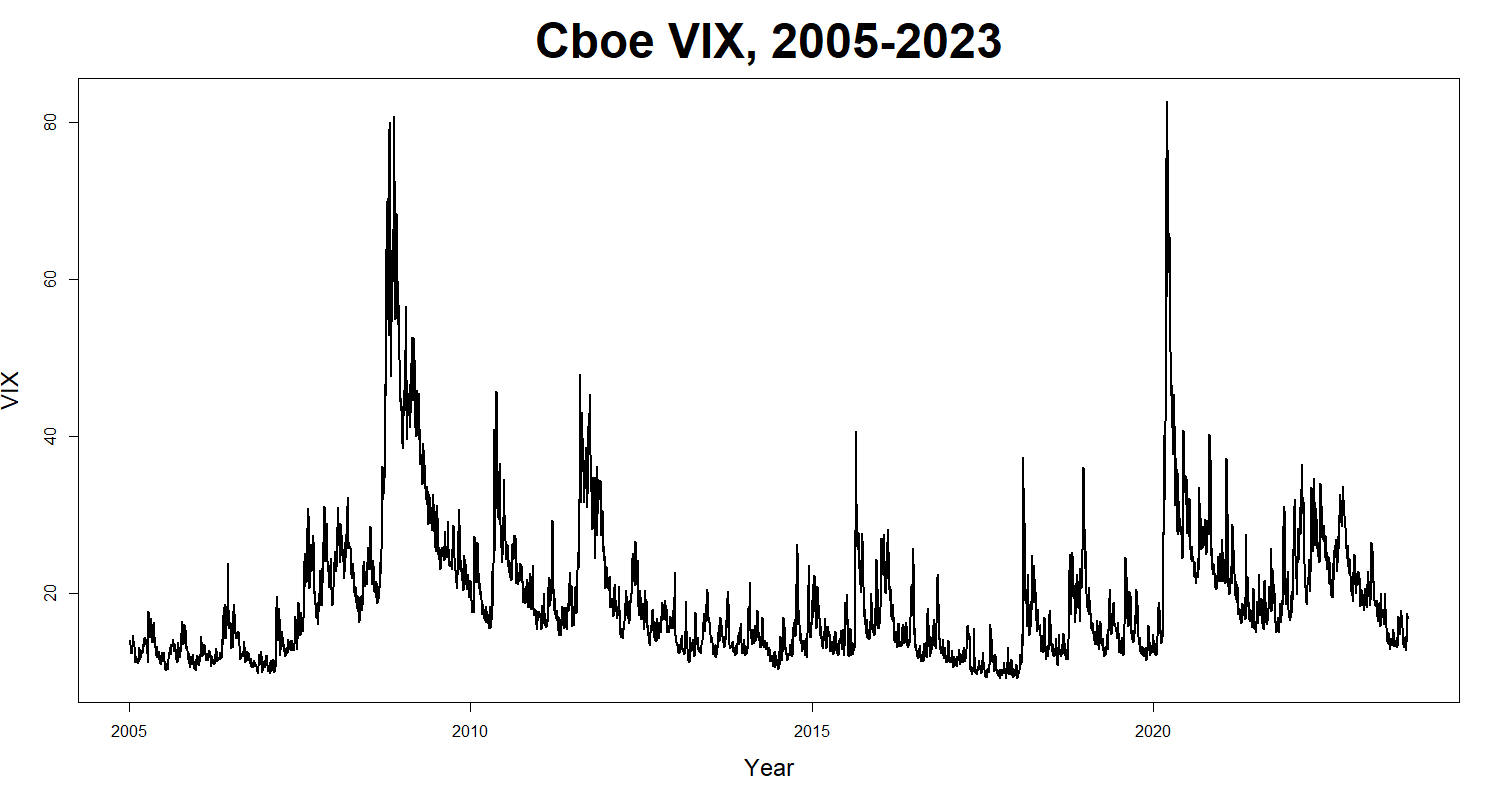}
    \caption{Daily values of the CBOE VIX index, 2005-2023. Note the two spikes in 2008 and 2020 which correspond to the respective economic crises. The data is retrieved from CBOE.com \cite{VIX_hist}.}
    \label{fig: VIX}
\end{figure}

As previously mentioned in the Introduction, volatility is an important tool for quantifying risk, in addition to being a critical variable in option pricing. Therefore, it is no surprise that market actors have been seeking dedicated financial instruments to hedge against drastic volatility changes and capitalize on overall volatility trends. Nowadays, derivatives written on various volatility proxies are extremely popular among investors: for instance, according to the Cboe Annual Report 2022 \cite{CBOE_2022_report}, the \textit{CBOE Volatility Index} (\textit{VIX}, see Fig.~\ref{fig: VIX}) is one of the most traded underlying assets of the Chicago Board Options Exchange, alongside the S\&P 500 (SPX) index.

Since volatility is not a directly observable value and, in some sense, it comes from the theoretical domain, its quantification is a necessary step before considering its use as an underlying asset. Naturally, various methodologies have been applied for such quantification; a detailed historical overview of this subject can be found in e.g. \cite[Section 3]{Carr_Lee_2009}. Here we mention the early approaches of Gastineau (1977) \cite{Gastineau_1977} and Cox \& Rubinstein (1985) \cite[Appendix 8A]{Cox_Rubinstein_1985}, who proposed indices based on the implied volatility, as well as Brenner \& Galai (1989) \cite{Brenner_Galai_1989}, who suggested a metric coming from the realized volatility and considered derivatives employing it as the underlying assets. Another index utilizing an average of S\&P100 option implied volatilities was introduced by Fleming et. al. \cite{Fleming_Ostdiek_Whaley_1995}, and this methodology was used by the Chicago Board Options Exchange for computation of their early version of VIX between 1993 and 2003. The general idea behind the modern computation of VIX can be traced back to Breeden \& Litzenberger (1978) \cite{Breeden_Litzenberger_1978}, but the exact formulas were crystalized in the works of Dupire \cite{Dupire_1993} and Carr \& Madan \cite{Carr_Madan_1998}.

Nowadays, VIX is the preeminent volatility proxy on a global scale, closely connected to a volatility modeling challenge occasionally referred to as ``\textit{The Holy Grail of volatility modeling}'' \cite{Gatheral_Jusselin_Rosenbaum_2020, Guyon_2021}. In order to outline this problem in more detail and prevent any ambiguity, let us first dedicate some time and delve into the specifics of VIX computation and interpretation.

\subsection{Intermezzo: VIX and its interpretation} 

Assume that the SPX forward price $S$ is a continuous martingale satisfying
\begin{equation}\label{eq: SPX model for VIX, equation}
    dS(t) = \sigma(t) S(t) dW(t),
\end{equation}
on a filtered probability space $(\Omega, \mathcal F, \{\mathcal F_t\}, \mathbb Q)$ under the risk-neutral probability measure $\mathbb Q$, where $W$ is a $\mathbb Q$-Brownian motion and $\sigma = \{\sigma(t),~t\ge 0\}$ is an adapted square integrable stochastic process. It is easy to verify that, under some mild assumptions on $\sigma$, \eqref{eq: SPX model for VIX, equation} has a unique solution of the form
\[
    S(t) = S(0) \exp\left\{ -\frac{1}{2}\int_0^t \sigma^2(s)ds + \int_0^t \sigma(s)dW(s) \right\}.
\]
Next, fix a period of time $T$ (in the case of CBOE VIX, $T = 30$ days) and consider the value
\begin{equation}\label{eq: theoretical VIX}
    -\frac{2}{T} \mathbb E_{\mathbb Q} \left[ \log \frac{S({t+T})}{S(t)}~\Big|~\mathcal F_t\right].
\end{equation}
One the one hand, since, by the martingale property,
\[
    \mathbb E_{\mathbb Q} \left[ \int_t^{t+T} \sigma(s) dW(s)~\Big|~\mathcal F_t\right] = 0,
\]
one can observe that
\begin{equation}\label{eq: VIX LHS}
\begin{aligned}
    -\frac{2}{T} \mathbb E_{\mathbb Q} \left[ \log \frac{S({t+T})}{S(t)}~\Big|~\mathcal F_t\right] & = -\frac{2}{T} \mathbb E_{\mathbb Q} \left[ -\frac{1}{2}\int_t^{t+T} \sigma^2(s) ds + \int_t^{t+T} \sigma(s) dW(s)~\Big|~\mathcal F_t\right]
    \\
    & = \frac{1}{T} \int_t^{t+T} \mathbb E_{\mathbb Q}\left[\sigma^2(s)~\big|~\mathcal F_t\right] ds.
\end{aligned}    
\end{equation}
On the other hand, by Taylor's formula with integral remainder,
\begin{align*}
    \log \frac{S(t+T)}{S(t)} & = \frac{S(t+T) - S(t)}{S(t)} - \int_{S(t)}^{S(t+T)} \frac{(S(t+T) - K)}{K^2}dK
    \\
    & = \frac{S(t+T) - S(t)}{S(t)} - \int_{0}^{S(t)} \frac{(K-S(t+T))_+}{K^2}dK -  \int_{S(t)}^{\infty} \frac{(S(t+T) - K)_+}{K^2}dK,
\end{align*}
and hence, since
\[
    \mathbb E_{\mathbb Q} \left[ \frac{S(t+T) - S(t)}{S(t)} ~\bigg|~\mathcal F_t\right] = 0
\]
by the martingale property, \eqref{eq: theoretical VIX} can be re-written as
\begin{equation}\label{eq: VIX RHS}
\begin{aligned}
    -\frac{2 }{T} \mathbb E_{\mathbb Q} \left[\log \frac{S({t+T})}{S(t)}~\Big|~\mathcal F_t\right]  &=\mathbb E_{\mathbb Q} \left[ \frac{2 }{T} \int_{0}^{S(t)} \frac{(K-S(t+T))_+}{K^2}dK~\Big|~\mathcal F_t\right] 
    \\
    &\qquad +  \mathbb E_{\mathbb Q} \left[\int_{S(t)}^{\infty} \frac{(S(t+T) - K)_+}{K^2}dK~\Big|~\mathcal F_t\right]
    \\
    & = \frac{2 }{T} \int_{0}^{S(t)} \frac{\mathbb E_{\mathbb Q} \left[ (K-S(t+T))_+~|~\mathcal F_t\right]}{K^2}dK 
    \\
    &\qquad +  \int_{S(t)}^{\infty} \frac{\mathbb E_{\mathbb Q} \left[(S(t+T) - K)_+~|~\mathcal F_t\right]}{K^2}dK
    \\
    & = \frac{2e^{rT}}{T} \left( \int_{0}^{S(t)} \frac{P_t(K,T)}{K^2}dK +  \int_{S(t)}^{\infty} \frac{C_t(K,T)}{K^2}dK \right),
\end{aligned}    
\end{equation}
where $r$ is the instantaneous interest rate and $P_t(K,T)$, $C_t(K,T)$ are, respectively, the prices (at moment $t$) of put and call options with strike $K$ and expiry date $t+T$. To summarize, under the setting specified above, \eqref{eq: VIX LHS} and \eqref{eq: VIX RHS} yield
\begin{equation}\label{eq: VIX LHS and RHS}
    {\frac{1}{T} \int_t^{t+T} \mathbb E_{\mathbb Q}\left[\sigma^2(s)~\big|~\mathcal F_t\right] ds} = {\frac{2e^{rT}}{T} \left( \int_{0}^{S(t)} \frac{P_t(K,T)}{K^2}dK +  \int_{S(t)}^{\infty} \frac{C_t(K,T)}{K^2}dK \right)}.
\end{equation}

VIX is computed (see e.g. \cite{Cboe} or \cite[Section 1.5.2]{Kwok_Zheng_2022}) by discretizing the right-hand side of \eqref{eq: VIX LHS and RHS} (note that the latter relies exclusively on the empirically observable values, SPX forward $S(t)$ and put/call option prices):
\[
    {\text{VIX}}^2_t(T) := \frac{2e^{rT}}{T} \left( \sum_{j=-M}^{-1}\frac{P_t(K_j, T)}{K_j^2}(K_{j+1} - K_{j}) + \sum_{i=1}^{N}\frac{C_t(K_i, T)}{K_i^2}(K_{i} - K_{i-1}) \right) - \frac{1}{T}\left( \frac{S({t})}{K_0} - 1 \right)^2,
\]
where $K_{-M} < K_{-(M-1)} < ... < K_0$ are payoffs of all listed out-of-the-money put options with maturity at $t+T$, $K_0 \le S(t)$, and $K_{1} < K_{2} < ... < K_N$ are payoffs of all listed out-of-the-money call options with maturity at $t+T$, $K_1 > S(t)$. In the meanwhile, the left-hand side of \eqref{eq: VIX LHS and RHS} is a standard notion of the forward variance, precisely the value one intends to capture. 

\subsection{VIX smile and the joint calibration puzzle} 

Despite any potential inaccuracies, VIX stands as the most popular proxy for market volatility in the world and, therefore, it is unsurprising that derivatives with VIX as an underlying are actively traded on the market. In particular, in March 2004, CBOE launched VIX futures and, in February 2006, VIX options were introduced -- and, naturally, a problem of pricing such derivatives emerged.

First of all, having VIX option prices, one can use the procedure described in Subsection \ref{subsec: IV and smile} and construct the corresponding implied volatility surface $\widehat\sigma(T,K)$, with $T$ denoting the time to maturity and $K$ being the strike. Contrary to the convex ``smiles'' for derivatives written on classical stocks and stock indices, the VIX implied volatility $\widehat\sigma(T,K)$ consistently exhibits concave behavior in $K$ for every fixed $T$ with positive slope at the money (see \cite[Figure 1]{Alòs_García-Lorite_Gonzalez_2022} as well as \cite{Baldeaux_Badran_2014, Fouque_Saporito_2018} or \cite{Kokholm_Stisen_2015}). 

It turns out that continuous stochastic volatility models have mixed success in reproducing this phenomenon, especially for shorter maturities -- in this regard, we recommend an excellent review \cite{Rømer_2022} on the topic. Alòs, García-Lorite \& Gonzalez \cite{Alòs_García-Lorite_Gonzalez_2022} (see also \cite[Chapter 10]{Alos_Garcia_Lorite_2021}) use Malliavin calculus tools and analyze selected models to check whether they produce the positive slope at the money for short maturities. Their results show that
\begin{itemize}
    \item  SABR model \eqref{eq: SABR} gives the flat skew, but its modification called the \textit{mixed SABR model}
    \[
        \sigma(t) = \theta_0 \left(\delta \exp\left\{2\theta_1 W(t) - \theta_1^2 t\right\} + (1-\delta)\exp\left\{2\theta_2 W(t) - \theta_2^2 t\right\} \right), 
    \]
    where $\delta\in[0,1]$, $\theta_0$, $\theta_1$, $\theta_2>0$, generates the required positive slope provided that $\delta\ne 0,1$ and $\theta_1\ne\theta_2$;

    \item Heston model \eqref{eq: Heston model} normally generates negative VIX skew (see also \cite[Figure 3]{Goutte_Ismail_Pham_2017}); however, under some conditions on coefficients (e.g. when Feller condition is violated, see \cite{Kokholm_Stisen_2015}), the slope can be positive;

    \item the rough Bergomi model \eqref{eq: RB} as well as its mixed modification \eqref{eq: MRB} both give the required slope (see also \cite{Jacquier_Martini_Muguruza_2018}).
\end{itemize}
Other notable continuous stochastic volatility that are able to capture the upward smirk include:
\begin{itemize}
    \item the 3/2-model \eqref{eq: 3/2 model} \cite[Section 2]{Baldeaux_Badran_2014};
    
    \item double CEV-model \eqref{eq: double CEV} \cite{Gatheral_2006};

    \item Heston with stochastic vol-of-vol model \eqref{eq: Heston vol-of-vol} \cite{Fouque_Saporito_2018};

    \item quintic Ornstein-Uhlenbeck model \cite{Jaber_Illand_Shaun_Li_2022}
    \begin{equation}\label{eq: quintic OU}
    \begin{aligned}
        \sigma(t) &= \sqrt{\theta_0(t)} \frac{p(X(t))}{\sqrt{\mathbb E\left[ p^2(X(t)) \right]}},
        \\
        X(t) &= \varepsilon^{-\theta} \int_0^t e^{-\frac{\theta}{\varepsilon}(t-s)}dB(s),
    \end{aligned}
    \end{equation}
    and, more generally, a class of Gaussian polynomial models \cite{Jaber_Illand_Shaun_Li_2022b}
    \begin{equation}\label{eq: Gaussian polynomial}
    \begin{aligned}
        \sigma(t) &= \sqrt{\theta_0(t)} \frac{p(X(t))}{\sqrt{\mathbb E\left[ p^2(X(t)) \right]}},
        \\
        X(t) &= \int_0^t \mathcal K(t-s)dB(s),
    \end{aligned}
    \end{equation}
    where $\theta>0$, $p$ is a polynomial and $\mathcal K$ is a square integrable Volterra kernel;

    \item quadratic rough Heston model \eqref{eq: quadratic rough Heston}.
\end{itemize}

As one can see, multiple continuous models, both standard Brownian diffusions and rough volatility models, are able to reproduce the shape of the VIX smile. However, there is an additional dimension to that problem: since there is a direct connection between SPX and VIX, models calibrated with respect to the SPX data should be consistent with VIX and vice versa. In other words, one seeks a market model that would \textit{jointly calibrate to both VIX and SPX simultaneously}. As noted by Guyon in \cite{Guyon_2021}, ``\textit{Without such models, financial institutions could possibly arbitrage each other, and even market making desks within the same institution could do so...}''.

It turns out that this problem, known as the ``\textit{joint calibration puzzle}'', is extremely difficult to solve; so much so that some researchers label it ``\textit{the Holy Grail of volatility modeling}'' \cite{Gatheral_Jusselin_Rosenbaum_2020, Guyon_2021}. For a long time, continuous-time stochastic volatility models without jumps failed to produce the perfect SPX-VIX joint calibration. Some notable attempts include the 3/2 model \cite{Baldeaux_Badran_2014}, double CEV-model \cite{Gatheral_2006}, Heston with stochastic vol-of-vol model \cite{Fouque_Saporito_2018} and
rough Berghomi model \cite{Jacquier_Martini_Muguruza_2018} -- in all cases, the joint fit was only partially successful, predominantly failing for short maturities (see also an excellent presentation by J. Guyon \cite{Guyon_RBM} on the topic). The most fruitful approaches in solving the joint calibration puzzle were outside of the continuous-time-continuous-price paradigm:
\begin{itemize}
    \item Cont \& Kokholm \cite{Cont_Kokholm_2013} use jump-diffusions to directly model forward variances and SPX; jumps indeed allow to de-couple the ATM S\&P500 skew and the ATM VIX implied volatility giving a good joint fit;

    \item Guyon \cite{Guyon_2019, Guyon_2021} utilizes nonparametric discrete time model building joint probability measure on SPX and VIX on a discrete set of dates; this allows to obtain an almost perfect joint calibration.
\end{itemize}

However, in recent years there have been several successful attempts inside the fully continuous framework, all with a higher number of parameters:
\begin{itemize}
    \item Gatheral, Jusselin \& Rosenbaum \cite{Gatheral_Jusselin_Rosenbaum_2020} and Rosenbaum \& Zhang \cite{Rosenbaum_Zhang_2021} show that quadratic rough Heston model \eqref{eq: quadratic rough Heston} calibrates very well to SPX and VIX smiles;

    \item Abi Jaber, Illand \& Li \cite{Jaber_Illand_Shaun_Li_2022b, Jaber_Illand_Shaun_Li_2022} provide a good fit with Gaussian polynomial models \eqref{eq: Gaussian polynomial};

    \item Bourgey \& Guyon \cite[Section 4]{Guyon_Bourgey_2022} extend the discrete time model of \cite{Guyon_2019, Guyon_2021} to a continuous-time setting;

    \item Guyon \& Mustapha \cite{Guyon_Mustapha_2022} achieve an impressive calibration with the model
    \begin{align*}
        dS_t &= -\frac{1}{2} \sigma^2_S(t, S_t, Y_t)dt + \sigma_S(t, S_t, Y_t)dW_t,
        \\
        dY_t &= \mu_Y(t, S_t, Y_t)dt + \sigma_Y(t, S_t, Y_t)\left( \rho(t, S_t, Y_t) dW_t + \sqrt{1 - \rho^2(t, S_t, Y_t)}dB_t \right),
    \end{align*}
    where the drift, volatility and correlation functions $\sigma_S$, $\mu_Y$, $\sigma_Y$ and $\rho$ are modeled with neural networks.
\end{itemize}

\subsection{VIX and continuous models: interpretation issues}

We finalize this section by highlighting a significant caveat related to the interpretation and tractability of VIX within the context of continuous stochastic volatility models. 

Despite the undoubted elegance of VIX computation methodology, the precision of VIX per se in capturing the volatility of SPX heavily depends on the underlying assumptions. Namely, it is crucial that SPX is assumed to follow the continuous dynamics \eqref{eq: SPX model for VIX, equation} and any deviations from this model lead to some deterioration of the VIX tractability. In particular, if there are jumps in the SPX forward dynamics, \eqref{eq: VIX LHS} does not hold with an additional term appearing on the right-hand side of the expression (see e.g. \cite{Ait-Sahalia_Karaman_Mancini_2020}). Therefore, if \eqref{eq: SPX model for VIX, equation} is violated, \eqref{eq: VIX LHS and RHS} does not hold either, and VIX loses its original interpretation. 

It is important to note that the viability of the model \eqref{eq: SPX model for VIX, equation} can be tested statistically by comparing VIX to \textit{variance swap contracts}. By definition, a \textit{variance swap} is a futures contract with a payoff at the moment $T$ of the form
\[
    \frac{1}{T}\sum_{k=1}^n \left(\log \frac{S(t_k)}{S(t_{k-1})}\right)^2 - VS,
\] 
where $0 = t_0 < t_1 < \cdots < t_n = T$ is a partition of $[0,T]$ and $VS$ denotes the corresponding swap rate that is determined by the market. Non-arbitrage arguments imply that, if the dynamics \eqref{eq: SPX model for VIX, equation} indeed holds and some mild assumptions on $\sigma$ are satisfied,
\begin{align*}
    VS &= \mathbb E_{\mathbb Q} \left[\frac{1}{T}\sum_{k=1}^n \left(\log \frac{S(t_k)}{S(t_{k-1})}\right)^2 \right] \to \frac{1}{T}\int_0^T E_{\mathbb Q} \left[\sigma^2(s)\right]ds
\end{align*}
as $\max_k|t_k - t_{k-1}| \to 0$. In other words, under the model \eqref{eq: SPX model for VIX, equation}, market swap rates $\sqrt{VS}$ should approximately coincide with VIX. However, as reported by Aït-Sahalia et. al. in their analysis \cite{Ait-Sahalia_Karaman_Mancini_2020} using the over-the-counter swap rate data, there is a statistically significant gap between $VS$ and $\text{VIX}^2$: the values of $VS - \text{VIX}^2$ are mostly positive, larger during market turmoils but sizable even in quiet times. For more details in that regard as well as for possible alternatives to the current VIX computation method, we also refer the reader to the discussions in \cite{Andersen_Bondarenko_Gonzalez-Perez_2015, Carr_Lee_Wu_2012, Martin_2011}. 

In other words, there is some evidence showing that the tractability of VIX as a volatility index is not that straightforward and the real-world observations of VIX are not sampled from
\begin{equation}\label{eq: theoretical VIX under continuous prices}
    \sqrt{\frac{1}{T} \int_t^{t+T} \mathbb E_{\mathbb Q}\left[\sigma^2(s)~\big|~\mathcal F_t\right] ds}
\end{equation}
as models with continuous price dynamics assume\footnote{As suggested in \cite{Martin_2011}, in the presence of jumps, VIX measures \textit{the risk-neutral entropy} of the simple return on the S\&P 500 index rather than \eqref{eq: theoretical VIX under continuous prices}.}. Therefore, fitting stochastic volatility models with the continuous price dynamics to VIX without any adjustments may in principle produce unstable results.

\section{Concluding remarks}\label{sec: conclusion}

In this survey, we outlined -- though with considerably broad brushstrokes -- the existing literature on stochastic volatility. Clearly, the number of approaches is very large and we acknowledge that we are far from describing all of them, even within the considerably narrower framework of continuous models. Nevertheless, we hope that the reader received a general understanding of the motivation behind this modeling paradigm. 

We conclude our presentation by saying that updating, refining and improving models still seems an endless race. Right now, new techniques and modeling paradigms are being developed -- in particular, we mention \cite{Guyon_Lekeufack_2023} which finds strong arguments for fully path-dependent volatility models as well as machine learning and signature methods such as in \cite{Buehler_Gonon_Teichmann_Wood_2019, Bühler_Horvath_Lyons_Arribas_Wood_2020} or \cite{Cuchiero_Gazzani_Möller_Svaluto-Ferro_2023, Cuchiero_Gazzani_Svaluto-Ferro_2023} (the model in the latter actually includes the Gaussian polynomial models \eqref{eq: Gaussian polynomial} as a special case). Perhaps these novel tools will become the new classics and open new frontiers in understanding the financial market.

\bibliographystyle{acm}
\bibliography{biblio.bib}

\begin{thebibliography}{100}

\bibitem{Rough_volatility_literature}
Rough volatility literature.
\newblock
  \url{https://sites.google.com/site/roughvol/home/rough-volatility-literature?authuser=0}.
\newblock Accessed: 2023-08-30.

\bibitem{VIX_hist}
{VIX} index historical data.
\newblock
  \url{https://www.cboe.com/tradable_products/vix/vix_historical_data/}.
\newblock Accessed: 2023-09-26.

\bibitem{AbiJaber_ElEuch_2019}
{\sc Abi~Jaber, E., and El~Euch, O.}
\newblock Multifactor approximation of rough volatility models.
\newblock {\em SIAM journal on financial mathematics 10}, 2 (2019), 309–349.

\bibitem{Jaber_Illand_Shaun_Li_2022b}
{\sc Abi~Jaber, E., Illand, C., and Li, S.}
\newblock Joint {SPX}-{VIX} calibration with {G}aussian polynomial volatility
  models: deep pricing with quantization hints.
\newblock {\em ArXiv:2212.08297\/} (2022).

\bibitem{Jaber_Illand_Shaun_Li_2022}
{\sc Abi~Jaber, E., Illand, C., and Li, S.}
\newblock The quintic {O}rnstein-{U}hlenbeck volatility model that jointly
  calibrates {SPX} \& {VIX} smiles.
\newblock {\em ArXiv:2212.10917\/} (2022).

\bibitem{Alos_Garcia_Lorite_2021}
{\sc Alòs, E., and Garcia~Lorite, D.}
\newblock {\em Malliavin calculus in finance: Theory and practice}.
\newblock CRC Press, London, England, 2021.

\bibitem{Alòs_García-Lorite_Gonzalez_2022}
{\sc Alòs, E., García-Lorite, D., and Gonzalez, A.~M.}
\newblock On smile properties of volatility derivatives: {U}nderstanding the
  {VIX} skew.
\newblock {\em SIAM journal on financial mathematics 13}, 1 (2022), 32–69.

\bibitem{Alos_Leon_Vives_2007}
{\sc Alòs, E., León, J.~A., and Vives, J.}
\newblock On the short-time behavior of the implied volatility for
  jump-diffusion models with stochastic volatility.
\newblock {\em Finance and stochastics 11}, 4 (2007), 571–589.

\bibitem{Alos_Rolloos_Shiraya_2022}
{\sc Alòs, E., Rolloos, F., and Shiraya, K.}
\newblock Forward start volatility swaps in rough volatility models.
\newblock {\em ArXiv:2207.10370\/} (2022).

\bibitem{Andersen_Piterbarg_2006}
{\sc Andersen, L. B.~G., and Piterbarg, V.~V.}
\newblock Moment explosions in stochastic volatility models.
\newblock {\em Finance and stochastics 11}, 1 (2006), 29–50.

\bibitem{Andersen_Bondarenko_Gonzalez-Perez_2015}
{\sc Andersen, T.~G., Bondarenko, O., and Gonzalez-Perez, M.~T.}
\newblock Exploring return dynamics via corridor implied volatility.
\newblock {\em The review of financial studies 28}, 10 (2015), 2902–2945.

\bibitem{Ayache_Peng_2012}
{\sc Ayache, A., and Peng, Q.}
\newblock Stochastic volatility and multifractional {B}rownian motion.
\newblock In {\em Stochastic Differential Equations and Processes\/} (2012),
  Springer Berlin Heidelberg, p.~211–237.

\bibitem{Ait-Sahalia_Fan_Li_2013}
{\sc Aït-Sahalia, Y., Fan, J., and Li, Y.}
\newblock The leverage effect puzzle: {D}isentangling sources of bias at high
  frequency.
\newblock {\em Journal of financial economics 109}, 1 (2013), 224–249.

\bibitem{AitSahalia_Jacod_2019}
{\sc Aït-Sahalia, Y., and Jacod, J.}
\newblock Testing whether volatility can be written as a function of the asset
  price.
\newblock Presented at Princeton-QUT-SJTU-SMU Conference on Econometrics, 2019.

\bibitem{Ait-Sahalia_Karaman_Mancini_2020}
{\sc Aït-Sahalia, Y., Karaman, M., and Mancini, L.}
\newblock The term structure of equity and variance risk premia.
\newblock {\em Journal of Econometrics 219}, 2 (2020), 204–230.

\bibitem{Bachelier_1900}
{\sc Bachelier, L.}
\newblock Théorie de la spéculation.
\newblock {\em Annales Scientifiques de l’Ecole Normale Superieure. Quatrieme
  Serie 17\/} (1900), 21–86.

\bibitem{Bachelier_1900_eng}
{\sc Bachelier, L., Davis, M., and Etheridge, A.}
\newblock {\em Louis {B}achelier’s theory of speculation: The origins of
  modern finance. Translated and with commentary by Mark Davis and Alison
  Etheridge}.
\newblock Princeton University Press, Princeton, NJ, 2006.

\bibitem{Bakshi_Ju_Ou-Yang_2006}
{\sc Bakshi, G., Ju, N., and Ou-Yang, H.}
\newblock Estimation of continuous-time models with an application to equity
  volatility dynamics.
\newblock {\em Journal of financial economics 82}, 1 (2006), 227–249.

\bibitem{Baldeaux_Badran_2014}
{\sc Baldeaux, J., and Badran, A.}
\newblock Consistent modelling of {VIX} and equity derivatives using a 3/2 plus
  jumps model.
\newblock {\em Applied Mathematical Finance 21}, 4 (2014), 299--312.

\bibitem{Barndorff-Nielsen_1997}
{\sc Barndorff-Nielsen, O.~E.}
\newblock Normal inverse {G}aussian distributions and stochastic volatility
  modelling.
\newblock {\em Scandinavian journal of statistics, theory and applications 24},
  1 (1997), 1–13.

\bibitem{Barndorff-Nielsen_Shephard_2001}
{\sc Barndorff-Nielsen, O.~E., and Shephard, N.}
\newblock Non-{G}aussian {O}rnstein-{U}hlenbeck-based models and some of their
  uses in financial economics.
\newblock {\em Journal of the Royal Statistical Society. Series B, Statistical
  methodology 63}, 2 (2001), 167–241.

\bibitem{Barndorff-Nielsen_Shephard_2011}
{\sc Barndorff-Nielsen, O.~E., and Shephard, N.}
\newblock Lévy driven volatility models.
\newblock \url{https://pure.au.dk/ws/files/193874249/levybook.pdf}, 2011.
\newblock Accessed: 2023-08-30.

\bibitem{Bayer_Friz_Gatheral_2016}
{\sc Bayer, C., Friz, P., and Gatheral, J.}
\newblock Pricing under rough volatility.
\newblock {\em Quantitative finance 16}, 6 (2016), 887–904.

\bibitem{Benhamou_Gobet_Miri_2010}
{\sc Benhamou, E., Gobet, E., and Miri, M.}
\newblock Time dependent {H}eston model.
\newblock {\em SIAM journal on financial mathematics 1}, 1 (2010), 289–325.

\bibitem{Bennedsen_Lunde_Pakkanen_2022}
{\sc Bennedsen, M., Lunde, A., and Pakkanen, M.~S.}
\newblock Decoupling the short- and long-term behavior of stochastic
  volatility.
\newblock {\em Journal of financial econometrics 20}, 5 (2022), 961–1006.

\bibitem{Beran}
{\sc Beran, J.}
\newblock {\em Statistics for Long-Memory Processes}.
\newblock Chapman and Hall/CRC, 1994.

\bibitem{Bergomi_2016}
{\sc Bergomi, L.}
\newblock {\em Stochastic volatility modeling}.
\newblock Chapman \& Hall/CRC, Oakville, MO, 2016.

\bibitem{BMdP2018}
{\sc Bezborodov, V., Persio, L.~D., and Mishura, Y.}
\newblock Option pricing with fractional stochastic volatility and
  discontinuous payoff function of polynomial growth.
\newblock {\em Methodology and Computing in Applied Probability 21}, 1 (Aug.
  2018), 331--366.

\bibitem{BGP2000}
{\sc Biagini, F., Guasoni, P., and Pratelli, M.}
\newblock Mean-variance hedging for stochastic volatility models.
\newblock {\em Mathematical Finance 10}, 2 (Apr. 2000), 109--123.

\bibitem{Black_1976}
{\sc Black, F.}
\newblock Studies of stock price volatility changes.
\newblock In {\em Proceedings of the Business and Economics Section of the
  American Statistical Association\/} (1976), p.~177–181.

\bibitem{BS_1987}
{\sc Black, F.}
\newblock Citation classic – the pricing of options and corporate
  liabilities.
\newblock {\em Citation Classic}, 33 (1987), 16.

\bibitem{BS_1973}
{\sc Black, F., and Scholes, M.}
\newblock The pricing of options and corporate liabilities.
\newblock {\em Journal of Political Economy 81}, 3 (1973), 637--654.

\bibitem{Bolko_Christensen_Pakkanen_Veliyev_2023}
{\sc Bolko, A.~E., Christensen, K., Pakkanen, M.~S., and Veliyev, B.}
\newblock A {GMM} approach to estimate the roughness of stochastic volatility.
\newblock {\em Journal of Econometrics 235}, 2 (2023), 745–778.

\bibitem{Bollerslev_1986}
{\sc Bollerslev, T.}
\newblock Generalized autoregressive conditional heteroskedasticity.
\newblock {\em Journal of Econometrics 31}, 3 (1986), 307–327.

\bibitem{Bollerslev_Chou_Kroner_1992}
{\sc Bollerslev, T., Chou, R.~Y., and Kroner, K.~F.}
\newblock {ARCH} modeling in finance.
\newblock {\em Journal of Econometrics 52}, 1–2 (1992), 5–59.

\bibitem{Bollerslev_Mikkelsen_1996}
{\sc Bollerslev, T., and Ole~Mikkelsen, H.}
\newblock Modeling and pricing long memory in stock market volatility.
\newblock {\em Journal of Econometrics 73}, 1 (1996), 151–184.

\bibitem{Boness_1964}
{\sc Boness, A.~J.}
\newblock Elements of a theory of stock-option value.
\newblock {\em Journal of Political Economy 72}, 2 (1964), 163--175.

\bibitem{Bouchaud_Potters_2000}
{\sc Bouchaud, J.-P., and Potters, M.}
\newblock {\em Theory of financial risks: From statistical physics to risk
  management}.
\newblock Cambridge University Press, Cambridge, England, 2000.

\bibitem{Breeden_Litzenberger_1978}
{\sc Breeden, D.~T., and Litzenberger, R.~H.}
\newblock Prices of state-contingent claims implicit in option prices.
\newblock {\em The journal of business 51}, 4 (1978), 621–651.

\bibitem{Breidt_Crato_Lima_1998}
{\sc Breidt, F.~J., Crato, N., and {de Lima}, P.}
\newblock The detection and estimation of long memory in stochastic volatility.
\newblock {\em Journal of Econometrics 83}, 1 (1998), 325--348.

\bibitem{Brenner_Galai_1989}
{\sc Brenner, M., and Galai, D.}
\newblock New financial instruments for hedging changes in volatility.
\newblock {\em Financial analysts journal 45}, 4 (1989), 61–65.

\bibitem{Buehler_Gonon_Teichmann_Wood_2019}
{\sc Buehler, H., Gonon, L., Teichmann, J., and Wood, B.}
\newblock Deep hedging.
\newblock {\em Quantitative finance 19}, 8 (2019), 1271–1291.

\bibitem{Buraschi_Jackwerth_2001}
{\sc Buraschi, A., and Jackwerth, J.}
\newblock The price of a smile: Hedging and spanning in option markets.
\newblock {\em The review of financial studies 14}, 2 (2001), 495–527.

\bibitem{Bauerle_Desmettre_2020}
{\sc Bäuerle, N., and Desmettre, S.}
\newblock Portfolio optimization in fractional and rough {H}eston models.
\newblock {\em SIAM journal on financial mathematics 11}, 1 (2020), 240–273.

\bibitem{Bühler_Horvath_Lyons_Arribas_Wood_2020}
{\sc Bühler, H., Horvath, B., Lyons, T., Arribas, I.~P., and Wood, B.}
\newblock A data-driven market simulator for small data environments.
\newblock {\em ArXiv:2006.14498\/} (2020).

\bibitem{Carr_Lee_2009}
{\sc Carr, P., and Lee, R.}
\newblock Volatility derivatives.
\newblock {\em Annual review of financial economics 1}, 1 (2009), 319–339.

\bibitem{Carr_Lee_Wu_2012}
{\sc Carr, P., Lee, R., and Wu, L.}
\newblock Variance swaps on time-changed lévy processes.
\newblock {\em Finance and stochastics 16}, 2 (2012), 335–355.

\bibitem{Carr_Madan_1998}
{\sc Carr, P., and Madan, D.}
\newblock Towards a theory of volatility trading.
\newblock In {\em Volatility}, R.~Jarrow, Ed. Risk Publications, 1998,
  pp.~417--427.

\bibitem{Carr_Sun_2007}
{\sc Carr, P., and Sun, J.}
\newblock A new approach for option pricing under stochastic volatility.
\newblock {\em Review of derivatives research 10}, 2 (2007), 87–150.

\bibitem{Carr_Wu_2003}
{\sc Carr, P., and Wu, L.}
\newblock The finite moment log stable process and option pricing.
\newblock {\em The journal of finance 58}, 2 (2003), 753–777.

\bibitem{Catalini_Pacchiarotti_2022}
{\sc Catalini, G., and Pacchiarotti, B.}
\newblock Asymptotics for multifactor {V}olterra type stochastic volatility
  models.
\newblock {\em Stochastic analysis and applications\/} (2022), 1–31.

\bibitem{CBOE_2022_report}
{\sc CBOE}.
\newblock Cboe {G}lobal {M}arkets {A}nnual {R}eport 2022, 2023.

\bibitem{Cboe}
{\sc CBOE}.
\newblock Volatility index methodology: Cboe volatility index.
\newblock
  \url{https://cdn.cboe.com/api/global/us_indices/governance/Volatility_Index_Methodology_Cboe_Volatility_Index.pdf},
  2023.
\newblock Accessed: 2023-08-30.

\bibitem{Cherny_Engelbert_2005}
{\sc Cherny, A.~S., and Engelbert, H.-J.}
\newblock {\em Singular Stochastic Differential Equations}.
\newblock Springer Berlin Heidelberg, Berlin, Heidelberg, 2005.

\bibitem{Cheung_Ng_1992}
{\sc Cheung, Y.-W., and Ng, L.~K.}
\newblock Stock price dynamics and firm size: {A}n empirical investigation.
\newblock {\em The journal of finance 47}, 5 (1992), 1985--1997.

\bibitem{Christie_1982}
{\sc Christie, A.}
\newblock The stochastic behavior of common stock variances: {V}alue, leverage
  and interest rate effects.
\newblock {\em Journal of financial economics 10}, 4 (1982), 407–432.

\bibitem{ChronopoulouViens2012}
{\sc Chronopoulou, A., and Viens, F.~G.}
\newblock Estimation and pricing under long-memory stochastic volatility.
\newblock {\em Annals of finance 8}, 2–3 (2012), 379–403.

\bibitem{Clark_1973}
{\sc Clark, P.~K.}
\newblock A subordinated stochastic process model with finite variance for
  speculative prices.
\newblock {\em Econometrica: journal of the Econometric Society 41}, 1 (1973),
  135.

\bibitem{Comte_Renault_1998}
{\sc Comte, F., and Renault, E.}
\newblock Long memory in continuous-time stochastic volatility models.
\newblock {\em Mathematical Finance 8}, 4 (1998), 291–323.

\bibitem{Cont_2001}
{\sc Cont, R.}
\newblock Empirical properties of asset returns: stylized facts and statistical
  issues.
\newblock {\em Quantitative finance 1}, 2 (2001), 223–236.

\bibitem{Cont_2005}
{\sc Cont, R.}
\newblock Long range dependence in financial markets.
\newblock In {\em Fractals in Engineering\/} (London, 2005), Springer-Verlag,
  p.~159–179.

\bibitem{Cont_2006}
{\sc Cont, R.}
\newblock Volatility clustering in financial markets: {E}mpirical facts and
  agent-based models.
\newblock In {\em Long Memory in Economics\/} (2006), Springer Berlin
  Heidelberg, p.~289–309.

\bibitem{Cont_Das_2022}
{\sc Cont, R., and Das, P.}
\newblock Rough volatility: fact or artefact?
\newblock {\em ArXiv:2203.13820\/} (2022).

\bibitem{Cont_Kokholm_2013}
{\sc Cont, R., and Kokholm, T.}
\newblock A consistent pricing model for index options and volatility
  derivatives.
\newblock {\em Mathematical Finance 23}, 2 (2013), 248–274.

\bibitem{Cont_Potters_Bouchaud_1997}
{\sc Cont, R., Potters, M., and Bouchaud, J.-P.}
\newblock Scaling in stock market data: Stable laws and beyond.
\newblock {\em SSRN Electronic Journal\/} (1997).

\bibitem{Tankov_Cont_2003}
{\sc Cont, R., and Tankov, P.}
\newblock {\em Financial modelling with jump processes}.
\newblock Chapman \& Hall/CRC, Philadelphia, PA, 2003.

\bibitem{Cootner_1967}
{\sc Cootner, P.}
\newblock {\em Random character of stock market prices}.
\newblock MIT Press, London, England, 1967.

\bibitem{CLV_2014}
{\sc Corlay, S., Lebovits, J., and Lévy~Véhel, J.}
\newblock Multifractional stochastic volatility models.
\newblock {\em Mathematical Finance 24}, 2 (2014), 364–402.

\bibitem{Cox_1997}
{\sc Cox, J.~C.}
\newblock Notes on option pricing {I}: the constant elasticity of variance
  option pricing model.
\newblock {\em Reprinted in The Journal of Portfolio Management 23}, 5 (1997),
  15–17.

\bibitem{Cox_Ingersoll_Ross_1985}
{\sc Cox, J.~C., Ingersoll, J.~E., and Ross, S.~A.}
\newblock A theory of the term structure of interest rates.
\newblock {\em Econometrica: {J}ournal of the Econometric Society 53}, 2
  (1985), 385--407.

\bibitem{Cox_Ross_1976}
{\sc Cox, J.~C., and Ross, S.~A.}
\newblock The valuation of options for alternative stochastic processes.
\newblock {\em Journal of financial economics 3}, 1–2 (1976), 145–166.

\bibitem{Cox_Rubinstein_1985}
{\sc Cox, J.~C., and Rubinstein, M.}
\newblock {\em Options Markets}.
\newblock Pearson, Upper Saddle River, NJ, 1985.

\bibitem{Cuchiero_Gazzani_Möller_Svaluto-Ferro_2023}
{\sc Cuchiero, C., Gazzani, G., Möller, J., and Svaluto-Ferro, S.}
\newblock Joint calibration to {SPX} and {VIX} options with signature-based
  models.
\newblock {\em ArXiv:2301.13235\/} (2023).

\bibitem{Cuchiero_Gazzani_Svaluto-Ferro_2023}
{\sc Cuchiero, C., Gazzani, G., and Svaluto-Ferro, S.}
\newblock Signature-based models: {T}heory and calibration.
\newblock {\em SIAM journal on financial mathematics 14}, 3 (2023), 910–957.

\bibitem{Cutler_Poterba_Summers_1989}
{\sc Cutler, D., Poterba, J., and Summers, L.}
\newblock What moves stock prices?
\newblock {\em Journal of Portfolio Management\/} (1989), 4–12.

\bibitem{Dandapani_Jusselin_Rosenbaum_2021}
{\sc Dandapani, A., Jusselin, P., and Rosenbaum, M.}
\newblock From quadratic {H}awkes processes to super-{H}eston rough volatility
  models with {Z}umbach effect.
\newblock {\em Quantitative finance 21}, 8 (2021), 1235–1247.

\bibitem{Das_Sundaram_1999}
{\sc Das, S.~R., and Sundaram, R.~K.}
\newblock Of smiles and smirks: {A} term structure perspective.
\newblock {\em Journal of financial and quantitative analysis 34}, 2 (1999),
  211.

\bibitem{Delbaen_Schachermayer_1994}
{\sc Delbaen, F., and Schachermayer, W.}
\newblock A general version of the fundamental theorem of asset pricing.
\newblock {\em Mathematische annalen 300}, 1 (1994), 463–520.

\bibitem{Delbaen_Schachermayer_1998}
{\sc Delbaen, F., and Schachermayer, W.}
\newblock The fundamental theorem of asset pricing for unbounded stochastic
  processes.
\newblock {\em Mathematische annalen 312}, 2 (1998), 215–250.

\bibitem{Delbaen_Schachermayer_2006}
{\sc Delbaen, F., and Schachermayer, W.}
\newblock {\em Mathematics of arbitrage}.
\newblock Springer, New York, NY, 2006.

\bibitem{Delemotte_Marco_Segonne_2023}
{\sc Delemotte, J., De~Marco, S., and Segonne, F.}
\newblock Yet another analysis of the {SP500} at-the-money skew: {C}rossover of
  different power-law behaviours.
\newblock {\em SSRN Electronic Journal\/} (2023).

\bibitem{Derman_Kani_1994}
{\sc Derman, E., and Kani, I.}
\newblock Riding on the smile.
\newblock {\em Risk 7}, 2 (1994), 32–39.

\bibitem{Derman_Miller_Park_2016}
{\sc Derman, E., Miller, M.~B., and Park, D.}
\newblock {\em The volatility smile}.
\newblock John Wiley \& Sons, Nashville, TN, 2016.

\bibitem{DiNunno_Mishura_Yurchenko-Tytarenko_2022}
{\sc Di~Nunno, G., Mishura, Y., and Yurchenko-Tytarenko, A.}
\newblock Option pricing in {V}olterra sandwiched volatility model.
\newblock {\em ArXiv:2209.10688\/} (2022).

\bibitem{DiNunno_Mishura_Yurchenko-Tytarenko_2023b}
{\sc Di~Nunno, G., Mishura, Y., and Yurchenko-Tytarenko, A.}
\newblock Drift-implicit {E}uler scheme for sandwiched processes driven by
  {H}ölder noises.
\newblock {\em Numerical algorithms 93}, 2 (2023), 459–491.

\bibitem{DiNunno_Yurchenko-Tytarenko_2022}
{\sc Di~Nunno, G., and Yurchenko-Tytarenko, A.}
\newblock Sandwiched {V}olterra {V}olatility model: {M}arkovian approximations
  and hedging.
\newblock {\em ArXiv:2209.13054\/} (2022).

\bibitem{Ding_Granger_1996}
{\sc Ding, Z., and Granger, C. W.~J.}
\newblock Modeling volatility persistence of speculative returns: A new
  approach.
\newblock {\em Journal of Econometrics 73}, 1 (1996), 185–215.

\bibitem{Ding_Granger_Engle_1993}
{\sc Ding, Z., Granger, C. W.~J., and Engle, R.~F.}
\newblock A long memory property of stock market returns and a new model.
\newblock {\em Journal of empirical finance 1}, 1 (1993), 83–106.

\bibitem{Duffee_1995}
{\sc Duffee, G.}
\newblock Stock returns and volatility a firm-level analysis.
\newblock {\em Journal of financial economics 37}, 3 (1995), 399–420.

\bibitem{Duffie_Pan_Singleton_2000}
{\sc Duffie, D., Pan, J., and Singleton, K.}
\newblock Transform analysis and asset pricing for affine jump-diffusions.
\newblock {\em Econometrica: journal of the Econometric Society 68}, 6 (2000),
  1343–1376.

\bibitem{Dumas_Fleming_Whaley_1998}
{\sc Dumas, B., Fleming, J., and Whaley, R.~E.}
\newblock Implied volatility functions: {E}mpirical tests.
\newblock {\em The journal of finance 53}, 6 (1998), 2059–2106.

\bibitem{Duong_Swanson_2011}
{\sc Duong, D., and Swanson, N.~R.}
\newblock Volatility in discrete and continuous-time models: {A} survey with
  new evidence on large and small jumps.
\newblock In {\em Missing Data Methods: Time-Series Methods and Applications\/}
  (2011), Emerald Group Publishing Limited, p.~179–233.

\bibitem{Dupire_1993}
{\sc Dupire, B.}
\newblock Arbitrage pricing with stochastic volatility.
\newblock \url{https://cims.nyu.edu/~essid/ctf/stochvol.pdf}, 1993.
\newblock Accessed: 2023-08-30.

\bibitem{Dupire_1994}
{\sc Dupire, B.}
\newblock Pricing with a smile.
\newblock {\em Risk 7\/} (1994), 18–20.

\bibitem{Edwards_Lazzara_2011}
{\sc Edwards, T., and Lazzara, C.}
\newblock Realized volatility indices: Measuring market risk.
\newblock Research for S\&P Dow Jones Indices, McGraww Hill Financial, 2016.

\bibitem{Einstein_1905}
{\sc Einstein, A.}
\newblock Über die von der molekularkinetischen {T}heorie der {W}ärme
  geforderte {B}ewegung von in ruhenden {F}lüssigkeiten suspendierten
  {T}eilchen.
\newblock {\em Annalen der Physik 322}, 8 (1905), 549–560.

\bibitem{El_Euch_Fukasawa_Rosenbaum_2018}
{\sc El~Euch, O., Fukasawa, M., and Rosenbaum, M.}
\newblock The microstructural foundations of leverage effect and rough
  volatility.
\newblock {\em Finance and stochastics 22}, 2 (2018), 241–280.

\bibitem{El_Euch_Gatheral_Radoicic_Rosenbaum_2020}
{\sc El~Euch, O., Gatheral, J., Radoičić, R., and Rosenbaum, M.}
\newblock The {Z}umbach effect under rough {H}eston.
\newblock {\em Quantitative finance 20}, 2 (2020), 235–241.

\bibitem{El_Euch_Rosenbaum_2019}
{\sc El~Euch, O., and Rosenbaum, M.}
\newblock The characteristic function of rough {H}eston models.
\newblock {\em Mathematical Finance 29}, 1 (2019), 3–38.

\bibitem{Emanuel_MacBeth_1982}
{\sc Emanuel, D.~C., and MacBeth, J.~D.}
\newblock Further results on the constant elasticity of variance call option
  pricing model.
\newblock {\em Journal of financial and quantitative analysis 17}, 4 (1982),
  533.

\bibitem{Engle_1982}
{\sc Engle, R.~F.}
\newblock Autoregressive conditional heteroscedasticity with estimates of the
  variance of {U}nited {K}ingdom inflation.
\newblock {\em Econometrica: journal of the Econometric Society 50}, 4 (1982),
  987.

\bibitem{Eom_Kaizoji_Scalas_2019}
{\sc Eom, C., Kaizoji, T., and Scalas, E.}
\newblock Fat tails in financial return distributions revisited: {E}vidence
  from the {K}orean stock market.
\newblock {\em Physica A 526}, 121055 (2019), 121055.

\bibitem{Fama_1965}
{\sc Fama, E.~F.}
\newblock The behavior of stock-market prices.
\newblock {\em The Journal of Business 38}, 1 (1965), 34--105.

\bibitem{Fengler_2005}
{\sc Fengler, M.}
\newblock {\em Semiparametric modeling of implied volatility}, 2005~ed.
\newblock Springer, Berlin, Germany, 2005.

\bibitem{Figlewski_Wang_2001}
{\sc Figlewski, S., and Wang, X.}
\newblock Is the “leverage effect” a leverage effect?
\newblock {\em SSRN Electronic Journal\/} (2001).

\bibitem{Fleming_Ostdiek_Whaley_1995}
{\sc Fleming, J., Ostdiek, B., and Whaley, R.~E.}
\newblock Predicting stock market volatility: {A} new measure.
\newblock {\em Journal of futures markets 15}, 3 (1995), 265–302.

\bibitem{Fouque_Papanicolaou_Sircar_2000}
{\sc Fouque, J.-P., Papanicolaou, G., and Sircar, K.~R.}
\newblock {\em Derivatives in financial markets with stochastic volatility}.
\newblock Cambridge University Press, Cambridge, England, 2000.

\bibitem{Fouque_Papanicolaou_Sircar_Solna_2003}
{\sc Fouque, J.-P., Papanicolaou, G., Sircar, R., and Sølna, K.}
\newblock Multiscale stochastic volatility asymptotics.
\newblock {\em Multiscale modeling \& simulation 2}, 1 (2003), 22–42.

\bibitem{Fouque_Papanicolaou_Sircar_Solna_2004}
{\sc Fouque, J.-P., Papanicolaou, G., Sircar, R., and Sølna, K.}
\newblock Maturity cycles in implied volatility.
\newblock {\em Finance and stochastics 8}, 4 (2004).

\bibitem{Fouque_Saporito_2018}
{\sc Fouque, J.-P., and Saporito, Y.~F.}
\newblock Heston stochastic vol-of-vol model for joint calibration of {VIX} and
  {S\&P} 500 options.
\newblock {\em Quantitative finance 18}, 6 (2018), 1003–1016.

\bibitem{Francq_2019}
{\sc Francq, C.}
\newblock {\em GARCH models: structure, statistical inference and financial
  applications}, 2~ed.
\newblock Wiley-Blackwell, Hoboken, NJ, 2019.

\bibitem{Frey_1997}
{\sc Frey, R.}
\newblock Derivative asset analysis in models with level-dependent and
  stochastic volatility.
\newblock {\em Research Papers in Economics\/} (1997).

\bibitem{Fukasawa_2021}
{\sc Fukasawa, M.}
\newblock Volatility has to be rough.
\newblock {\em Quantitative finance 21}, 1 (2021), 1–8.

\bibitem{Fukasawa_Gatheral_2022}
{\sc Fukasawa, M., and Gatheral, J.}
\newblock A rough {SABR} formula.
\newblock {\em Frontiers of Mathematical Finance 1}, 1 (2022), 81.

\bibitem{Fukasawa_Takabatake_Westphal_2019}
{\sc Fukasawa, M., Takabatake, T., and Westphal, R.}
\newblock Is volatility rough?
\newblock {\em ArXiv:1905.04852\/} (2019).

\bibitem{Fukasawa_Takabatake_Westphal_2022}
{\sc Fukasawa, M., Takabatake, T., and Westphal, R.}
\newblock Consistent estimation for fractional stochastic volatility model
  under high‐frequency asymptotics.
\newblock {\em Mathematical Finance. An International Journal of Mathematics,
  Statistcs and Financial Economics 32}, 4 (2022), 1086–1132.

\bibitem{Funahashi_Kijima_2017}
{\sc Funahashi, H., and Kijima, M.}
\newblock Does the {H}urst index matter for option prices under fractional
  volatility?
\newblock {\em Annals of finance 13}, 1 (2017), 55–74.

\bibitem{Funahashi_Kijima_2017-1}
{\sc Funahashi, H., and Kijima, M.}
\newblock A solution to the time-scale fractional puzzle in the implied
  volatility.
\newblock {\em Fractal and fractional 1}, 1 (2017), 14.

\bibitem{Garnier_Sølna_2020}
{\sc Garnier, J., and Sølna, K.}
\newblock Optimal hedging under fast-varying stochastic volatility.
\newblock {\em SIAM journal on financial mathematics 11}, 1 (2020), 274–325.

\bibitem{Gastineau_1977}
{\sc Gastineau, G.~L.}
\newblock An index of listed option premiums.
\newblock {\em Financial analysts journal 33}, 3 (1977), 70–75.

\bibitem{Gatheral_2006}
{\sc Gatheral, J.}
\newblock {\em The volatility surface: A practitioner’s guide}.
\newblock John Wiley \& Sons, Nashville, TN, 2006.

\bibitem{Gatheral2008ConsistentMO}
{\sc Gatheral, J.}
\newblock Consistent modeling of {SPX} and {VIX} options.
\newblock Presented at The 5th World Congress of the Bachelier Finance Society,
  2008.

\bibitem{Gatheral_Jaisson_Rosenbaum_2014}
{\sc Gatheral, J., Jaisson, T., and Rosenbaum, M.}
\newblock Volatility is rough.
\newblock {\em Quantitative finance 18}, 6 (2018), 933–949.

\bibitem{Gatheral_Jusselin_Rosenbaum_2020}
{\sc Gatheral, J., Jusselin, P., and Rosenbaum, M.}
\newblock The quadratic rough {H}eston model and the joint {S\&P 500/VIX} smile
  calibration problem.
\newblock {\em SSRN Electronic Journal\/} (2020).

\bibitem{Ghysels_Harvey_Renault_1996}
{\sc Ghysels, E., Harvey, A.~C., and Renault, E.}
\newblock Stochastic volatility.
\newblock In {\em Handbook of Statistics\/} (1996), Elsevier, p.~119–191.

\bibitem{Gihman_Skorokhod_2004}
{\sc Gihman, I., and Skorokhod, A.}
\newblock {\em The theory of stochastic processes I}.
\newblock Springer Berlin Heidelberg, Berlin, Heidelberg, 2004.

\bibitem{Goutte_Ismail_Pham_2017}
{\sc Goutte, S., Ismail, A., and Pham, H.}
\newblock Regime-switching stochastic volatility model: estimation and
  calibration to {VIX} options.
\newblock {\em Applied mathematical finance 24}, 1 (2017), 38–75.

\bibitem{Guillaume_Dacorogna_Dave_Muller_Olsen_Pictet_1997}
{\sc Guillaume, D.~M., Dacorogna, M.~M., Davé, R.~R., Müller, U.~A., Olsen,
  R.~B., and Pictet, O.~V.}
\newblock From the bird’s eye to the microscope: {A} survey of new stylized
  facts of the intra-daily foreign exchange markets.
\newblock {\em Finance and stochastics 1}, 2 (1997), 95–129.

\bibitem{Guyon_2019}
{\sc Guyon, J.}
\newblock The joint {S\&P} 500/{VIX} smile calibration puzzle solved.
\newblock {\em SSRN Electronic Journal\/} (2019).

\bibitem{Guyon_RBM}
{\sc Guyon, J.}
\newblock On the joint calibration of {SPX} and {VIX} options: A
  dispersion-constrained martingale transport approach.
\newblock Presented at Research in Options 2019, IMPA, Rio de Janeiro, 2019.

\bibitem{Guyon_2021}
{\sc Guyon, J.}
\newblock Dispersion-constrained martingale {S}chrödinger problems and the
  exact joint {S\&P} 500/{VIX} smile calibration puzzle.
\newblock {\em SSRN Electronic Journal\/} (2021).

\bibitem{Guyon_Bourgey_2022}
{\sc Guyon, J., and Bourgey, F.}
\newblock Fast exact joint {S\&P} 500/{VIX} smile calibration in discrete and
  continuous time.
\newblock {\em SSRN Electronic Journal\/} (2022).

\bibitem{Guyon_El_Amrani_2022}
{\sc Guyon, J., and El~Amrani, M.}
\newblock Does the term-structure of equity at-the-money skew really follow a
  power law?
\newblock {\em SSRN Electronic Journal\/} (2022).

\bibitem{Guyon_Lekeufack_2023}
{\sc Guyon, J., and Lekeufack, J.}
\newblock Volatility is (mostly) path-dependent.
\newblock {\em Quantitative finance 23}, 9 (2023), 1221–1258.

\bibitem{Guyon_Mustapha_2022}
{\sc Guyon, J., and Mustapha, S.}
\newblock Neural joint {S\&P} 500/{VIX} smile calibration.
\newblock {\em SSRN Electronic Journal\/} (2022).

\bibitem{Hagan}
{\sc Hagan, P., Kumar, D., Lesniewski, A., and Woodward, D.}
\newblock Managing smile risk.
\newblock {\em Wilmott Magazine 1\/} (01 2002), 84--108.

\bibitem{Harms_Stefanovits_2019}
{\sc Harms, P., and Stefanovits, D.}
\newblock Affine representations of fractional processes with applications in
  mathematical finance.
\newblock {\em Stochastic processes and their applications 129}, 4 (2019),
  1185–1228.

\bibitem{Harrison_Kreps_1979}
{\sc Harrison, J.~M., and Kreps, D.~M.}
\newblock Martingales and arbitrage in multiperiod securities markets.
\newblock {\em Journal of Economic Theory 20}, 3 (1979), 381–408.

\bibitem{Harrison_Pliska_1981}
{\sc Harrison, J.~M., and Pliska, S.~R.}
\newblock Martingales and stochastic integrals in the theory of continuous
  trading.
\newblock {\em Stochastic processes and their applications 11}, 3 (1981),
  215–260.

\bibitem{Hasanhodzic_Lo_2019}
{\sc Hasanhodzic, J., and Lo, A.~W.}
\newblock On {B}lack’s leverage effect in firms with no leverage.
\newblock {\em The Journal of Portfolio Management 46}, 1 (2019), 106–122.

\bibitem{Heston_1993}
{\sc Heston, S.~L.}
\newblock A closed-form solution for options with stochastic volatility with
  applications to bond and currency options.
\newblock {\em Review of Financial Studies 6}, 2 (1993), 327--43.

\bibitem{Hull_White_1987}
{\sc Hull, J., and White, A.}
\newblock The pricing of options on assets with stochastic volatilities.
\newblock {\em The journal of finance 42}, 2 (1987), 281–300.

\bibitem{Jackwerth_Rubinstein_1996}
{\sc Jackwerth, J.~C., and Rubinstein, M.}
\newblock Recovering probability distributions from option prices.
\newblock {\em The journal of finance 51}, 5 (1996), 1611.

\bibitem{Jacquier_Martini_Muguruza_2018}
{\sc Jacquier, A., Martini, C., and Muguruza, A.}
\newblock On {VIX} futures in the rough {B}ergomi model.
\newblock {\em Quantitative finance 18}, 1 (2018), 45–61.

\bibitem{Javaheri_2004}
{\sc Javaheri, A.}
\newblock {\em The volatility process: A study of stock market dynamics via
  para- metric stochastic volatility models and a comparison to the information
  embedded in option prices}.
\newblock PhD thesis, Ecole de Mines de Paris, Paris, 2004.

\bibitem{Jovanovic_Le_Gall_2001}
{\sc Jovanovic, F., and Le~Gall, P.}
\newblock Does god practice a random walk? {T}he “financial physics” of a
  nineteenth-century forerunner, {J}ules {R}egnault.
\newblock {\em The European journal of the history of economic thought 8}, 3
  (2001), 332–362.

\bibitem{Kokholm_Stisen_2015}
{\sc Kokholm, T., and Stisen, M.}
\newblock Joint pricing of {VIX} and {SPX} options with stochastic volatility
  and jump models.
\newblock {\em Journal of Risk Finance 16}, 1 (2015), 27--48.

\bibitem{Kolmogorov_1940}
{\sc Kolmogorov, A.}
\newblock Wiener spirals and some other interesting curves in {H}ilbert space.
\newblock {\em Doklady AN SSSR 26}, 2 (1940).

\bibitem{Kreps_1981}
{\sc Kreps, D.~M.}
\newblock Arbitrage and equilibrium in economies with infinitely many
  commodities.
\newblock {\em Journal of mathematical economics 8}, 1 (1981), 15–35.

\bibitem{Kubilius_Mishura_Ralchenko_2017}
{\sc Kubilius, K., Mishura, Y., and Ralchenko, K.}
\newblock {\em Parameter estimation in fractional diffusion models}.
\newblock Springer International Publishing, Cham, 2017.

\bibitem{Kwok_Zheng_2022}
{\sc Kwok, Y.~K., and Zheng, W.}
\newblock {\em Pricing models of volatility products and exotic variance
  derivatives}.
\newblock Chapman \& Hall/CRC, Philadelphia, PA, 2022.

\bibitem{Lacombe_Muguruza_Stone_2021}
{\sc Lacombe, C., Muguruza, A., and Stone, H.}
\newblock Asymptotics for volatility derivatives in multi-factor rough
  volatility models.
\newblock {\em Mathematics and financial economics 15}, 3 (2021), 545–577.

\bibitem{Latane_Rendleman_1976}
{\sc Latané, H.~A., and Rendleman, Richard~J., J.}
\newblock Standard deviations of stock price ratios implied in option prices.
\newblock {\em The journal of finance 31}, 2 (1976), 369–381.

\bibitem{Lee_2006}
{\sc Lee, R.~W.}
\newblock Implied volatility: {S}tatics, dynamics, and probabilistic
  interpretation.
\newblock In {\em Recent Advances in Applied Probability\/} (Boston, 2006),
  Kluwer Academic Publishers, p.~241–268.

\bibitem{Lewis_2000}
{\sc Lewis, A.~L.}
\newblock {\em Option Valuation under Stochastic Volatility}.
\newblock Finance Press, 2000.

\bibitem{Linetsky_Mendoza_2010}
{\sc Linetsky, V., and Mendoza, R.}
\newblock Constant elasticity of variance ({CEV}) diffusion model.
\newblock {\em Encyclopedia of Quantitative Finance\/} (Feb 2010).

\bibitem{Lobato_Velasco_2000}
{\sc Lobato, I.~N., and Velasco, C.}
\newblock Long memory in stock-market trading volume.
\newblock {\em Journal of business \& economic statistics: a publication of the
  {A}merican {S}tatistical {A}ssociation 18}, 4 (2000), 410--427.

\bibitem{Mandelbrot_1963}
{\sc Mandelbrot, B.}
\newblock The variation of certain speculative prices.
\newblock {\em The Journal of Business 36}, 4 (1963), 394--419.

\bibitem{Mandelbrot_Van_Ness_1968}
{\sc Mandelbrot, B.~B., and Van~Ness, J.~W.}
\newblock Fractional brownian motions, fractional noises and applications.
\newblock {\em SIAM review 10}, 4 (1968), 422–437.

\bibitem{Mariani_2020}
{\sc Mariani, M.~C., and Florescu, I.}
\newblock {\em Quantitative Finance}.
\newblock John Wiley \& Sons, Nashville, TN, 2020.

\bibitem{Markowitz_1952}
{\sc Markowitz, H.}
\newblock Portfolio selection.
\newblock {\em The journal of finance 7}, 1 (1952), 77.

\bibitem{Martin_2011}
{\sc Martin, I.}
\newblock Simple variance swaps.
\newblock NBER working paper series, 2011.

\bibitem{Melino_Turnbull_1990}
{\sc Melino, A., and Turnbull, S.~M.}
\newblock Pricing foreign currency options with stochastic volatility.
\newblock {\em Journal of Econometrics 45}, 1–2 (1990), 239–265.

\bibitem{Melino_Turnbull_1995}
{\sc Melino, A., and Turnbull, S.~M.}
\newblock Misspecification and the pricing and hedging of long-term foreign
  currency options.
\newblock {\em Journal of international money and finance 14}, 3 (1995),
  373–393.

\bibitem{Merino_Pospisil_Sobotka_Sottinen_Vives_2021}
{\sc Merino, R., Pospíšil, J., Sobotka, T., Sottinen, T., and Vives, J.}
\newblock Decomposition formula for rough {V}olterra stochastic volatility
  models.
\newblock {\em International journal of theoretical and applied finance 24}, 02
  (2021), 2150008.

\bibitem{Merton_1973}
{\sc Merton, R.~C.}
\newblock Theory of rational option pricing.
\newblock {\em The Bell journal of economics and management science 4}, 1
  (1973), 141.

\bibitem{Mikosch_Starica_2000}
{\sc Mikosch, T., and Starica, C.}
\newblock Is it really long memory we see in financial returns?
\newblock In {\em Extremes and Integrated Risk Management\/} (2000), Risk
  Books, pp.~149--168.

\bibitem{Mishura_2007}
{\sc Mishura, Y.}
\newblock {\em Stochastic calculus for fractional {B}rownian motion and related
  processes}, 2008~ed.
\newblock Springer, Berlin, Germany, 2007.

\bibitem{MYuT2018}
{\sc Mishura, Y., and Yurchenko-Tytarenko, A.}
\newblock Fractional {C}ox–{I}ngersoll–{R}oss process with non-zero
  ``mean''.
\newblock {\em Modern Stochastics: Theory and Applications 5}, 1 (2018),
  99--111.

\bibitem{MYuT2019}
{\sc Mishura, Y., and Yurchenko-Tytarenko, A.}
\newblock Fractional {C}ox–{I}ngersoll–{R}oss process with small {H}urst
  indices.
\newblock {\em Modern Stochastics: Theory and Applications 6}, 1 (2018),
  13--39.

\bibitem{MYT2020}
{\sc Mishura, Y., and Yurchenko-Tytarenko, A.}
\newblock Approximating expected value of an option with non-{L}ipschitz payoff
  in fractional {H}eston-type model.
\newblock {\em International Journal of Theoretical and Applied Finance 23}, 05
  (July 2020), 2050031.

\bibitem{Nelson_1991}
{\sc Nelson, D.~B.}
\newblock Conditional heteroskedasticity in asset returns: A new approach.
\newblock {\em Econometrica: journal of the Econometric Society 59}, 2 (1991),
  347--370.

\bibitem{Nualart_2006}
{\sc Nualart, D.}
\newblock {\em The Malliavin calculus and related topics}.
\newblock Springer-Verlag, Berlin/Heidelberg, 2006.

\bibitem{Ocone_Karatzas_1991}
{\sc Ocone, D.~L., and Karatzas, I.}
\newblock A generalized {C}lark representation formula, with application to
  optimal portfolios.
\newblock {\em Stochastics and stochastics reports 34}, 3–4 (1991),
  187–220.

\bibitem{Osborne_1956}
{\sc Osborne, M. F.~M.}
\newblock Brownian motion in the stock market.
\newblock {\em Operations Research 7}, 2 (1959), 145--173.

\bibitem{Prause_1998}
{\sc Prause, K.}
\newblock The generalized hyperbolic model: {E}stimation, financial
  derivatives, and risk measures.

\bibitem{Rachev_Kim_Bianchi_Fabozzi_2011}
{\sc Rachev, S.~T., Kim, Y.~S., Bianchi, M.~L., and Fabozzi, F.~J.}
\newblock {\em Financial models with Lévy processes and volatility
  clustering}.
\newblock John Wiley \& Sons, Chichester, England, 2011.

\bibitem{Regnault_1863}
{\sc Regnault, J.}
\newblock {\em Calcul des Chances et Philosophie de la Bourse}.
\newblock Mallet Bachelier and Castel, 1863.

\bibitem{Renault_Touzi_1996}
{\sc Renault, E., and Touzi, N.}
\newblock Option hedging and implied volatilities in a stochastic volatility
  model.
\newblock {\em Mathematical Finance 6}, 3 (1996), 279–302.

\bibitem{Revuz_Yor_1999}
{\sc Revuz, D., and Yor, M.}
\newblock {\em Continuous Martingales and Brownian Motion}, 3~ed.
\newblock Springer, 1999.

\bibitem{Rogers_2019}
{\sc Rogers, L. C.~G.}
\newblock Things we think we know (working paper).
\newblock
  \url{https://www.skokholm.co.uk/wp-content/uploads/2019/11/TWTWKpaper.pdf},
  2019.

\bibitem{Rosenbaum_2008}
{\sc Rosenbaum, M.}
\newblock Estimation of the volatility persistence in a discretely observed
  diffusion model.
\newblock {\em Stochastic processes and their applications 118}, 8 (2008),
  1434–1462.

\bibitem{Rosenbaum_Zhang_2021}
{\sc Rosenbaum, M., and Zhang, J.}
\newblock Deep calibration of the quadratic rough {H}eston model.
\newblock {\em ArXiv:2107.01611\/} (2021).

\bibitem{Ross_1978}
{\sc Ross, S.~A.}
\newblock A simple approach to the valuation of risky streams.
\newblock {\em The Journal of Business 51}, 3 (1978), 453--475.

\bibitem{Ross_1987}
{\sc Ross, S.~A.}
\newblock Finance.
\newblock In {\em The New Palgrave Dictionary of Economics\/} (London, 1987),
  J.~Eatwell, M.~Milgate, and P.~Newman, Eds., vol.~2, Macmillan, pp.~322--336.

\bibitem{Rømer_2022}
{\sc Rømer, S.~E.}
\newblock Empirical analysis of rough and classical stochastic volatility
  models to the {SPX} and {VIX} markets.
\newblock {\em Quantitative finance 22}, 10 (2022), 1805–1838.

\bibitem{Samorodnitsky_2016}
{\sc Samorodnitsky, G.}
\newblock {\em Stochastic processes and long range dependence}.
\newblock Springer International Publishing, Cham, 2016.

\bibitem{Samuelson_1965}
{\sc Samuelson, P.~A.}
\newblock Rational theory of warrant pricing.
\newblock {\em Industrial Management Review 6}, 2 (1959), 13--39.

\bibitem{Samuelson_1973}
{\sc Samuelson, P.~A.}
\newblock Mathematics of speculative price.
\newblock {\em SIAM review 15}, 1 (1973), 1–42.

\bibitem{Samuelson_Merton_1972}
{\sc Samuelson, P.~A., and Merton, R.~C.}
\newblock A complete model of warrant pricing that maximizes utility.
\newblock {\em Industrial Management Review 10\/} (1972), 17--46.

\bibitem{Schachermayer_2013}
{\sc Schachermayer, W.}
\newblock The fundamental theorem of asset pricing.
\newblock In {\em Handbook of the Fundamentals of Financial Decision Making\/}
  (2013), World Scientific, p.~31–48.

\bibitem{SZh1999}
{\sc Schöbel, R., and Zhu, J.}
\newblock Stochastic volatility with an {O}rnstein–{U}hlenbeck process: An
  extension.
\newblock {\em Review of finance 3}, 1 (1999), 23–46.

\bibitem{Scott_1987}
{\sc Scott, L.~O.}
\newblock Option pricing when the variance changes randomly: Theory,
  estimation, and an application.
\newblock {\em Journal of financial and quantitative analysis 22}, 4 (1987),
  419.

\bibitem{Shephard_Andersen_2009}
{\sc Shephard, N., and Andersen, T.~G.}
\newblock Stochastic volatility: Origins and overview.
\newblock In {\em Handbook of Financial Time Series\/} (Berlin, Heidelberg,
  2009), Springer Berlin Heidelberg, p.~233–254.

\bibitem{Shiryaev_1999}
{\sc Shiryaev, A.~N.}
\newblock {\em Essentials of stochastic finance: {F}acts, models, theory}.
\newblock World Scientific Publishing, Singapore, 1999.

\bibitem{Skiadopoulos_2001}
{\sc Skiadopoulos, G.}
\newblock Volatility smile consistent option models: {A} survey.
\newblock {\em International journal of theoretical and applied finance 04}, 03
  (2001), 403–437.

\bibitem{Sprenkle_1961}
{\sc Sprenkle, C.~M.}
\newblock Warrant prices as indicators of expectations and preferences.
\newblock {\em Yale economic essays 1}, 2 (1959), 179--231.

\bibitem{Stein_Stein_1991}
{\sc Stein, E.~M., and Stein, J.~C.}
\newblock Stock price distributions with stochastic volatility: {A}n analytic
  approach.
\newblock {\em Review of Financial Studies 4\/} (1991), 727--752.

\bibitem{Thorp_Kassouf_1967}
{\sc Thorp, E.~O., and Kassouf, S.~T.}
\newblock {\em Beat the Market: A Scientific Stock Market System}.
\newblock Random House, New York, 1967.

\bibitem{Wiggins_1987}
{\sc Wiggins, J.~B.}
\newblock Option values under stochastic volatility: Theory and empirical
  estimates.
\newblock {\em Journal of financial economics 19}, 2 (1987), 351–372.

\bibitem{Willinger_Taqqu_Teverovsky_1999}
{\sc Willinger, W., Taqqu, M.~S., and Teverovsky, V.}
\newblock Stock market prices and long-range dependence.
\newblock {\em Finance and stochastics 3}, 1 (1999), 1–13.

\bibitem{YW1970}
{\sc Yamada, T., and Watanabe, S.}
\newblock On the uniqueness of solutions of stochastic differential equations.
\newblock {\em Kyoto journal of mathematics 11}, 1 (1971), 155–167.

\bibitem{Zumbach_2010}
{\sc Zumbach, G.}
\newblock Volatility conditional on price trends.
\newblock {\em Quantitative finance 10}, 4 (2010), 431–442.

\bibitem{Zumbach_2012}
{\sc Zumbach, G.}
\newblock {\em Discrete time series, processes, and applications in finance},
  2013~ed.
\newblock Springer, Berlin, Germany, 2012.

\end{thebibliography}

\end{document}